\documentclass[a4paper,twosides]{article}
\usepackage{CJK,multicol,multirow,graphics,fancyhdr,epstopdf}
\usepackage{amsmath,amsfonts,amssymb,bm,upgreek,mathrsfs,ccmap,mathcomp}
\usepackage[pagewise,switch,columnwise]{lineno}
\usepackage[compress,nospace]{cite}
\usepackage[dvipsnames]{xcolor}
\usepackage{graphicx}

\usepackage[
  a4paper,
  left=1.8cm,
  right=1.8cm
]{geometry}

\usepackage{tocloft}

\newcommand{\ucite}[1]{\cite{#1}}

\newcommand{\fl}[1]{\label{fig:#1}}
\newcommand{\figcaption}[3]{%
  \par\medskip
  \begin{center}
    \small\textbf{FIG.~#2.}~#3
  \end{center}
  \label{fig:#1}%
  \par\medskip
}

\begin{document}
\begin{CJK}{GBK}{song}\vspace*{-4mm}
\begin{center}
%-------------------------Title----------------------------
\large\bf{\boldmath{Universal Drude Weights in One-dimensional Repulsive Fermi Gas}}
%------------------------Footnote--------------------------
\footnotetext{\hspace*{-5.4mm}$^{*}$Corresponding authors. Emails: $\dagger$ ybzhang@zstu.edu.cn; $*$ xwe105@wipm.ac.cn}
\\[5mm]
%------------------------Authors----------------------------
\normalsize \rm{}Xiangguo Yin$^{1,2}$, Dawei Niu$^{1}$, Shuang Zhang$^{1}$, Shizhong Zhang$^{3}$, Yunbo Zhang$^{4\dagger}$, and Xi-Wen Guan$^{5,6*}$
%----------------------COM. or University-------------------
\\[3mm]\small\sl $^{1}$Institute of Theoretical Physics and State Key Laboratory of Quantum Optics and Quantum Optics Devices, Shanxi University, Taiyuan 030006, China

$^{2}$Collaborative Innovation Center of Extreme Optics, Shanxi University, Taiyuan, Shanxi 030006, China

$^{3}$Department of Physics and State Key Laboratory of Optical Quantum Materials, University of Hong Kong, Hong Kong, China;
 Hong Kong Institute of Quantum Science and Technology, University of Hong Kong, Hong Kong, China

$^{4}$Department of Physics and Key Laboratory of Optical Field Manipulation of Zhejiang Province, Zhejiang Sci-Tech University, Hangzhou 310018, China

$^{5}$Innovation Academy for Precision Measurement Science and Technology, Chinese Academy of Sciences, Wuhan 430071, China

$^{6}$Department of Fundamental and Theoretical Physics, Research School of Physics, Australian National University, Canberra ACT 0200, Australia
%------------------------Received date----------------------
\\[4mm]\normalsize\rm{}(Dated: \today)
\end{center}
\end{CJK}
%----------------------Abstract and PACS--------------------
\vskip 1.5mm

\small{
\narrower\space
Building on the Bethe ansatz and generalized hydrodynamics, we rigorously establish universal relations between Drude weights governing charge, spin, and energy transport and thermodynamic properties for the one-dimensional repulsive two-component continuum Fermi gas - the paradigmatic integrable Yang-Gaudin model.
We analytically derive thermodynamic expressions for charge Drude weights:
 $D_{nn}$, $D_{nm}$, $D_{ne}$ (responses of particle, magnetization and energy currents to a chemical potential gradient) equal the particle density, magnetization, and temperature times entropy density, respectively.
These relations constitute a hallmark of ballistic transport, and can be generalized to other 1D continuum integrable systems.
Furthermore, we investigate spin Drude weight ($D_{mm}$, $D_{me}$) at zero and low temperatures.
They characterize the magnetization and energy current responses to a magnetic field gradient, revealing an essential spin-charge coupling feature in quantum transport.
We find that the Drude weights $D_{nm}$, $D_{mm}$, and $D_{me}$ across the $\mu\text{-}H$ plane remarkably map out the zero-temperature phase boundary.
This work fills a critical gap in the understanding of transport within integrable quantum gases and furnishes a direct theoretical foundation for future ultracold-atom experiments.

\par
}

\vskip 5mm
%-------------------TEXT TEXT TEXT TEXT---------------------
% ====== 从 Introduction 起切换为双栏 ======

\begin{multicols}{2}
{\it Introduction.} 
The Drude Weight (DW) represents a fundamental transport coefficient in condensed matter physics, characterizing the ballistic (dissipationless) component of charge, spin, energy or other conserve charge  transport in quantum many-body systems\ucite{1}. 
Originally developed to describe charge stiffness in solid-state materials, it quantifies the ratio of mobile charge carrier density to its effective mass. Crucially, the DW serves as an important diagnostic for distinguishing metallic and insulating phases: metals exhibit finite DWs in the thermodynamic limit, while insulators show vanishing values\ucite{2,3}. 
In recent years, its properties have been extensively investigated within quantum integrable models\ucite{4}, such as the one-dimensional (1D) XXZ spin chain\ucite{5,6,7,8,9}, the 1D Hubbard model\ucite{10,11,12,13}, and the sine-Gordon model \ucite{14}.

Significant progress in understanding DWs has been made, particularly with the development of generalized hydrodynamics (GHD) for integrable systems\ucite{15,16,17,18,19,20,21,22}. 
For 1D interacting systems such as the Lieb-Liniger Bose gas, theoretical progress using the Thermodynamic Bethe Ansatz (TBA) has revealed elegant integral representations of DWs\ucite{23}. 
These representations connect DWs to quasiparticle rapidity distributions and are equivalent to current derivatives, reflecting linear response in bipartition systems subjected to small bias potentials.
 Experimental studies with ultracold gases, employing protocols such as constant force or bipartition setups, have validated these theoretical predictions for 1D Bose systems\ucite{24}. 
 These experiments confirm that DWs provide a comprehensive characterization of ballistic dynamics. 
 Furthermore, Ref. \cite{24} experimentally demonstrated that the DW characterizing the linear response of particle current to the variation of chemical potential  is directly proportional to the particle density. 
 Theoretical advances have also established a key  property: in monoatomic gases, charge DW exhibits temperature independence and scale linearly with particle density, which follows from Galilei invariance \ucite{25,Chandak:2026,Liu:2026}.

Despite these advancements, a significant gap remains in the comprehensive understanding of DWs for 1D multi-component continuous  Fermi gases, particularly those described by the Yang-Gaudin model ($\delta$-function interacting fermions)\ucite{26,27,28,29,30,31,32,33,34,35}. 
Although the two-component Yang-Gaudin Fermi gas is solved by Bethe ansatz, existing literature primarily focuses on Bose gases or addresses only particle-current-related DWs\ucite{23}. 
A systematic derivation of DWs encompassing spin, particle, and energy transport for these Fermionic  systems is notably lacking. 
Moreover, the  behavior of these DWs across canonical and grand canonical ensembles, specifically, their dependence on interaction strength, spin polarization, and chemical potential, has not been thoroughly characterized. 
This represents a significant gap in the understanding of  quantum transport in strongly correlated fermions.

This paper aims to fill this gap  through a comprehensive investigation of DWs in 1D repulsive Fermi gases.
 First, we establish a general  theoretical framework by deriving charge DW as thermodynamic quantities using two complementary approaches: a linear response  form under small chemical potential gradients (via generalized hydrodynamics) and an integral representation in rapidity space (via the TBA equations). 
This combined approach ensures the consistency and reliability of our theoretical results. 
Second, we systematically characterize the spin DWs in both canonical and grand canonical ensembles, revealing how key parameters, including interaction strength, spin polarization, and chemical potential, govern the particle, magnetization, and energy related DWs in these strongly correlated Fermi systems.

{\it The Yang-Gaudin Model and its equilibrium state.}
We consider the 1D $\delta$-function interacting spin-$1/2$ Fermi gas, widely known as the Yang-Gaudin model\ucite{26,27}. This exactly solvable system describes $N$ fermions interacting via a point-like potential with Hamiltonian:
\begin{equation}
H=-\frac{\hbar^{2}}{2m}\sum_{i=1}^{N}\frac{\partial^{2}}{\partial x_{i}^{2}}+g_{\textrm{1D}}\sum_{1\leq i<j\leq N}\delta\left(x_{i}-x_{j}\right)-\mu N-HM^{z},
\label{eq:H}
\end{equation}
in which the first term is the kinetic energy, the second term is the interaction energy between particles with different spins, the last two terms are the potential energies, $m$ is the particle mass, $x_i$ denotes spatial coordinate, and $g_{\textrm{1D}}=\hbar^{2}c/m$ characterizes the interaction strength.

The system comprises $N_{\uparrow}$ spin-up and $N_{\downarrow}$ spin-down fermions, with total particle number $N=N_{\uparrow}+N_{\downarrow}$ and magnetization $M^{z}=(N_{\uparrow}-N_{\downarrow})/2$.
 The chemical potential $\mu$ and magnetic field $H$ serve as external thermodynamic potentials that couple to particle number and magnetization, respectively. 
 This Bethe ansatz solvable model provides a powerful framework for analyzing many-body properties. Setting $\hbar=1$, we define the particle density $n=N/L$ and magnetization density $m^{z}=M^{z}/L$, where $L$ is the system length. 
 These densities are essential for characterizing transport properties in quantum many-body systems.
 
In the thermodynamic limit, the equilibrium state is characterized by the pseudoenergies $\epsilon_1(k)$ and $\epsilon_{2,l}(k)$, which satisfy the Yang--Yang integral equations \ucite{28}
\begin{equation}\label{eq:YY_intro}
\begin{aligned}
\epsilon_{1}(k) &= \frac{k^{2}}{2m}-\mu-\frac{H}{2} - T\sum_{l=1}^{\infty}\phi_{l}*\ln\!\bigl(1+e^{-\epsilon_{2,l}/T}\bigr)(k), \\[2mm]
\epsilon_{2,l}(k) &= lH - T\phi_{l}*\ln\!\bigl(1+e^{-\epsilon_{1}/T}\bigr)(k) \\
&\quad + T\sum_{m=1}^{\infty}\Phi_{l,m}*\ln\!\bigl(1+e^{-\epsilon_{2,m}/T}\bigr)(k),
\end{aligned}
\end{equation}
with kernels $\phi_l(k)=\dfrac{2lc}{\pi[(lc)^2+4k^2]}, \quad l=1,2,\ldots$ and $\Phi_{l,m}$ as defined in the Supplemental Material (SM)\ucite{36}. Here $k$ is the rapidity, which can also be interpreted as a generalized quasi-momentum or wave number. The occupation numbers follow from the pseudoenergies through the Fermi--Dirac form
\begin{equation}
\label{eq:theta_intro}
\theta_{\alpha}(k) = \left[1 + e^{\epsilon_{\alpha}(k)/T}\right]^{-1},
\end{equation}
where \(\alpha\) labels the quasiparticle species: \(\alpha=1\) for the charge branch and \(\alpha=(2,l)\) for the length-\(l\) spin string (\(l=1,2,\ldots\)). The particle density \(n \) and the magnetization density \(m^z \) are expressed in terms of the rapidity distribution functions as
\begin{equation}
\label{eq:n_mz}
n = \int_{-\infty}^{\infty} \mathrm{d}k \, \rho_{1}(k),
\quad
m^{z} = \frac{n}{2} - \sum_{l=1}^{\infty} l \int_{-\infty}^{\infty} \mathrm{d}k \, \rho_{2,l}(k),
\end{equation}
where \(\rho_{1}(k)\) is the distribution function of the charge rapidity \(k\), and \(\rho_{2,l}(k)\) denotes the distribution function of the length-\(l\) spin string. The rapidity distribution functions are obtained from the total densities of states via \(\rho_{\alpha}(k)=\theta_{\alpha}(k)\rho_{\alpha}^{\mathrm{tot}}(k)\), with the total densities defined through the dressed relation
\begin{equation}
\label{eq:rho_tot_intro}
\vec{q}_n^{\,\mathrm{dr}} = 2\pi\vec{\rho}^{\,\mathrm{tot}},
\end{equation}
where $\vec{q}_{n} = \{1, 0\}$ and \(\vec{\rho}^{\,\mathrm{tot}}=\{\rho_1^{\mathrm{tot}},\rho_{2,l}^{\mathrm{tot}}\}\). For any set of bare charges \(\vec{q}=\{q_1(k),q_{2,l}(k)\}\), the corresponding interaction dressed charges \(\vec{q}^{\,\mathrm{dr}}=\{q_1^{\mathrm{dr}}(k),q_{2,l}^{\mathrm{dr}}(k)\}\) are determined by the linear integral equations
\begin{equation}
\label{eq:dressed_intro}
\begin{aligned}
q_{1}^{\mathrm{dr}}(k) &= q_{1}(k) + \sum_{l=1}^{\infty}\phi_{l} \ast \bigl(q_{2,l}^{\mathrm{dr}} \theta_{2,l}\bigr)(k), \\[2mm]
q_{2,l}^{\mathrm{dr}}(k) &= q_{2,l}(k) + \phi_{l} \ast \bigl(q_{1}^{\mathrm{dr}} \theta_{1}\bigr)(k) \\
&\quad - \sum_{m=1}^{\infty}\Phi_{l,m} \ast \bigl(q_{2,m}^{\mathrm{dr}} \theta_{2,m}\bigr)(k).
\end{aligned}
\end{equation}
Finally, the derivatives of the pseudoenergies with respect to the thermodynamic potentials are directly related to the corresponding conserved quantities, namely, dressed charge density \( \vec{q}_n^{\,\mathrm{dr}}=\partial_{-\mu}\vec{\epsilon} \),  dressed magnetization \(\vec{q}_m^{\,\mathrm{dr}}=\partial_{-H}\vec{\epsilon} \), dressed momentum \(\vec{q}_k^{\,\mathrm{dr}}=m \partial_k\vec{\epsilon} \), and dressed energy \(\vec{q}_e^{\,\mathrm{dr}}=\partial_{1/T}\vec{\epsilon}/T \) with $\vec{q}_{m}=\{1/2, -l\}$,  
$\vec{q}_{k}=\{k, 0\}$,  $\vec{q}_{e}=\{k^2/2m - \mu - H/2, lH\}$. These relations connect the microscopic Bethe ansatz data to the macroscopic transport coefficients.

{\it Thermodynamic Expression of  Charge DW from Tilted Chemical Potential.}
The charge DW characterizes ballistic transport in integrable systems and can be derived through the tilted chemical potential approach\ucite{6}. Consider a system initially in thermal equilibrium with homogeneous chemical potential, subsequently subjected to a linear potential $\mu(x) = \phi_n x$ that induces constant force driving particles transport. The DW quantifies the linear current response\ucite{24}:
\begin{equation}
D_{ni} = \lim_{\phi_n \rightarrow 0} \lim_{t \rightarrow \infty} \frac{j_i(t)}{t \phi_n},
\label{eq:Dni}
\end{equation}
where $j_i$ denotes particle ($i=n$), magnetization ($i=m$), or energy ($i=e$) currents and $t$ is the time.

The time evolution of the system follows GHD\ucite{11}
\begin{equation}
\left(\partial_t +v_\alpha \partial_x +a_\alpha\, \partial_{k/m}\right) \theta_\alpha(t, k)= 0,
\label{eq:GHD}
\end{equation}
 describing quasiparticle occupation numbers $\theta_\alpha$ in phase ($x,k$) space.  Here $v_\alpha$ is the effective velocity, and $a_\alpha$ is the effective acceleration. Since the occupation number is uniform at every position in coordinate space for the initial state, and  GHD Eq. \eqref{eq:GHD} preserve spatial translational invariance, the occupation number is independent of the spatial coordinate $x$ at arbitrary time, such that GHD Eq. \eqref{eq:GHD} simplify to:
\begin{equation}
\left(\partial_t + a\, \partial_{k/m}\right) \theta_\alpha(t, k)= 0,
\label{eq:GHD_simple}
\end{equation}
with the effective acceleration ($a = \phi_n/m$), representing uniform force on all quasiparticles. The above partial differential Eq. \eqref{eq:GHD_simple} can be transformed into an ordinary differential equation $\mathrm{d} \theta_\alpha(t, k(t))/\mathrm{d}t=0$ along characteristics $k(t) = k(0) + \phi_n t$, indicating quasiparticles undergo uniform acceleration while preserving occupation numbers:
\begin{equation}
\theta_\alpha(t, k) = \theta_\alpha(0, k - \phi_n t).
\label{eq:theta_evol}
\end{equation}
This constitutes a distinct manifestation of ballistic transport. From the relation between the occupation numbers at $t=0$ and arbitrary time $t$ in the above Eq. \eqref{eq:theta_evol}, the particle current at time $t$ relates to initial state occupation numbers, yielding purely ballistic contribution:
\begin{equation}
j_n(t) = \frac{n}{m} t \phi_n,
\end{equation}
and the charge DW:
\begin{equation}
D_{nn} = \frac{n}{m}. 
\label{eq:Dnn_tilted}
\end{equation}
Detailed derivations are given in SM \ucite{36}.
This reflects the ballistic transport behavior of charge, i.e., a dissipationless fluid of quasiparticles that is independent of interaction strength and temperature.
On the other hand, the Drude weight \(D_{nn}\) can also be equivalently defined via the second derivative of the pressure \ucite{25}
\begin{equation}
D_{nn} = 4\, \partial_{\lambda}^2 p \big|_{\lambda=0,\mu,H}
\label{eq:Dnnp}
\end{equation}
in the grand canonical ensemble in which $p$ is the pressure and $\lambda$ is the potential coupled to total momentum. This expression is equivalent to the second derivative of energy in the canonical ensemble under gauge field. Following a similar procedure as in Ref.~\cite{25}, we obtain the identical result given in Eq.~\eqref{eq:Dnn_tilted}. Our  system features a uniform external potential and contact interactions of the delta-function, which guarantees Galilei invariance. Consequently, the system possesses continuous translational symmetry, leading to the conservation of total momentum and ballistic transport behavior, see  detailed discussion in SM  \ucite{36}. 

The magnetization current, describing spin transport, gives:
\begin{equation}
D_{nm} = \frac{m^z}{m}.
\label{eq:Dnm_tilted}
\end{equation}
This spin  DW also matches non-interacting systems and exhibits interaction- and temperature-independent ballistic transport.

For energy transport (the sum of kinetic energy, interaction energy, and potential energy), the energy current leads to:
\begin{equation}
D_{ne} = \frac{Ts}{m},
\label{eq:Dne_tilted}
\end{equation}
where $T$ is the temperature and $s$ is the entropy density, revealing that finite-temperature energy transport fundamentally relates to entropy flow. At zero temperature, $D_{ne}$ vanishes. Increasing temperature enhances entropy, consequently increasing $D_{ne}$. For fixed particle density, enhanced interaction strength and polarization ($P=2m^z/n$) reduce available microstates, and therefore decrease the  entropy and$D_{ne}$. These results demonstrate profound connections between ballistic transport coefficients and thermodynamic quantities in integrable systems in the continuous pace.

{\it Thermodynamic Expression of  Charge DW from Rapidity-Space Integration.} 
Starting from the current-current correlation function, the DW admits an alternative formulation in rapidity space as a symmetric matrix\ucite{23}:
\begin{equation}
D_{ij} = \frac{1}{T} \sum_{\alpha} \int \mathrm{d}k\, \rho_\alpha(1-\theta_\alpha) v_\alpha^2 (\vec{q}_i^{\mathrm{dr}})_\alpha (\vec{q}_j^{\mathrm{dr}})_\alpha,
\label{eq:Dij_rapidity}
\end{equation}
with $i,j=n,m,e$.
Here we suppress the explicit dependence on the rapidity $k $  and the integration limits ($-\infty$  to $+\infty$) in the notation. 
The expression (\ref{eq:Dij_rapidity}) also can be derived rigorously from the second derivative of the energy with respect to  spin and charge fluxes\ucite{13}. 
Incidentally, for $D_{nn}$, the second-order derivative of pressure  \eqref{eq:Dnnp} is equivalent to the integral formulation \eqref{eq:Dij_rapidity} in rapidity space\ucite{36}.

Using the thermal equilibrium relation
$-\partial\theta_{\alpha}/\partial k = 2\pi\rho_{\alpha}(1-\theta_{\alpha})v_{\alpha}/T$, the above expression can be rewritten as an integral incorporating derivatives of the occupation numbers:
\begin{equation}
D_{ij}=-\sum_{\alpha}\frac{1}{2\pi }\int \mathrm{d}k\,\frac{\partial\theta_{\alpha}}{\partial k}v_{\alpha}(\vec{q}_{i}^{\mathrm{dr}})_{\alpha}(\vec{q}_{j}^{\mathrm{dr}})_{\alpha}.
\label{eq:Dij_dtheta}
\end{equation}

For charge transport $D_{nn}$ with $\vec{q}_{n} = \{1, 0\}$, applying the effective velocity relation $v_{\alpha} = (\vec{q}_{k}^{\mathrm{dr}})_{\alpha}/m(\vec{q}_{n}^{\mathrm{dr}})_{\alpha}$, the expression (Eq.~\eqref{eq:Dij_dtheta}) reduces to the following simple form:
\begin{equation}
D_{nn}=-\sum_{\alpha}\frac{1}{2\pi m}\int \mathrm{d}k\,\frac{\partial\theta_{\alpha}}{\partial k}(\vec{q}_{n}^{\mathrm{dr}})_{\alpha}(\vec{q}_{k}^{\mathrm{dr}})_{\alpha}.
\label{eq:Dnn_intermediate}
\end{equation}

 The derivatives of occupation numbers in Eq.~\eqref{eq:Dnn_intermediate} can transform into derivatives of rapidity distribution functions by differentiating the TBA integral equations related to $\rho_{\alpha}^{\mathrm{tot}}$ from Eq.~\eqref{eq:rho_tot_intro}. After interchanging the order of integration, applying the dressed equations \eqref{eq:dressed_intro}, and integrating by parts, we obtain:
\begin{equation}
D_{nn} = -\frac{1}{m}\int \mathrm{d}k\,\frac{\partial\rho_{1}}{\partial k}k = \frac{1}{m}\int \mathrm{d}k\,\rho_{1} = \frac{n}{m},
\label{eq:Dnn_rapidity_final}
\end{equation}
which exactly matches the result derived from the tilted potential (Eq.~\eqref{eq:Dnn_tilted}). This confirms the universal, interaction- and temperature-independent nature of charge transport in integrable systems. Detailed derivations are given in SM\ucite{36}. 

For spin transport ($D_{nm}$), analogous derivation yields
$ D_{nm} = m^z/m$,
reproducing Eq.~\eqref{eq:Dnm_tilted}. For energy transport, 
$D_{ne} = Ts/m$,
reproduces Eq.~\eqref{eq:Dne_tilted}. Detailed derivations are given in SM\ucite{36}.  Unlike charge and spin transport, energy transport depends on temperature and interactions through entropy. The rapidity-space formulation provides microscopic justification for tilted potential results, confirming the inherently ballistic nature of charge and spin transport in integrable systems, while energy transport reflects thermodynamic entropy flow. 

{\it Ground State.}
At zero temperature, only the length  $l=1$ string contribution remains significant for spin degrees of freedom. 
The occupation number $\theta_\alpha$ equals unity within the Fermi sea and vanishes outside, rendering its derivative with respect to $k$ a positive (negative) delta function at the negative (positive) Fermi point. Consequently, the DW (Eq.~\eqref{eq:Dij_dtheta}) receives contributions exclusively from these Fermi points
\begin{equation}
D_{ij}=\frac{1}{\pi}\left(v_{c}Z_{i1}Z_{j1}+v_{s}Z_{i2}Z_{j2}\right),
\end{equation}
in which $v_{c}=v_1(k_0)$  and $v_{s}=v_{2,1}(\lambda_0)$ are the sound velocities of charge and spin degrees of freedom, respectively, $Z_{i1}=(\vec{q}_i^{\mathrm{dr}})_1 (k_0)$ and $Z_{i2}=(\vec{q}_i^{\mathrm{dr}})_{2,1}(\lambda_0)$. Here $k_0$ and $\lambda_0$ denote the positive boundaries of the rapidity distribution functions $\rho_1(k)$ and $\rho_{2,1}(k)$ at zero temperature, respectively, determined by pseudoenergy $\epsilon_1(k_0) = \epsilon_{2,1}(\lambda_0) = 0$.

 Furthermore, the Yang-Yang integral equations \eqref{eq:YY_intro} imply $\vec{\epsilon} = \vec{q}_e^{\mathrm{dr}}$ at  zero temperature, which implies energy related DWs $D_{ne}=D_{me}=D_{ee} = 0$ at zero temperature. Acoording to Eqs.~\eqref{eq:Dnn_tilted} and ~\eqref{eq:Dnm_tilted}, $D_{nn}=n/m$ and $D_{nm}=m^z/m$ still work at zero temperature, and then one obtain the following two identities
\begin{equation}
\begin{aligned}
v_{c}Z_{n1}Z_{n1}+v_{s}Z_{n2}Z_{n2} &= \pi n/m,   \\
v_{c}Z_{n1}Z_{m1}+v_{s}Z_{n2}Z_{m2} &= \pi m^z/m.
 \label{eq:Z_n_m}
\end{aligned}
\end{equation}

The compressibility matrix is defined as
\begin{equation}
\boldsymbol{\kappa}=\left(\begin{array}{cc}
\kappa_{nn} & \kappa_{nm}\\
\kappa_{mn} & \kappa_{mm}
\end{array}\right)=\left(\begin{array}{cc}
\frac{\partial n}{\partial\mu} & \frac{\partial n}{\partial H}\\
\frac{\partial m^z}{\partial\mu} & \frac{\partial m^z}{\partial H}
\end{array}\right).
 \label{eq:kappa}
\end{equation}
According to the dressed equation at zero temperature, the matrix
element of compressibility can be expressed in terms of the sound velocities and $Z_{i\alpha}$
\begin{equation}
\kappa_{ij}=\frac{1}{\pi}\left(\frac{Z_{i1}Z_{j1}}{v_{c}}+\frac{Z_{i2}Z_{j2}}{v_{s}}\right),\;i,j\in\{n,m\}.
 \label{eq:kappa_i_j}
\end{equation}
In matrix form, the DW and compressibility have the form $\boldsymbol{D}  =\boldsymbol{Z}\boldsymbol{v}\boldsymbol{Z}^{\mathrm{T}}/\pi$ and $\boldsymbol{\kappa}=\boldsymbol{Z}\boldsymbol{v}^{-1}\boldsymbol{Z}^{\textrm{T}}/\pi$, in which the matrix $\boldsymbol{D},\boldsymbol{Z}, \boldsymbol{v}$ are defined in SM\ucite{36}. The relation between DW and compressibility is given by sound velocity irrelated form $\boldsymbol{D} =(\boldsymbol{Z}\boldsymbol{Z}^{T})\boldsymbol{\kappa}^{-1}(\boldsymbol{Z}\boldsymbol{Z}^{T})/\pi^{2}$ and the concise form $\boldsymbol{D} \boldsymbol{\kappa}^{-1}=\boldsymbol{Z}\boldsymbol{v}^{2}\boldsymbol{Z}^{-1}$.

{\it Spin DW.}
Analogous to the tilted chemical potential approach, the spin DW can be derived using a tilted magnetic field. The system is initially prepared in an equilibrium state, after which a linear spatial dependence magnetic field $H(x) = \phi_m x$ is imposed. The corresponding DW characterizes the asymptotic current response $D_{mi} = \lim_{\phi_m \rightarrow 0} \lim_{t \rightarrow \infty} j_i(t)/t \phi_m$.

While the GHD Eq.~\eqref{eq:GHD_simple} maintain spatial homogeneity of occupation numbers, the effective accelerations become rapidity-dependent:
$ a_\alpha= \phi_m (\vec{q}_m^{\mathrm{dr}})_\alpha/m(\vec{q}_n^{\mathrm{dr}})_\alpha$.
This contrasts with the tilted chemical potential case where accelerations is constant. Numerical computation via the method of characteristics confirms consistency with the rapidity-space integral formulation (Eq.~\eqref{eq:Dij_rapidity}), validating both approaches for spin transport coefficients\ucite{36}.

\vskip 4mm

\fl{1}\centerline{\includegraphics[width=9cm,keepaspectratio]{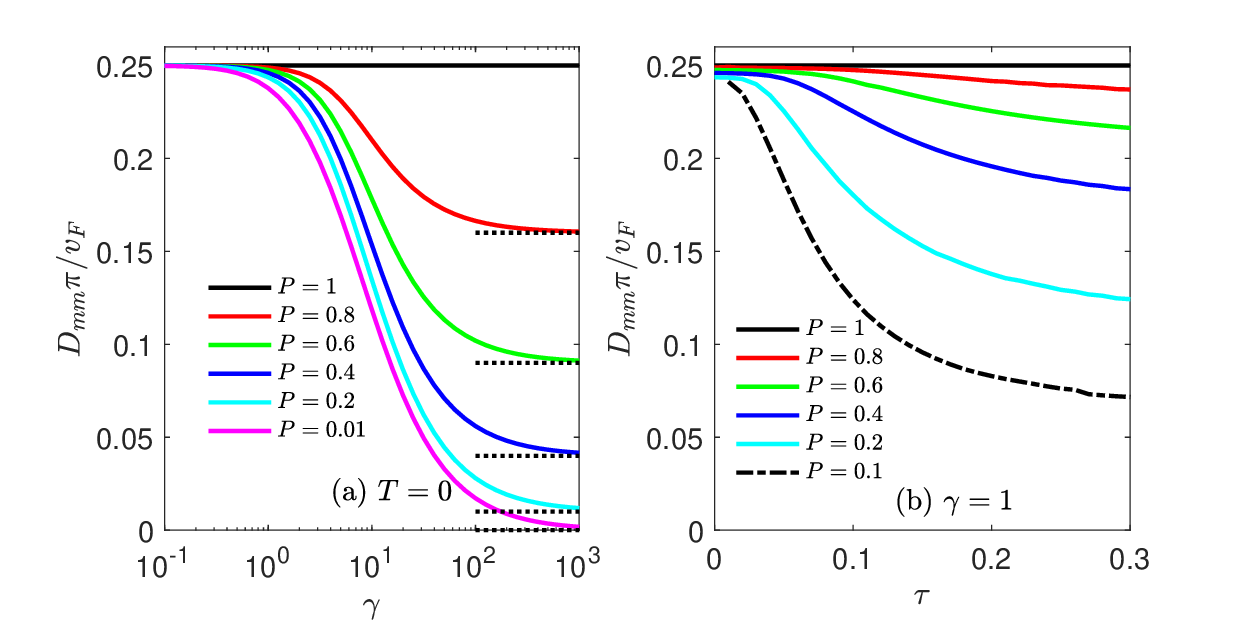}}

\vskip 2mm

\figcaption{8}{1}{DWs $D_{mm}$    for the ground state (a)  and low temperature (b). Here $\gamma = c/n$ is the dimensionless interaction strength, $P$ is the polarization, and $\tau=T/E_F$ is the dimensionless temperature with Fermi energy $E_F=\pi^2 n^2/2m $. Dotted lines in (a) represent analytical results in the strong interaction limits. The  dimensionless  interaction  $\gamma$  in (b) is set to be 1.}

Unlike $D_{nn}$, the spin Drude weight (DW) $D_{mm}$ depends sensitively on interactions and polarization at zero temperature.
For a non-interacting Fermi gas, $D_{mm}=v_F/4\pi$ is independent of polarization, with the Fermi velocity given by $v_F=\pi n/m$.
The dimensionless DW $D_{mm}/v_F$ decreases as the dimensionless interaction strength $\gamma$ increases at fixed polarization (see Fig.~1(a)), demonstrating that repulsive interactions suppress spin transport.
At fixed interaction strength, $D_{mm}/v_F$ also grows with both polarization $P$ and magnetization density $m^z$.
Notably, in the limit of infinite interaction strength the spin DW approaches $D_{mm}=v_F P^2/4\pi$.
In this limit, the spin velocity vanishes,  nevertheless, charge transport effects still yield a non-zero spin Drude weight for the polarized gas.
This feature exposes a subtle spin-charge coupling within quantum transport.

% The most notable point is that for the fully paired state with infinitely strong interactions, $D_{mm}$ equals zero, indicating that spin transport is completely suppressed. 

Fig.~1(b) shows that $D_{mm}$ decreases at low temperatures except ideal Fermi gas, demonstrating that temperature, similar to interactions, moderately suppresses spin transport. 
%A smaller polarization gives a larger absolute slope of \( D_{mm} \), whereas a higher temperature generally yields a smaller absolute slope of \( D_{mm} \). 
For strong interactions, $D_{mm}$ is approximately given by $ \int \mathrm{d}k\, \rho_1 (\vec{q}_m^{\mathrm{dr}})_1^2 /m (\vec{q}_n^{\mathrm{dr}})_1^2$\ucite{36}.

For the fully polarized (ideal) Fermi gas with $P=1$, the DW expression \eqref{eq:Dij_rapidity} contains only the one integral term with $(\vec{q}_m^{\mathrm{dr}})_1 = 1/2$, leading to $D_{mi} = D_{ni}/2$. This gives $\quad D_{mm} = n/4m$, and $\quad D_{me} = Ts/2m$.
Applying the Sommerfeld expansion at low temperatures ($\tau = T/E_F \ll 1$) with Fermi energy $E_F=mv_F^2/2$, one obtain$
\quad D_{ne} = \left( \pi^2\tau^2 /6+ \pi^4\tau^4 /45\right)E_Fn/m $,
indicating temperature-enhanced energy-related DW. 

Fig.~2 demonstrates that for both very large and very small polarization, $D_{me}$ is greater than zero and increases monotonically with temperature. In contrast, for moderate polarization, $D_{me}$ exhibits a non-monotonic behavior: it first decreases (even becoming negative at intermediate temperatures) before increasing as temperature rises further. Compared with positive spin transport, a negative  $D_{me}$  indicates a reversal  energy transport. 
Notably, at relatively high temperatures, $D_{me}$ increases monotonically with polarization, this trend is consistent with that of $D_{mm}$ and stands in complete contrast to that of $D_{ne}$. The DW $D_{me}$ can also be defined as the linear response of the magnetization current with respect to the temperature gradient
\begin{equation}
D_{me} = T\lim_{\phi_T \rightarrow 0} \lim_{t \rightarrow \infty} \frac{j_m(t)}{t \phi_T},
\label{eq:Dme_tilded_T}
\end{equation}
where the linear temperature is described by $T=\phi_T x$. The negative  $D_{me}$  means a reversal  magnetization transport, which manifests the inverse counterpart of the Seebeck effect\ucite{37,38}.

%\begin{figure*}[t!]
%\center
%\includegraphics[width=0.45\textwidth]{fig_Dme.eps}% 固定宽度8cm，保持宽高比
%\caption{DWs  $D_{me}$ as functions of dimensionless temperature $\tau = T/E_F$ for different polarizations $P$ at fixed interaction strength $\gamma=1$. Dotted lines represent analytical results for the ideal Fermi gas at low temperature.}
%\end{figure*}

\vskip 4mm
\fl{2}\centerline{\includegraphics[width=8cm,keepaspectratio]{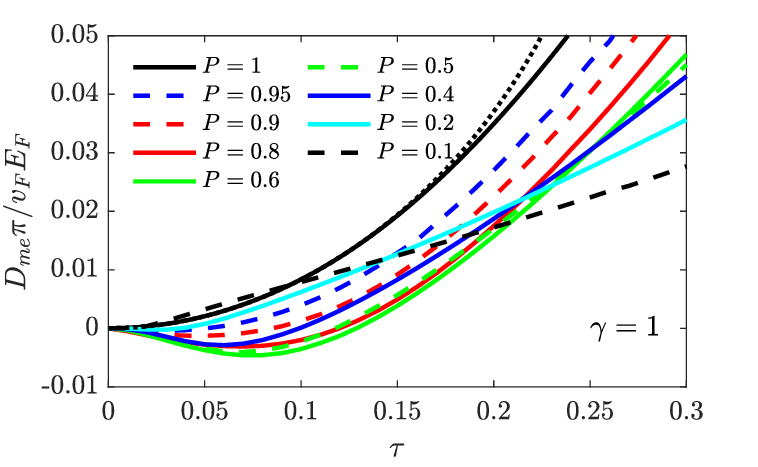}} % 固定宽度8cm，保持宽高比
\vskip 2mm
\figcaption{8}{2}{DWs  $D_{me}$ as functions of dimensionless temperature $\tau = T/E_F$ for different polarizations $P$ at fixed interaction strength $\gamma=1$. Dotted lines represent analytical results for the ideal Fermi gas at low temperature.}
\medskip

 \vskip 4mm

\fl{3}\centerline{\includegraphics[width=9cm,keepaspectratio]{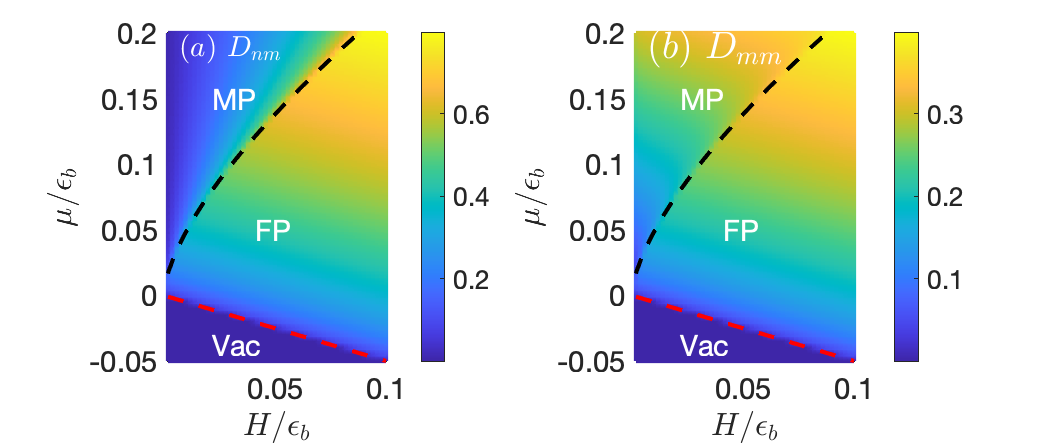}}

\vskip 2mm

\figcaption{8}{3}{DW (a) $D_{nm}$ and (b) $D_{mm}$ in $\mu$-$H$ space. The black and red dashed-lines dominate the boundaries of three quantum phases: MP, FP, and Vac at zero temperature. Here $\epsilon_b=\hbar^2c^2/2m$, $T=0.02$, and $c=10$.}

\medskip

{\it DW as indicators  of zero-temperature  phase boundary.} In the grand canonical ensemble, Fig.~3 presents  $D_{nm}$ and $D_{mm}$ in the $\mu$-$H$ space at low temperatures.
 In the vacuum state (Vac) , since there are almost no particles, both $D_{nm}$ and $D_{mm}$ are zero. As $\mu$ and $H$ increase, the particle number density rises accordingly, thereby driving $D_{nm}$ and $D_{mm}$ to increase from zero to non-zero values. 
 This indicates a phase transition (red dashed line) from Vac to the fully polarized phase  (FP). 
 For a fixed magnetic field, as the chemical potential increases further, DW exhibits a peak, which signifies the phase transition (black dashed line) from FP to the mixed phase (MP). 
 In MP, for a fixed magnetic field, as the chemical potential increases, the particle number density rises accordingly while the polarization decreases. This causes $D_{nm}$ to decrease, whereas $D_{mm}$ first decreases and then increases.

\vskip 4mm

\fl{4}\centerline{\includegraphics[width=9cm,keepaspectratio]{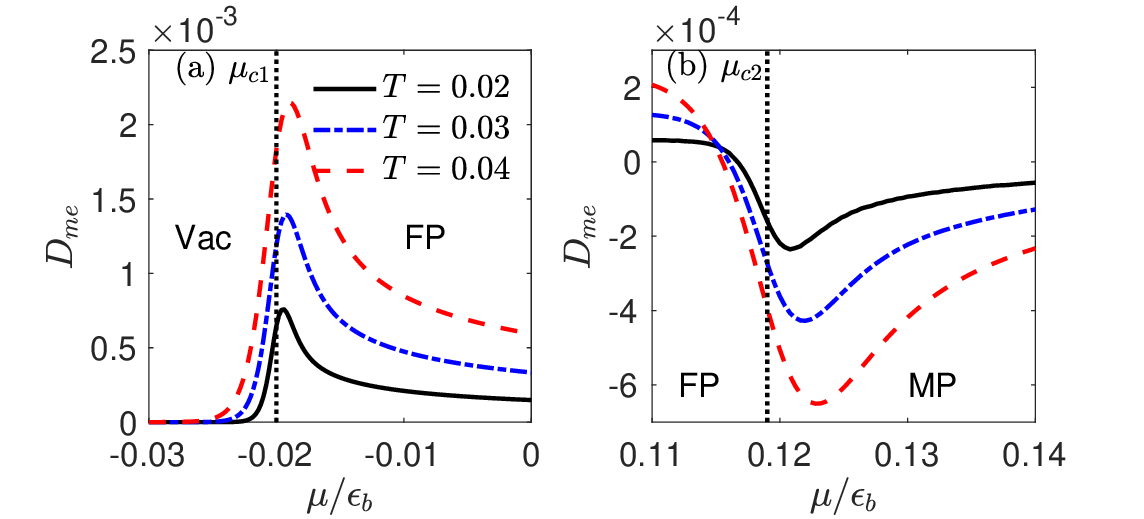}}

\vskip 2mm

\figcaption{8}{4}{DW $D_{me}$ as functions of dimensional chemical potential $\mu/\epsilon_b$ for different temperatures $T$ at $H=2, c=10$. The critical chemical potential $\mu_{c1}$, $\mu_{c2}$, indicated as vertical black dashed lines, are boundaries for Vac-FP and FP-MP at zero temperature, respectively.}

\medskip

As is well known, the entropy at low temperatures exhibits a peak behavior near phase transitions. Consequently, $D_{ne}$ and $D_{me}$  show a similar behavior. 
In Fig. 4(a), the peak position of $D_{me} $  at low temperatures is consistent with the zero-temperature Vac-FP transition point ($\mu_{c1}=-H/2$), where $D_{me} = D_{ne}/2 = Ts/(2m) $. 
%
% The spin current direction does not reverse under the Seebeck effect near Vac-FP phase boundary. 
%
Fig. 4(b) shows that the valley position of $D_{me} $ at low temperatures corresponds to the zero-temperature FP-MP transition point.
However, % the spin current direction reverse under the Seebeck effect for a large polarization close to FP-MP phase transition. 
on the right side of the  FP-MP phase boundary, a large negative value of  $D_{me}$  emerges, indicating the existence of a counterpropagating magnetization current under a temperature gradient, which is consistent with the results presented in Fig. 2.

At low temperatures, low-energy transport is dominated by real-$k$ quasiparticle excitations coupled with length-one string excitations. In the vicinity of FP-MP phase transition, quasiparticles with spin degrees of freedom occupy only the region where the rapidity $\lambda$ is small. This allows for a Taylor expansion of the pseudoenergies, the total density of states, and the dressed quantities with respect to small $\lambda$ near the critical regime. One thus obtains the expression of the DWs for critical regime
\begin{equation}
D_{ij}=D_{\mathrm{bg}}+\left(\mathcal{A}_{1}+\mathcal{A}_{2}\right)T^{1/2}\,\mathrm{Li}_{\frac{1}{2}}\!\left(-e^{-\epsilon_{2,1}(0)/T}\right),
\end{equation}
with $i,j\in\{n,m\}. $Here, $D_{\mathrm{bg}}$ denotes the background contribution, which consists of the zero-temperature leading term and its analytic finite-temperature corrections. The coefficients $\mathcal{A}_{1}$ and $\mathcal{A}_{2}$, on the other hand, characterize the contributions from the charge and spin sectors, respectively. It follows that, upon subtracting the background term, the DWs $D_{nn}$, $D_{nm}$, and $D_{mm}$ exhibit a distinct $T^{1/2}$ scaling behavior in the vicinity of the phase transition. The rescaled DWs, $[D_{ij}-D_{\mathrm{bg}}]T^{-1/2}$ at different temperatures intersect at the phase transition boundary, which is determined by dressed energies satisfying $\epsilon_{2,1}(0)=0$. The detailed derivation and calculation  are provided in the SM VII\ucite{36}.

For the strongly interacting regime ($\gamma \gg 1$) at low temperature, we asymptotically expand the kernels of the TBA equation  up to the order of $O(1/c^3)$. 
By iteratively solving the dressed equations \eqref{eq:dressed_intro}, we can derive explicit analytical expressions for the full Drude weight matrix ($D_{nn}, D_{nm}, D_{mm}, D_{ne}, D_{me}, D_{ee}$) at low temperatures in terms of polylogarithm functions. 
%In the limit of infinite interaction strength, the Drude weight receives contributions exclusively from charge excitations. By contrast, both charge and spin excitations contribute to the Drude weight in the strong-interaction regime.
%
Detailed derivations and the complete analytical formulas are provided in the SM VIII\ucite{36}.

{\it Experimental Protocol.} 
Motivated by the recent experiment measuring the charge DW of a 1D spinless Bose gas\ucite{24}, we propose two experimental protocols to measure the DWs.
The first protocol adopts the aforementioned tilted potential field and  the second one takes the  bipartitioning quench protocol. In the tilted chemical potential scheme, all particles are subjected to the same external force, which permits direct measurement of the charge DW. In the tilted magnetic field scheme, particles with opposite spins suffer opposite forces, and this setup is suitable for measuring the spin DW.  
On the other hand, the second protocol examines charge and energy transport between two initially equilibrium quasi-one-dimensional subsystems with different chemical potentials, magnetic fields or temperatures.
 After connecting the two parts in this bipartitioning quench protocol, the corresponding DWs can be extracted by measuring particle, magnetization and energy currents.
The DW is defined as the asymptotic current growth rate with respect to the  vanishing bias, mathematically:
\begin{equation}
\begin{split}
D_{ni} &= \lim_{\delta\mu \rightarrow 0} \frac{\partial}{\partial \delta\mu} \lim_{t \rightarrow \infty} \frac{J_i(t)}{t},\\
D_{mi} &= \lim_{\delta H \rightarrow 0} \frac{\partial}{\partial \delta H} \lim_{t \rightarrow \infty} \frac{J_i(t)}{t},
\end{split}
\label{eq:D_bias}
\end{equation}
where $\delta\mu$ and $\delta H$ are the chemical potential difference and magnetic field difference between the two subsystems, respectively, and \(J_i(t)\) is the integral of the corresponding current over coordinate space.
Using the GHD method, we obtain results consistent with the above integral form \eqref{eq:Dij_rapidity}, validated via analytical derivation and numerical simulation, see SM   \ucite{36}.

{\it Conclusion.}
In summary, we have analytically investigated universal spin and charge quantum transport behavior in the 1D  repulsive Fermi gas known as the Yang-Gaudin model.
Using the tilted chemical potential method and the generalized hydrodynamics, we have rigorously derived thermodynamic expressions for the charge and spin Drude weights.
Specifically, we find that the Drude weights $D_{nn}$, $D_{nm}$, and $D_{ne}$, which describe particle, spin, and energy current responses to a chemical potential gradient, equal the particle number density, magnetization, and the product of temperature and entropy density, respectively, per unit mass.
These general relations also hold for other 1D integrable continuum systems with multicomponent internal spin degrees of freedom.
Furthermore, we have also examined the spin Drude weight $D_{mm}$ at zero and low temperatures, which characterizes the magnetization-current response to a magnetic field gradient,  revealing an essential spin-charge coupling feature in quantum transport.
We have observed  that $D_{mm}$  decreases substantially with increasing interparticle interaction strength and temperature.
By contrast, the energy Drude weight $D_{me}$  (energy-current response to a magnetic-field gradient) is negative at low temperatures and turns positive at higher temperatures under moderate polarization. 
This sign reversal indicates inverted energy transport at low temperatures, a signature of the spin Seebeck effect.
In the extreme low-temperature limit, the behaviors of $D_{nm}$, $D_{mm}$, and $D_{me}$ across the $\mu\text{-}H$ plane effectively trace the zero-temperature phase boundary.
Finally, we propose two experimental protocols for measuring these universal relations between Drude weights and thermodynamic quantities. 
The relations are supported by generalized hydrodynamics calculations, analytical derivations, and numerical simulations.

\textit{Acknowledgements.} We  acknowledge support from the Natural Science Foundation of China (NSFC)  grants No. 12461160324,  No. U25D8013, No. 92365202, No. 12474492, and the Innovation Program for Quantum Science and Technology Grant No.  2021ZD0302000 and 2023ZD0300404.
We also acknowledges partial support from the National Key R\&D Program of China under Grant No. 2022YFA1404104 and the Science Challenge Project  Grant No.TZ2025017. 
SZ acknowledges support from HK GRF (Grant No. 17306024), CRF (Grants No. C6009-20G, No. C7012-21G, No. C4050-23GF), CRS HKU701/24 and a RGC Fellowship Award No. HKU RFS2223-7S03.

\end{multicols}  % 双栏结束\end{multicols}  % 双栏结束

\tableofcontents

\section{Thermodynamic Bethe Ansatz and Generalized Hydrodynamic}
\label{sec:model}

In the thermodynamic limit, where $N \rightarrow \infty$, $M^z \rightarrow \infty$, and $L \rightarrow \infty$ while the densities $n=N/L$ and $m^z=M^z/L$ remain finite, the thermodynamic quantities can be expressed as integral forms in rapidity space. Specifically, the total particle number density ($n$), magnetization density ($m^z$), the total momentum per length $K/L$, internal energy densities $e_0$ and total energy density $e$ are given by:
\begin{align}
\label{eq:nm}
n &= \int_{-\infty}^{\infty} \textrm{d}k \, \rho_{1}(k), \\
m^{z} &= n/2 - \sum_{l=1}^{\infty} l \int_{-\infty}^{\infty} \textrm{d}k \, \rho_{2,l}(k),\\
K/L&=\int_{-\infty}^{\infty}\textrm{d}k \, k\rho_{1}(k),\\
e_{0} &= E/L = \int_{-\infty}^{\infty} \textrm{d}k \, \frac{k^{2}}{2m} \rho_{1}(k), \\
e &= e_{0} - \mu n - Hm^{z},
\label{eq:e}
\end{align}
respectively, where $\rho_{1}(k)$ is the distribution function of rapidity $k$ in the charge sector, and $\rho_{2,l}(k)$ are the distribution functions for the spin sectors corresponding to bound states of length $l$. Here, $m$ is the mass of the particle, and we set $\hbar=1$.

From these expressions, we can identify the bare eigenvalues associated with each conserved quantity. These are the coefficients multiplying the distribution functions $\rho_1(k)$ and $\rho_{2,l}(k)$ when the quantity is written in the form $\sum_i q_i \rho_i$. Explicitly, these bare eigenvalues are defined as:
\begin{align}
\vec{q}_{n} &= \{1, 0\}, \\
\vec{q}_{m} &= \{1/2, -l\}, \\
\vec{q}_{k} &= \{k, 0\}, \\
\vec{q}_{e_0} &= \{k^2/2m, 0\}, \\
\vec{q}_{e} &=\vec{q}_{e_0}-\mu \vec{q}_{n}-H \vec{q}_{m}=\{k^2/2m - \mu - H/2, lH\}.
\end{align}
Here, the first component of the vector corresponds to the charge sector (multiplying $\rho_1(k)$) and the second component to the spin sector (multiplying $\rho_{2,l}(k)$). These bare eigenvalues will later be influenced by interactions to become the eigenvalues of quasi-particles.

The rapidity distribution functions $\rho_{1}(k)$ and $\rho_{2,l}(k)$ are not independent but are self-consistently determined by the thermodynamic Bethe Ansatz (TBA) integral equations. These equations describe the thermodynamic state of the system and are given by:
\begin{equation} \label{eq:TBA}
\begin{aligned}
\rho_{1}^{\textrm{tot}}(k) &= \frac{1}{2\pi} + \sum_{l=1}^{\infty} \phi_{l} \ast \rho_{2,l}(k) \\
\rho_{2,l}^{\textrm{tot}}(k) &= \phi_{l} \ast \rho_{1}(k) - \sum_{m=1}^{\infty} \Phi_{l,m} \ast \rho_{2,m}(k), \quad l=1, 2, \ldots
\end{aligned}
\end{equation}
The integral kernels involved in these equations are defined as:
\begin{equation}
\phi_{l}(k) = \frac{1}{\pi} \frac{2lc}{(lc)^{2}+4k^{2}}
\end{equation}
and
\begin{equation}
\Phi_{l,m} = (1-\delta_{l,m})\phi_{|l-m|} + 2\sum_{j=1}^{\min(l,m)-1} \phi_{|l-m|+2j} + \phi_{l+m}.
\end{equation}
Here $\min(l,m)$ denote the minimum of $l$ and $m$, and the convolution notation $\phi_n \ast f(k)$ implies $\int_{-\infty}^{\infty} \textrm{d}y \, \phi_n(k-y)f(y)$.

The total distribution functions, $\rho_{1}^{\textrm{tot}}(k)$ and $\rho_{2,l}^{\textrm{tot}}(k)$, are composed of contributions from both occupied states and hole states: $\rho_{1}^{\textrm{tot}}(k)=\rho_{1}(k)+\rho_{1}^{\textrm{hole}}(k)$ and $\rho_{2,l}^{\textrm{tot}}(k)=\rho_{2,l}(k)+\rho_{2,l}^{\textrm{hole}}(k)$. To solve the TBA equations (\ref{eq:TBA}), we typically introduce occupation numbers
\begin{equation}\label{eq:theta}
\theta_{1}(k)=\rho_{1}(k)/\rho_{1}^{\textrm{tot}}(k), \  \theta_{2,l}(k)=\rho_{2,l}(k)/\rho_{2,l}^{\textrm{tot}}(k).
\end{equation}
If the occupation numbers $\theta_{1}(k)$ and $\theta_{2,l}(k)$ are known, the rapidity distribution functions $\rho_{1}(k)$ and $\rho_{2,l}(k)$ can be obtained by solving this coupled system of integral equations.

The occupation numbers $\theta_1(k)$ and $\theta_{2,l}(k)$ are derived from the Yang-Yang integral equations, which self-consistently determine the pseudoenergies of the system. In a generalized thermodynamic setting, the Gibbs free energy can be expressed as:
\begin{equation}
\label{eq:G}
G=L\left(e_{0}-\mu n-Hm^{z}-g\right)-TS,
\end{equation}
where $T$ is the temperature and $S$ is the entropy. Here, $g$ represents a sum of other conserved quantities multiplied by their corresponding Lagrange multipliers (potentials). For example, if we consider the total momentum $K$ as an additional conserved quantity, $g$ would include a term like $\lambda K/2mL$, where $\lambda$ is the corresponding potential.
The Yang-Yang integral equations for the pseudoenergies $\epsilon_1(k)$ and $\epsilon_{2,l}(k)$ are obtained by minimizing the generalized Gibbs free energy ($\delta G=0$) under TBA equations (\ref{eq:TBA}) . These equations describe the thermal state that maximizes entropy under the constraints of fixed conserved quantities\ucite{2}:
\begin{equation} \label{eq:YangYang}
\begin{aligned}
\epsilon_{1}(k) &= \frac{k^{2}}{2m}-\mu-\frac{H}{2}-\epsilon_{g}(k)-T\sum_{l=1}^{\infty}\phi_{l}*\ln\left(1+\textrm{e}^{-\epsilon_{2,l}/T}\right)(k), \\
\epsilon_{2,l}(k) &= lH-T\phi_{l}*\ln\left(1+\textrm{e}^{-\epsilon_{1}/T}\right)(k)+T\sum_{m=1}^{\infty}\Phi_{l,m}*\ln\left(1+\textrm{e}^{-\epsilon_{2,m}/T}\right)(k), \quad l=1,2,\ldots.
\end{aligned}
\end{equation}
In these equations, $\epsilon_{g}(k)$ is the sum of the bare single-particle eigenvalues for any additional conserved quantities, multiplied by their respective potentials. For instance, if only total momentum is considered, $\epsilon_{g}(k)=\lambda k/2m$. For a system in thermal equilibrium, we set $\epsilon_{g}(k)=0$.
Once the pseudoenergies are determined by solving Eqs. \eqref{eq:YangYang}, the occupation numbers are given by Fermi-Dirac like distribution functions:
\begin{equation} \label{eq:theta_epsilon}
\theta_{1}(k) = \left[1+\textrm{e}^{\epsilon_{1}(k)/T}\right]^{-1}, \ \theta_{2,l}(k) = \left[1+\textrm{e}^{\epsilon_{2,l}(k)/T}\right]^{-1}.
\end{equation}
These occupation numbers, in turn, allow for the calculation of the rapidity distribution functions $\rho_1(k)$ and $\rho_{2,l}(k)$ through the TBA equations \eqref{eq:TBA}.

The pressure $p$ of the system can be calculated from the pseudoenergy $\epsilon_1(k)$ via the following integral in rapidity space:
\begin{equation}
\label{eq:p}
p=\frac{T}{2\pi}\int_{-\infty}^{\infty}\ln\left(1+\textrm{e}^{-\epsilon_{1}(k)/T}\right)\textrm{d}k.
\end{equation}
For the grand canonical ensemble, key thermodynamic densities are directly related to derivatives of the pressure:
\begin{equation}
n = \frac{\partial p}{\partial\mu}, \ m^{z} = \frac{\partial p}{\partial H}, \ S/L = \frac{\partial p}{\partial T}.
\end{equation}

The ballistic transport of charge, magnetization, and energy is characterized by their respective currents. In rapidity space, the particle current ($j_n$), magnetization current ($j_m$), and energy currents ($j_{e_0}$, $j_e$) are defined as:
\begin{align}
j_{n} &= \int_{-\infty}^{\infty}\textrm{d}k \, v_{1}(k)\rho_{1}(k), \label{eq:jn} \\
j_{m} &= j_{n}/2 - \sum_{l=1}^{\infty} l \int_{-\infty}^{\infty}\textrm{d}k \, v_{2,l}(k)\rho_{2,l}(k), \label{eq:jm} \\
j_{e_{0}} &= \int_{-\infty}^{\infty}\textrm{d}k \, \frac{k^{2}}{2m}v_{1}(k)\rho_{1}(k), \label{eq:je0} \\
j_{e} &= j_{e_{0}}-\mu j_{n}-Hj_{m}, \label{eq:je}
\end{align}
respectively. Here, $\vec{v}(k)=\{v_{1}(k), v_{2,l}(k)\}$ represents the effective velocities of the quasi-particles in the charge and spin sectors. These effective velocities are defined as the ratio of the derivative of energy to the derivative of momentum, both taken with respect to rapidity, which  can be expressed in terms of the dressed functions
\begin{equation}
\label{eq:v}
\vec{v}(k)=\frac{(\partial_{k}\vec{q}_{e})^{\textrm{dr}}(k)}{(\partial_{k}\vec{q}_{k})^{\textrm{dr}}(k)}=\frac{\vec{q}_{k}^{\textrm{dr}}(k)/m}{\vec{q}_{n}^{\textrm{dr}}(k)},
\end{equation}
where $\partial_{k}\vec{q}_{e}=\vec{q}_{k}/m=\{k/m,0\}$ and $\partial_{k}\vec{q}_{k}=\vec{q}_{n}=\{1,0\}$.
The concept of "dressed" functions is central to integrable systems, accounting for interactions between particles. A dressed function $\vec{f}^{\textrm{dr}}=\{f_{1}^{\textrm{dr}}(k), f_{2,l}^{\textrm{dr}}(k)\}$ is the solution to a system of linear integral equations:
\begin{equation}\label{eq:dressed_equ}
\begin{aligned}
f_{1}^{\textrm{dr}}(k) &= f_{1}(k) + \sum_{l=1}^{\infty}\phi_{l} \ast (f_{2,l}^{\textrm{dr}} \theta_{2,l})(k) \\
f_{2,l}^{\textrm{dr}}(k) &= f_{2,l}(k) + \phi_{l} \ast (f_{1}^{\textrm{dr}} \theta_{1})(k) - \sum_{m=1}^{\infty}\Phi_{l,m} \ast (f_{2,m}^{\textrm{dr}} \theta_{2,m})(k), \quad l=1,2,\ldots.
\end{aligned}
\end{equation}
Here, $f_1(k)$ and $f_{2,l}(k)$ are the bare functions corresponding to the observable.

Crucially, the derivatives of the pseudoenergies $\vec{\epsilon}=\{\epsilon_1, \epsilon_{2,l}\}$ with respect to the thermodynamic potentials in the Yang-Yang integral equations \eqref{eq:YangYang}  are directly related to the dressed functions of the corresponding conserved quantities. For example:
\begin{equation}
\label{eq:qn}
\partial_{-\mu}\vec{\epsilon} = \vec{q}_{n}^{\textrm{dr}} , \ \partial_{-H}\vec{\epsilon} = \vec{q}_{m}^{\textrm{dr}}, \
\partial_{-\lambda}\vec{\epsilon} = \vec{q}_{k}^{\textrm{dr}}/(2m),
\end{equation}
where $\vec{\rho}^{\textrm{tot}}=\{\rho_{1}^{\textrm{tot}},\rho_{2,l}^{\textrm{tot}}\}$. Similarly, for a system in thermal equilibrium, the derivatives with respect to inverse temperature and rapidity yield:
\begin{equation}
\label{eq:qe}
\partial_{1/T}\vec{\epsilon}/T = \vec{q}_{e}^{\textrm{dr}}, \
\partial_{k}\vec{\epsilon} = \vec{q}_{k}^{\textrm{dr}}/m.
\end{equation}
It is worth noting that TBA Eq. ~\eqref{eq:TBA} can be written in dressed equation form
\begin{equation}
\label{eq:qn_rho_tot}
\vec{q}_{n}^{\textrm{dr}} = 2\pi\vec{\rho}^{\textrm{tot}},
\end{equation}
in which we used the definition of occupation number \eqref{eq:theta} and dressed equation \eqref{eq:dressed_equ}. According to above expression \eqref{eq:qe} and \eqref{eq:qn_rho_tot}, the effective velocity \eqref{eq:v} can be expressed by the derivative of pseudoenergy and total distribution function
\begin{equation}
\label{eq:v_epsilon}
\vec{v}(k)=\frac{\partial_{k}\vec{\epsilon}}{2\pi\vec{\rho}^{\textrm{tot}}}.
\end{equation}
These relationships are fundamental for connecting the microscopic Bethe ansatz description to macroscopic transport coefficients like the Drude weight.

For this integrable system, there are infinite conserved quantities, and then the time evolution of the system is governed by the generalized hydrodynamics (GHD) equations\ucite{1,2}:
\begin{equation}\label{eq:GHD}
\begin{aligned}
\left(\partial_t + v_1 \partial_x + a_1 \partial_{k/m}\right) \theta_1 &= 0, \\
\left(\partial_t + v_{2,l} \partial_x + a_{2,l} \partial_{k/m}\right) \theta_{2,l} &= 0, \quad l=1, 2, \ldots.
\end{aligned}
\end{equation}
in which $t$ is the time, and $x$ is the position-space coordinates.
Here the effective accelerations $a_1, a_{2,l}$ are given by
\begin{equation}
 \{ a_1, a_{2,l} \} =  \frac{\vec{q}_n^{\mathrm{dr}}\partial_x \mu(x) +\vec{q}_m^{\mathrm{dr}}\partial_x H(x) }{m\vec{q}_n^{\mathrm{dr}}}=\frac{\{ \partial_x \mu(x) + \frac{1}{2}\partial_x H(x),\, -l\partial_x H(x) \}^{\mathrm{dr}}}{m\vec{q}_n^{\mathrm{dr}}}.
\label{eq:a}
\end{equation}

%===============================================================================================
\section{Charge Drude Weight from Tilted Chemical Potential}

The Drude weight can be derived using the tilted chemical potential approach\ucite{3}. Initially, the system is prepared in an equilibrium state with homogeneous chemical potential $\mu(x)$ and magnetic field $H(x)$. Subsequently, a linear spatial dependence is introduced to the chemical potential, $\mu(x) = \phi_n x$, where $\phi_n$ is the gradient. The Drude weight can then be defined as the asymptotic growth rate of the local current in response to this static, spatially linear potential\ucite{3}:
\begin{equation}
D_{ni} = \lim_{\phi_n \rightarrow 0} \lim_{t \rightarrow \infty} \frac{j_i(x,t)}{t \phi_n},
\label{eq:Dni_tilted_mu}
\end{equation}
where $j_i(x,t)$ denotes the particle current ($i=n$), magnetization current ($i=m$), or energy current ($i=e_0,e$).

The time evolution of the system is governed by the GHD equations \eqref{eq:GHD}.
For this specific case with linear chemical potential $\partial_x \mu(x) = \phi_n$ and a homogeneous magnetic field ($\partial_x H=0$), according to \eqref{eq:a},  the effective accelerations are constant:
\begin{equation}
a_1 = a_{2,l} = a = \phi_n / m.
\label{eq:a2}
\end{equation}
Since the occupation number is uniform at every position in coordinate space for the initial state, and  GHD equations \eqref{eq:GHD} preserve spatial translational invariance, the occupation number is independent of the spatial coordinate $x$ at arbitrary time,  $\partial_x \theta_1 = \partial_x \theta_{2,l} = 0$,  such that GHD equations \eqref{eq:GHD} simplify to:
\begin{equation}
\left(\partial_t + a\, \partial_{k/m}\right) \theta_i = 0, \quad i=1 \text{ or } 2,l.
\label{eq:GHD3}
\end{equation}
Applying the method of characteristics, we define curves $k(t)$ satisfying $\mathrm{d}k(t)/\mathrm{d}t = m a$. Along these characteristics, the partial differential equations \eqref{eq:GHD3} reduce to ordinary differential equations
\begin{equation} \label{eq:GHD_t}
\frac{\mathrm{d}\theta_i(t, k(t))}{\mathrm{d}t} = 0,
\end{equation}
by virtue of the chain rule.
Given the constant effective acceleration, the characteristic curves are linear in time: $k(t) = k(0) + \phi_n t$. Consequently, the  ordinary differential equations \eqref{eq:GHD_t} implies that the occupation number at time $t$ is simply related to its initial value:
\begin{equation}
\theta_i(t, k) = \theta_i(0, k - \phi_n t).
\label{eq:theta_t_0}
\end{equation}

We now evaluate the particle current. Using the definition of the effective velocity \eqref{eq:v} and the occupation number \eqref{eq:theta} as well as the relation $\vec{q}_{n}^{\textrm{dr}} = 2\pi\vec{\rho}^{\textrm{tot}}$, the particle current at the time $t$ (Eq.~\eqref{eq:jn}) can be expressed as
\begin{equation}
j_n(t) = \frac{1}{2\pi} \int_{-\infty}^{\infty} \mathrm{d}k\, \theta_1(t,k) \left(\partial_k \vec{e}\,\right)^{\mathrm{dr}}_1 (t,k).
\label{eq:jn2}
\end{equation}
Based on the dressed Eqs.~\eqref{eq:dressed_equ}, the first equation of the dressed derivative of the energy at time $t$ reads
\begin{equation}\label{eq:e_k_dr}
\begin{aligned}
\left(\partial_{k}\vec{e}\right)_{1}^{\textrm{dr}}\left(t,k\right) & =\frac{k}{m}+\sum_{l=1}^{\infty}\left[\phi_{l}*\left(\partial_{k}\vec{e}\right)_{2,l}^{\textrm{dr}}\left(t,k\right)\theta_{2,l}\left(t,k\right)\right]\\
 & =\frac{k}{m}+\sum_{l=1}^{\infty}\int_{-\infty}^{+\infty} \mathrm{d}q\,\phi_{l}\left(k-q\right)\left(\partial_{k}\vec{e}\right)_{2,l}^{\textrm{dr}}\left(t,q\right)\theta_{2,l}\left(t,q\right)\\
 & =\frac{k}{m}+\sum_{l=1}^{\infty}\int_{-\infty}^{+\infty} \mathrm{d}q\,\phi_{l}\left(k-q\right)\left(\partial_{k}\vec{e}\right)_{2,l}^{\textrm{dr}}\left(t,q\right)\theta_{2,l}\left(0,q-t\phi_{n}\right)\\
 & =\frac{\tilde{k}}{m}+\frac{t\phi_{n}}{m}+\sum_{l=1}^{\infty}\int_{-\infty}^{+\infty} \mathrm{d}\tilde{q}\,\phi_{l}\left(\tilde{k}-\tilde{q}\right)\left(\partial_{k}\vec{e}\right)_{2,l}^{\textrm{dr}}\left(t,q\right)\theta_{2,l}\left(0,\tilde{q}\right)
\end{aligned}
\end{equation}
in which we employed Eq. ~\eqref{eq:theta_t_0} in the third equality and carried out a variable substitution $q-t\phi_{n}=\tilde{q}$ and $k-t\phi_{n}=\tilde{k}$ in the fourth one. For the initial time $t=0$, the first dressed equation for $\left(\partial_{k}\vec{e}\right)_{1}^{\textrm{dr}}\left(0,\tilde{k}\right)$ and $\left(\vec{q}_n\right)_{1}^{\textrm{dr}}\left(0,\tilde{k}\right)$ read
\begin{equation}\label{eq:e_k_dr0}
\begin{aligned}
\left(\partial_{k}\vec{e}\right)_{1}^{\textrm{dr}}\left(0,\tilde{k}\right)=\frac{\tilde{k}}{m}+\sum_{l=1}^{\infty}\left[\phi_{l}*\left(\partial_{k}\vec{e}\right)_{2,l}^{\textrm{dr}}\theta_{2,l}\left(0,\tilde{k}\right)\right],
\end{aligned}
\end{equation}
and 
\begin{equation}\label{eq:qn_dr0}
\begin{aligned}
\left(\vec{q}_n\right)_{1}^{\textrm{dr}}\left(0,\tilde{k}\right)=1+\sum_{l=1}^{\infty}\left[\phi_{l}*\left(\vec{q}_n\right)_{2,l}^{\textrm{dr}}\theta_{2,l}\left(0,\tilde{k}\right)\right],
\end{aligned}
\end{equation}
respectively. Due to the linearity of the dressed equation \eqref{eq:dressed_equ}, the superposition form $\left(\partial_{k}\vec{e}\right)^{\textrm{dr}}\left(0,\tilde{k}\right)+\frac{t \phi_n}{m}\left(\vec{q}_n\right)^{\textrm{dr}}\left(0,\tilde{k}\right)$ of above two equations satisfy the following equation
\begin{equation}\label{eq:e_k_qn}
\begin{aligned}
\left(\partial_{k}\vec{e}\right)_{1}^{\mathrm{dr}}\left(0,\tilde{k}\right)
+\frac{t \phi_n}{m}\left(\vec{q}_n\right)_{1}^{\mathrm{dr}}\left(0,\tilde{k}\right)
=\frac{\tilde{k}}{m}+\frac{t \phi_n}{m}
+\sum_{l=1}^{\infty}\phi_{l}*\left[\left(\partial_{k}\vec{e}\right)_{2,l}^{\mathrm{dr}}
+\frac{t \phi_n}{m}\left(\vec{q}_n\right)_{2,l}^{\mathrm{dr}}\right]\theta_{2,l}\left(0,\tilde{k}\right),
\end{aligned}
\end{equation}
For the spin sector, $\left(\partial_{k}\vec{e}\right)_{2}^{\textrm{dr}}\left(t,k\right) $, $\left(\partial_{k}\vec{e}\right)_{2}^{\textrm{dr}}\left(0,\tilde{k}\right)$ and $\left(\vec{q}_n\right)_{2}^{\textrm{dr}}\left(0,\tilde{k}\right)$ has the similar form. Through comparison between the dressed equation \eqref{eq:e_k_qn} and the last equation of \eqref{eq:e_k_dr}, one thus obtain the relation connecting $\left(\partial_{k}\vec{e}\right)^{\textrm{dr}}$ at an arbitrary time $t$ with that at the initial time $t=0$, namely,
\begin{equation}
\left(\partial_k \vec{e}\,\right)^{\mathrm{dr}} (t,k) = \left(\partial_k \vec{e}\,\right)^{\mathrm{dr}} (0, k - \phi_n t) + \frac{1}{m} t \phi_n \, \vec{q}_n^{\mathrm{dr}} (0, k - \phi_n t).
\label{eq:kdr}
\end{equation}
Substituting Eqs.~\eqref{eq:theta_t_0} and \eqref{eq:kdr} into Eq.~\eqref{eq:jn2} yields the particle current in terms of the initial state:
\begin{equation}
j_n(t) = \frac{1}{2\pi} \int_{-\infty}^{\infty} \mathrm{d}k\, \theta_1(0,k) \left(\partial_k \vec{e}\,\right)^{\mathrm{dr}}_1 (0,k) + \frac{1}{m} t \phi_n \frac{1}{2\pi} \int_{-\infty}^{\infty} \mathrm{d}k\, \theta_1(0,k) \left(\vec{q}_n^{\mathrm{dr}}\right)_1 (0,k).
\label{eq:jn3}
\end{equation}
For a thermally equilibrated initial state, $\theta_1(0,k)$ is an even function, while $\left(\partial_k \vec{e}\,\right)^{\mathrm{dr}}_1 (0,k)$ is an odd function. Hence, the first integral vanishes, leaving only the second term contributing to the current:
\begin{equation}
j_n(t) = \frac{1}{m} \, t \, \phi_n \, \frac{1}{2\pi} \int_{-\infty}^{\infty} \mathrm{d}k\, \theta_1(0,k) \left(\vec{q}_n^{\mathrm{dr}}\right)_1 (0,k) \label{eq:jn4}.
\end{equation}
Combining the expressions \eqref{eq:qn} and \eqref{eq:theta}, we obtain
\begin{equation}
\frac{1}{2\pi}\theta_1(0,k) \left(\vec{q}_n^{\mathrm{dr}}\right)_1 (0,k) =  \rho_1(0,k) .
\label{eq:rho_1}
\end{equation}
The current thereby simplifies to an expression proportional to the particle density $n$:
\begin{equation}
j_n(t) = \frac{1}{m} \, t \, \phi_n \int_{-\infty}^{\infty} \mathrm{d}k\, \rho_1(0,k) = \frac{n}{m} \, t \, \phi_n,
\label{eq:jn5}
\end{equation}
where the relation \eqref{eq:nm} has been utilized in the derivation.
The Drude weight is thus obtained as
\begin{equation}
D_{nn} = \lim_{\phi_n \rightarrow 0} \lim_{t \rightarrow \infty} \frac{j_n(x,t)}{t \phi_n} = \frac{n}{m},
\label{eq:Dnn_mu}
\end{equation}
where the limits are not necessary under the frame of GHD.

Following an analogous calculation, the magnetization current \eqref{eq:jm} is simplified into the following form
\begin{align}
j_m(t) &= \frac{j_n(t)}{2} - \frac{1}{2\pi} \sum_{l=1}^{\infty} l \int_{-\infty}^{\infty} \mathrm{d}k\, \theta_{2,l}(t,k) \left(\partial_k \vec{e}\,\right)^{\mathrm{dr}}_{2,l} (t,k), \label{eq:jm1} \\
&= \frac{n}{2m} t \phi_n -\frac{1}{2\pi}  \sum_{l=1}^{\infty} l \frac{1}{m} t \phi_n  \int_{-\infty}^{\infty} \mathrm{d}k\, \theta_{2,l}(0,k) \left(\vec{q}_n^{\mathrm{dr}}\right)_{2,l} (0,k), \label{eq:jm2}\\
&= \frac{n}{2m} t \phi_n - \frac{1}{m} t \phi_n\sum_{l=1}^{\infty} l  \int_{-\infty}^{\infty} \mathrm{d}k\, \rho_{2,l}(0,k), \label{eq:jm3} \\
&= \frac{m^z}{m} t \phi_n, \label{eq:jm4}
\end{align}
yielding the Drude weight
\begin{equation}
D_{nm} = \lim_{\phi_n \rightarrow 0} \lim_{t \rightarrow \infty} \frac{j_m(x,t)}{t \phi_n} = \frac{m^z}{m}.
\label{eq:Dnm_mu}
\end{equation}

The energy current \eqref{eq:je0} associated with the internal energy $e_0$ can be calculated as follows
\begin{equation}\label{eq:je0_s}
\begin{aligned}
j_{e_{0}}\left(t\right)&=\frac{1}{2\pi}\int_{-\infty}^{\infty}\textrm{d}k\theta_{1}\left(t,k\right)\left(\partial_{k}\vec{e}\right)_{1}^{\textrm{dr}}\left(t,k\right)\frac{k^{2}}{2m},\\
&=\frac{1}{2\pi}\int_{-\infty}^{\infty}\textrm{d}k\theta_{1}\left(0,k-t\phi_{n}\right)\left[\left(\partial_{k}\vec{e}\right)_{1}^{\textrm{dr}}\left(0,k-t\phi_{n}\right)+\frac{1}{m}t\phi_{n}\left(\vec{q}_{n}\right)_{1}^{\textrm{dr}}\left(0,k-t\phi_{n}\right)\right]\frac{k^{2}}{2m},\\
&=\frac{1}{2\pi}\int_{-\infty}^{\infty}\textrm{d}k\theta_{1}\left(0,k\right)\left[\left(\partial_{k}\vec{e}\right)_{1}^{\textrm{dr}}\left(0,k\right)+\frac{1}{m}t\phi_{n}\left(\vec{q}_{n}\right)_{1}^{\textrm{dr}}\left(0,k\right)\right]\frac{\left(k+t\phi_{n}\right)^{2}}{2m},\\
&=\frac{1}{4\pi m}\int_{-\infty}^{\infty}\textrm{d}k\theta_{1}\left(0,k\right)\left(\partial_{k}\vec{e}\right)_{1}^{\textrm{dr}}\left(0,k\right)\left(k+t\phi_{n}\right)^{2}+\frac{t\phi_{n}}{4\pi m^{2}}\int_{-\infty}^{\infty}\textrm{d}k\theta_{1}\left(0,k\right)\left(\vec{q}_n^{\mathrm{dr}}\right)_{1}\left(0,k\right)\left(k+t\phi_{n}\right)^{2},\\
&=\frac{t\phi_{n}}{2\pi m}\int_{-\infty}^{\infty}\textrm{d}k\theta_{1}\left(0,k\right)\left(\partial_{k}\vec{e}\right)_{1}^{\textrm{dr}}\left(0,k\right)k+\frac{t\phi_{n}}{4\pi m^{2}}\int_{-\infty}^{\infty}\textrm{d}k\theta_{1}\left(0,k\right)\left(\vec{q}_n^{\mathrm{dr}}\right)_{1}\left(0,k\right)\left(k^{2}+t^{2}\phi_{n}^{2}\right),
\end{aligned}
\end{equation}
in which the dressed functions at time $t$ are transformed to their counterparts at $t=0$ at the second equality, the variable substitution $k\to k+t\phi_n$ is adopted at the third equality, and the parity property of the integrand is utilized at the fifth equality.  Based on integration by parts, the pressure \eqref{eq:p} can be expressed in the following form:
      \begin{equation}\label{eq:p_k}
      p=\frac{1}{2\pi}\int_{-\infty}^{\infty}\textrm{d}k\,k\,\theta_{1}\left(k\right)\frac{\partial \epsilon_{1}\left(k\right)}{\partial k}.
      \end{equation}
According to Eq. \eqref{eq:qe}, the pressure can be further expressed as
    \begin{equation}\label{eq:p_k_dr}
      p=\frac{1}{2\pi m}\int_{-\infty}^{\infty}\textrm{d}k\,k\,\theta_{1}\left(k\right)\left(\vec{q}_{k}^{\textrm{dr}}\right)_{1}\left(k\right).
      \end{equation}
Recall that $\vec{q}_{k}=m\partial_{k}\vec{e}$, thus, the first term in the last equation of Eq. \eqref{eq:je0_s} can be expressed as $t\phi_{n}p/m$. Based on Eq. \eqref{eq:rho_1}, the second term in the last equation of Eq. \eqref{eq:je0_s} can be rewritten as
      \begin{equation}
      \frac{t\phi_{n}}{2 m^{2}}\int_{-\infty}^{\infty}\textrm{d}k\,\rho_{1}\left(k^{2}+t^{2}\phi_{n}^{2}\right)=\frac{1}{m}t\phi_{n}\left(e_{0}+\frac{t^{2}\phi_{n}^{2}}{2m}n\right),
      \end{equation}
 where we have utilized the definitions of the particle density \eqref{eq:nm} and energy density \eqref{eq:e}.
The energy current thereby simplifies to the form
      \begin{equation}\label{eq:je0_final}
      j_{e_{0}}\left(t\right)=\frac{1}{m}t\phi_{n}\left(p+e_{0}+\frac{t^{2}\phi_{n}^{2}}{2m}n\right),
      \end{equation}
and the corresponding Drude weight, in the limit $t\phi_n \rightarrow 0$, is the enthalpy per unit length per unit mass:
      \begin{equation}
      D_{ne_0} = \lim_{\phi_n \rightarrow 0} \lim_{t \rightarrow \infty} \frac{j_{e_0}(x,t)}{t \phi_n} = \frac{1}{m} (p + e_0)=\frac{h}{m},
      \label{eq:Dne0_mu}
      \end{equation}
in which $h$ is enthalpy density.

Combining with the result of particle current \eqref{eq:jn5}, magnetization current  \eqref{eq:jm4}, and energy current \eqref{eq:je0_final}, the total energy current (including chemical potential and magnetic field contributions) is
\begin{equation}
j_e = j_{e_0} - \mu j_n - H j_m=\frac{1}{m}t\phi_{n}\left(p+e_{0}-\mu n-H m^z+\frac{t^{2}\phi_{n}^{2}}{2m}n\right),
\end{equation}
leading to the Drude weight
\begin{equation}
D_{ne} = \lim_{\phi_n \rightarrow 0} \lim_{t \rightarrow \infty} \frac{j_e(x,t)}{t \phi_n} = D_{ne_0} - \mu D_{nn} - H D_{nm} =  \frac{e_{0} + p - \mu n - H m^{z}}{m}.
\label{eq:Dne_mu}
\end{equation}
In thermal equilibrium, the Gibbs free energy \(G = L(e_{0} - \mu n - H m^{z}) - TS = -pL\), which implies that the Drude weight  $D_{ne}$ is exactly equal to the product of temperature and entropy density per unit mass
\begin{equation}
D_{ne} = \frac{Ts}{m}\label{eq:Dne_final}
\end{equation}
with entropy density $s=S/L$. 

%==============================================================================================
\section{Charge Drude Weight from Integral Form in Rapidity Space}

The Drude weight can be expressed as an integral in rapidity space\ucite{4}. For convenience, we suppress the explicit dependence on the rapidity variable \( k \) and the integration limits (\(-\infty\) to \(+\infty\)) in the notation. The general form of the Drude weight matrix elements is given by\cite{4}:
\begin{equation}
D_{ij}=\frac{1}{T}\int \mathrm{d}k\,\rho_{1}(1-\theta_{1})v_{1}^{2}(\vec{q}_{i}^{\mathrm{dr}})_{1}(\vec{q}_{j}^{\mathrm{dr}})_{1} + \frac{1}{T}\sum_{l}\int \mathrm{d}k\,\rho_{2,l}(1-\theta_{2,l})v_{2,l}^{2}(\vec{q}_{i}^{\mathrm{dr}})_{2,l}(\vec{q}_{j}^{\mathrm{dr}})_{2,l},
\label{eq:Dij}
\end{equation}
where the indices \(i, j\) can represent \(n, m, k, e_0, e\), etc. From Eq.~\eqref{eq:Dij}, it is evident that the Drude weight matrix is symmetric, i.e., \(D_{ij} = D_{ji}\).

In thermal equilibrium, the derivative of the occupation number with respect to \(k\) is related to the first three terms of the integrand in above expression
\begin{equation}\label{eq:dtheta_k}
\begin{aligned}
-\frac{\partial\theta_{i}\left(k\right)}{\partial k}
&=\frac{\partial}{\partial k}\left(\frac{-1}{1+\textrm{e}^{\epsilon_{i}\left(k\right)/T}}\right)\\
&=\left(1-\theta_{i}\left(k\right)\right)\theta_{i}\left(k\right)\frac{1}{T}\frac{\partial\epsilon_{i}\left(k\right)}{\partial k}\\
&=\left(1-\theta_{i}\left(k\right)\right)\rho_{i}\left(k\right)\frac{1}{T}\frac{\frac{\partial\epsilon_{i}\left(k\right)}{\partial k}}{\rho_{i}^{\textrm{tot}}\left(k\right)}\\
&=\frac{2\pi}{T}\left(1-\theta_{i}\left(k\right)\right)\rho_{i}\left(k\right)\frac{\left(\vec{q}_{k}^{\textrm{dr}}\right)_{i}\left(k\right)/m}{\left(\vec{q}_{n}^{\textrm{dr}}\right)_{i}\left(k\right)}\\
&=\frac{2\pi}{T}\left(1-\theta_{i}\left(k\right)\right)\rho_{i}\left(k\right)\frac{(\partial_{k}\vec{q}_{e})_{i}^{\textrm{dr}}(k)}{(\partial_{k}\vec{q}_{k})_{i}^{\textrm{dr}}(k)}\\
&=\frac{2\pi}{T}\left(1-\theta_{i}\left(k\right)\right)\rho_{i}\left(k\right)v_{i}\left(k\right)
\end{aligned}
\end{equation}
in which the definition of the pseudoenergy \eqref{eq:theta_epsilon} is adopted for the first equality, the definition of the occupation number \eqref{eq:theta} is used for the third equality, and the definition of the effective velocity \eqref{eq:v} is invoked for the sixth equality. Using the result from Eq. \eqref{eq:dtheta_k}, the Drude weight can be rewritten as
\begin{equation}
D_{ij}=-\frac{1}{2\pi}\int \mathrm{d}k\,\frac{\partial\theta_{1}}{\partial k}v_{1}(\vec{q}_{i}^{\mathrm{dr}})_{1}(\vec{q}_{j}^{\mathrm{dr}})_{1} - \frac{1}{2\pi}\sum_{l}\int \mathrm{d}k\,\frac{\partial\theta_{2,l}}{\partial k}v_{2,l}(\vec{q}_{i}^{\mathrm{dr}})_{2,l}(\vec{q}_{j}^{\mathrm{dr}})_{2,l}.
\label{eq:Dij_dtheta}
\end{equation}
Based on the definition of the effective velocity given by Eq. \eqref{eq:v}, the Drude weight $D_{ni}$ simplifies to
\begin{equation}\label{eq:Dni}
D_{ni}=-\frac{1}{2\pi m}\int \mathrm{d}k\,\frac{\partial\theta_{1}}{\partial k}(\vec{q}_{k}^{\mathrm{dr}})_{1}(\vec{q}_{i}^{\mathrm{dr}})_{1} - \frac{1}{2\pi m}\sum_{l}\int \mathrm{d}k\,\frac{\partial\theta_{2,l}}{\partial k}(\vec{q}_{k}^{\mathrm{dr}})_{2,l}(\vec{q}_{i}^{\mathrm{dr}})_{2,l}.
\end{equation}

The Drude weight $D_{nn}$ characterizing the linear response of the particle current to a field conjugate to the particle density is:
\begin{equation}
D_{nn}=-\frac{1}{2\pi m}\int \mathrm{d}k\,\frac{\partial\theta_{1}}{\partial k}(\vec{q}_{k}^{\mathrm{dr}})_{1}(\vec{q}_{n}^{\mathrm{dr}})_{1} - \frac{1}{2\pi m}\sum_{l}\int \mathrm{d}k\,\frac{\partial\theta_{2,l}}{\partial k}(\vec{q}_{k}^{\mathrm{dr}})_{2,l}(\vec{q}_{n}^{\mathrm{dr}})_{2,l}.
\label{eq:Dnn}
\end{equation}
Since the dressed function $\vec{q}_{n}^{\mathrm{dr}}$ is related to the total distribution function $\rho^\mathrm{tot}$ via Eq. \eqref{eq:qn_rho_tot}, we obtain
\begin{equation}\label{eq:Dnn_rho_tot}
D_{nn}=-\frac{1}{m}\int \mathrm{d}k\,\frac{\partial\theta_{1}}{\partial k}\rho_{1}^{\mathrm{tot}}(\vec{q}_{k}^{\mathrm{dr}})_{1} - \frac{1}{m}\sum_{l}\int \mathrm{d}k\,\frac{\partial\theta_{2,l}}{\partial k}\rho_{2,l}^{\mathrm{tot}}(\vec{q}_{k}^{\mathrm{dr}})_{2,l}.
\end{equation}
Differentiating TBA Eqs.~\eqref{eq:TBA} with respect to \(k\) yields the integral equations:
\begin{equation}\label{eq:rho_d}
\begin{aligned}
\frac{\partial\rho_{1}}{\partial k} &= \rho_{1}^{\mathrm{tot}}\frac{\partial\theta_{1}}{\partial k} + \theta_{1}\sum_{l=1}^{\infty}\phi_{l} * \frac{\partial}{\partial k}\rho_{2,l}, \\
\frac{\partial\rho_{2,l}}{\partial k} &= \rho_{2,l}^{\mathrm{tot}}\frac{\partial\theta_{2,l}}{\partial k} + \theta_{2,l}\left(\phi_{l} * \frac{\partial}{\partial k}\rho_{1} - \sum_{m=1}^{\infty}\Phi_{l,m} * \frac{\partial}{\partial k}\rho_{2,m}\right).
\end{aligned}
\end{equation}
Using the above integral equations, the derivative of the occupation number in Eq.~\eqref{eq:Dnn_rho_tot} can be transformed into derivatives of the rapidity distribution functions, leading to:
\begin{align}
D_{nn} &= -\frac{1}{m}\int \mathrm{d}k\,\frac{\partial\rho_{1}}{\partial k}(\vec{q}_{k}^{\mathrm{dr}})_{1} - \frac{1}{m}\sum_{l}\int \mathrm{d}k\,\frac{\partial\rho_{2,l}}{\partial k}(\vec{q}_{k}^{\mathrm{dr}})_{2,l} \label{eq:Dnn4} \\
&\quad + \frac{1}{m}\int \mathrm{d}k\,\theta_{1}(\vec{q}_{k}^{\mathrm{dr}})_{1} \sum_{l=1}^{\infty}\phi_{l} * \frac{\partial}{\partial k}\rho_{2,l} \nonumber \\
&\quad + \frac{1}{m}\sum_{l}\int \mathrm{d}k\,\theta_{2,l}(\vec{q}_{k}^{\mathrm{dr}})_{2,l} \left(\phi_{l} * \frac{\partial}{\partial k}\rho_{1} - \sum_{m=1}^{\infty}\Phi_{l,m} * \frac{\partial}{\partial k}\rho_{2,m}\right). \nonumber
\end{align}
Interchanging the order of integration, we obtain:
\begin{align}
D_{nn} &= -\frac{1}{m}\int \mathrm{d}k\,\frac{\partial\rho_{1}}{\partial k} \left[(\vec{q}_{k}^{\mathrm{dr}})_{1} - \sum_{l}\phi_{l} * \left(\theta_{2,l}(\vec{q}_{k}^{\mathrm{dr}})_{2,l}\right)\right] \nonumber \\
&\quad - \frac{1}{m}\sum_{l}\int \mathrm{d}k\,\frac{\partial\rho_{2,l}}{\partial k} \left[(\vec{q}_{k}^{\mathrm{dr}})_{2,l} - \phi_{l} * \left(\theta_{1}(\vec{q}_{k}^{\mathrm{dr}})_{1}\right) + \sum_{m}\Phi_{l,m} * \left(\theta_{2,m}(\vec{q}_{k}^{\mathrm{dr}})_{2,m}\right)\right]. \label{eq:Dnn5}
\end{align}
Applying the dressed equation for \(\vec{q}_{k}^{\mathrm{dr}}\) with $\vec{q}_{k}=\{k,0\}$, we find that the Drude weight \(D_{nn}\) is equal to the ratio of the particle density \(n\) to the mass:
\begin{equation}
D_{nn} = -\frac{1}{m}\int \mathrm{d}k\,\frac{\partial\rho_{1}}{\partial k}k = \frac{1}{m}\int \mathrm{d}k\,\rho_{1} = \frac{n}{m},
\label{eq:Dnn6}
\end{equation}
which is the same as that for a non-interacting system and does not depend on the interaction strength or temperature. This result is consistent with the conclusion for monoatomic gases derived from the second-order derivatives of pressure \cite{5}.

The Drude weight $D_{nm}$ characterizing the linear response of the magnetization current to a field conjugate to the particle density is:
\begin{equation}
D_{nm}=-\frac{1}{2\pi m}\int \mathrm{d}k\,\frac{\partial\theta_{1}}{\partial k}(\vec{q}_{m}^{\mathrm{dr}})_{1}(\vec{q}_{k}^{\mathrm{dr}})_{1} - \frac{1}{2\pi m}\sum_{l}\int \mathrm{d}k\,\frac{\partial\theta_{2,l}}{\partial k}(\vec{q}_{m}^{\mathrm{dr}})_{2,l}(\vec{q}_{k}^{\mathrm{dr}})_{2,l}.
\label{eq:Dnm}
\end{equation}
which follows from the expression \eqref{eq:Dni}.
Defining $\sigma_{1} \equiv (\vec{q}_{m}^{\mathrm{dr}})_{1}\theta_{1}$ and $\sigma_{2,l} \equiv (\vec{q}_{m}^{\mathrm{dr}})_{2,l}\theta_{2,l}$, and differentiating with respect to $k$ (taking into account the derivative of the dressed equation for $\vec{q}_{m}^{\mathrm{dr}}$), we obtain:
\begin{equation}\label{q_m_k}
\begin{aligned}
\frac{\partial\sigma_{1}}{\partial k} &= \frac{\partial\theta_{1}}{\partial k}(\vec{q}_{m}^{\mathrm{dr}})_{1} + \theta_{1}\sum_{l=1}^{\infty}\phi_{l} * \frac{\partial}{\partial k}\sigma_{2,l},\\
\frac{\partial\sigma_{2,l}}{\partial k} & = \frac{\partial\theta_{2,l}}{\partial k}(\vec{q}_{m}^{\mathrm{dr}})_{2,l} + \theta_{2,l}\left(\phi_{l} * \frac{\partial}{\partial k}\sigma_{1} - \sum_{m=1}^{\infty}\Phi_{l,m} * \frac{\partial}{\partial k}\sigma_{2,m}\right).
\end{aligned}
\end{equation}
Transforming the derivative of the occupation number in expression \eqref{eq:Dnm} into derivatives of $\sigma_{1}$ and $\sigma_{2,l}$ leads to:
\begin{equation}\label{eq:Dnm_sigma}
\begin{aligned}
D_{nm} &= -\frac{1}{2\pi m}\int \mathrm{d}k\,\frac{\partial\sigma_{1}}{\partial k} \left[(\vec{q}_{k}^{\mathrm{dr}})_{1} - \sum_{l}\phi_{l} * \left(\theta_{2,l}(\vec{q}_{k}^{\mathrm{dr}})_{2,l}\right)\right]  \\
&\quad - \frac{1}{2\pi m}\sum_{l}\int \mathrm{d}k\,\frac{\partial\sigma_{2,l}}{\partial k} \left[(\vec{q}_{k}^{\mathrm{dr}})_{2,l} - \phi_{l} * \left(\theta_{1}(\vec{q}_{k}^{\mathrm{dr}})_{1}\right) + \sum_{m}\Phi_{l,m} * \left(\theta_{2,m}(\vec{q}_{k}^{\mathrm{dr}})_{2,m}\right)\right].
\end{aligned}
\end{equation}
Using the dressed equation for $\vec{q}_{k}^{\mathrm{dr}}$, we find that $D_{nm}$ is proportional to an integral involving $\sigma_{1}$:
\begin{equation}
D_{nm} = -\frac{1}{2\pi m}\int \mathrm{d}k\,\frac{\partial\sigma_{1}}{\partial k}k = \frac{1}{2\pi m}\int \mathrm{d}k\,\sigma_{1} = \frac{1}{2\pi m}\int \mathrm{d}k\,(\vec{q}_{m}^{\mathrm{dr}})_{1}\theta_{1}.
\label{eq:Dnm4}
\end{equation}
Employing the identity (valid for arbitrary functions $f_{i}$ and $F_{i}$ with $i=1$ or $2,l$)
\begin{equation}
\int \mathrm{d}k\,f_{1}^{\mathrm{dr}}\theta_{1}F_{1} + \sum_{l=1}^{\infty}\int \mathrm{d}k\,f_{2,l}^{\mathrm{dr}}\theta_{2,l}F_{2,l} = \int \mathrm{d}k\,F_{1}^{\mathrm{dr}}\theta_{1}f_{1}+ \sum_{l=1}^{\infty}\int \mathrm{d}k\,F_{2,l}^{\mathrm{dr}}\theta_{2,l}f_{2,l},
\label{eq:identity}
\end{equation}
The dressed operator is shifted from $\vec{q}_{m}^{\mathrm{dr}}$ in \eqref{eq:Dnm4}  to $\vec{q}_{n}^{\mathrm{dr}}$, yielding Drude weight $D_nm$ proportional to the magnetization density $m^{z}$:
\begin{equation}
D_{nm} = \frac{1}{2\pi m}\left[\frac{1}{2}\int \mathrm{d}k\,\theta_{1}(\vec{q}_{n}^{\mathrm{dr}})_{1} - \sum_{l=1}^{\infty}l\int \mathrm{d}k\,\theta_{2,l}(\vec{q}_{n}^{\mathrm{dr}})_{2,l}\right] = \frac{1}{m}\left[\frac{1}{2}\int \mathrm{d}k\,\rho_{1} - \sum_{l=1}^{\infty}l\int \mathrm{d}k\,\rho_{2,l}\right] = \frac{m^{z}}{m}.
\label{eq:Dnm5}
\end{equation}
The Drude weight $D_{nm}$ is also identical to that of a non-interacting system and does not depend on interaction strength and temperature, exhibiting the characteristic properties of ballistic transport.

The Drude weight $D_{ne_{0}}$ characterizing the linear response of the energy current $j_{e_0}$ to a field conjugate to the particle density is:
\begin{equation}
D_{ne_{0}} = -\frac{1}{2\pi m}\int \mathrm{d}k\,\frac{\partial\theta_{1}}{\partial k}(\vec{q}_{e_{0}}^{\mathrm{dr}})_{1}(\vec{q}_{k}^{\mathrm{dr}})_{1} - \frac{1}{2\pi m}\sum_{l}\int \mathrm{d}k\,\frac{\partial\theta_{2,l}}{\partial k}(\vec{q}_{e_{0}}^{\mathrm{dr}})_{2,l}(\vec{q}_{k}^{\mathrm{dr}})_{2,l}. \label{eq:Dne0}
\end{equation}
By performing a calculation analogous to that for $D_{nm}$, we obtain:
\begin{align}
D_{ne_{0}} &= \frac{1}{2\pi m}\int \mathrm{d}k\,(\vec{q}_{e_{0}}^{\mathrm{dr}})_{1}\theta_{1} + \frac{1}{2\pi m^{2}}\int \mathrm{d}k\,\theta_{1}k(\vec{q}_{k}^{\mathrm{dr}})_{1} \nonumber \\
&= \frac{1}{m}\int \mathrm{d}k\,\frac{k^{2}}{2m}\rho_{1}(k) + \frac{1}{2\pi m^{2}}\int \mathrm{d}k\,\theta_{1}k(\vec{q}_{k}^{\mathrm{dr}})_{1} \nonumber \\
&= \frac{e_{0} + p}{m}. \label{eq:Dne02}
\end{align}
The Drude weight $D_{ne_{0}}$  is  exactly the enthalpy density (enthalpy per unit length)  per unit mass. Just as in classical fluid dynamics, the corresponding energy flow consists not only of the advection of internal energy $e_0$ but also the rate of work done against the surrounding pressure $p$. For fixed particle density, the Drude weight \(D_{ne_{0}}\) increases with interaction strength, polarization  $P = 2m^z/n$ and temperature.

Finally, we consider the Drude weight $D_{ne}$ characterizing the linear response of the total energy current $j_e$ (including chemical potential and magnetic field contributions) to a field conjugate to the particle density:
\begin{equation}
D_{ne}=\frac{1}{T}\int \mathrm{d}k\,\rho_{1}(1-\theta_{1})v_{1}^{2}(\vec{q}_{n}^{\mathrm{dr}})_{1}(\vec{q}_{e}^{\mathrm{dr}})_{1} + \frac{1}{T}\sum_{l}\int \mathrm{d}k\,\rho_{2,l}(1-\theta_{2,l})v_{2,l}^{2}(\vec{q}_{n}^{\mathrm{dr}})_{2,l}(\vec{q}_{e}^{\mathrm{dr}})_{2,l}. \label{eq:Dne}
\end{equation}
Due to the linearity of the dressed equations, we have \(\vec{q}_{e}^{\mathrm{dr}} = \vec{q}_{e_{0}}^{\mathrm{dr}} - \mu\vec{q}_{n}^{\mathrm{dr}} - H\vec{q}_{m}^{\mathrm{dr}}\), leading to:
\begin{equation}
D_{ne} = D_{ne_{0}} - \mu D_{nn} - H D_{nm} = \frac{e_{0} + p - \mu n - H m^{z}}{m}. \label{eq:Dne2}
\end{equation}
In thermal equilibrium, the Gibbs free energy \(G = L(e_{0} - \mu n - H m^{z}) - TS = -pL\), which implies that the Drude weight  $D_{ne}$ is exactly equal to the product of temperature and entropy density per unit mass
\begin{equation}
D_{ne} = \frac{Ts}{m}\label{eq:Dne3}
\end{equation}
with entropy density $s=S/L$. At zero temperature, \(D_{ne}\) vanishes. An increase in temperature leads to an increase in entropy, which in turn causes  \(D_{ne}\) to increase. Given a constant particle density, increases of interaction strength and polarization reduce the number of available microstates, which in turn leads to a decrease in entropy and, consequently, a reduction in \(D_{ne}\).

\section{Charge Drude Weight as the Second Derivative of Pressure}

By incorporating the momentum conservation term into the Yang-Yang equations (Eqs.~\eqref{eq:YangYang}, the Drude weight $D_{nn}$ can be equivalently defined through the second derivative of the pressure\ucite{5}
\begin{equation}
D_{nn} = 4\, \partial_{\lambda}^2 p \big|_{\lambda=0,\mu,H}
\label{eq:Dnnp}
\end{equation}
in the grand canonical ensemble which is equivalent to the second derivative of energy in the canonical ensemble under gauge field.

Following the derivation approach of Drude weight of the Lieb-Liniger model\ucite{5}, we present the calculation process for the two-component Fermi gas.
%Completing the square in the momentum term of the first Yang-Yang equation  (Eqs.~\eqref{eq:YangYang}) allows the chemical potential $\mu$ and the parameter $\lambda$ to be combined into an effective chemical potential $\tilde{\mu} = \mu + \frac{\lambda^2}{8m}$, leading to the relation about pressure:
%\begin{equation}
%p(\mu, \lambda) = p(\tilde{\mu}, 0).
%\end{equation}
The first Yang-Yang equation in Eqs.~\eqref{eq:YangYang} contains the single-particle energy term
\begin{equation}
\frac{k^{2}}{2m}-\frac{\lambda k}{2m}-\mu,
\end{equation}
where the term $-\lambda k/(2m)$ accounts for the total momentum conservation via the corresponding Lagrange multiplier $\lambda$.  This quadratic expression can be rearranged by completing the square:
\begin{equation}
\frac{k^{2}}{2m}-\frac{\lambda k}{2m}-\mu
= \frac{1}{2m}\left(k-\frac{\lambda}{2}\right)^{2}
-\left(\mu+\frac{\lambda^{2}}{8m}\right).
\end{equation}
Thus, defining the shifted rapidity variable
\begin{equation}
k' = k - \frac{\lambda}{2},
\qquad\text{and}\qquad
\tilde{\mu} = \mu + \frac{\lambda^{2}}{8m},
\end{equation}
the argument of the single-particle energy becomes $\frac{k'^{2}}{2m}-\tilde{\mu}$, exactly as in the $\lambda=0$ case but with the chemical potential replaced by $\tilde{\mu}$.

Crucially, the integral kernels $\phi_{l}(k-q)$ and $\Phi_{l,m}(k-q)$ depend only on differences of rapidities, and all rapidity integrations extend over the entire real line.  Consequently, the shift $k\to k+\lambda/2$ (and similarly for all integration variables in the convolutions) leaves the full system of Yang-Yang integral equations invariant, provided that $\mu$ is simultaneously replaced by $\tilde{\mu}$.  In other words, if $\{\epsilon_{1}(k;\mu,\lambda),\epsilon_{2,l}(k;\mu,\lambda)\}$ denotes the solution of Eqs.~\eqref{eq:YangYang} at given $\mu,\lambda$, then
\begin{equation}
\epsilon_{1}\!\left(k+\frac{\lambda}{2};\mu,\lambda\right)
= \epsilon_{1}(k;\tilde{\mu},0),\qquad
\epsilon_{2,l}\!\left(k+\frac{\lambda}{2};\mu,\lambda\right)
= \epsilon_{2,l}(k;\tilde{\mu},0),
\end{equation}
for all $l\ge 1$.  This invariance follows from the translational invariance of the integral operators and the fact that the spin-sector equations are independent of $\lambda$.

The pressure of the system is given by the integral
\begin{equation}
p(\mu,\lambda)
= \frac{T}{2\pi}\int_{-\infty}^{\infty}\mathrm{d}k\,
\ln\!\left(1+\exp\!\left[-\frac{\epsilon_{1}(k;\mu,\lambda)}{T}\right]\right).
\end{equation}
Performing the same change of variables $k = k' + \lambda/2$ and using the above identity for the pseudoenergy, we obtain
\begin{equation}
p(\mu,\lambda)
= \frac{T}{2\pi}\int_{-\infty}^{\infty}\mathrm{d}k'\,
\ln\!\left(1+\exp\!\left[-\frac{\epsilon_{1}(k';\tilde{\mu},0)}{T}\right]\right)
= p(\tilde{\mu},0).
\end{equation}
Thus, the momentum potential $\lambda$ enters the thermodynamics solely through the shifted chemical potential $\tilde{\mu}$, leading to the compact relation
\begin{equation}
p(\mu,\lambda) = p\!\left(\mu+\frac{\lambda^{2}}{8m},\,0\right).
\end{equation}

Applying the chain rule, the first derivative of the pressure with respect to $\lambda$ is:
\begin{equation}
\partial_{\lambda} p(\mu, \lambda) = \frac{\lambda}{4m} \partial_{\tilde{\mu}} p(\tilde{\mu}, 0),
\end{equation}
and the second derivative is:
\begin{equation}
\partial_{\lambda}^2 p(\mu, \lambda) = \frac{1}{4m} \partial_{\tilde{\mu}} p(\tilde{\mu}, 0) + \left( \frac{\lambda}{4m} \right)^2 \partial_{\tilde{\mu}}^2 p(\tilde{\mu}, 0).
\end{equation}
Evaluating at $\lambda = 0$ yields the Drude weight in terms of the particle density $n$:
\begin{equation}
D_{nn} = \frac{1}{m} \partial_{\mu} p(\mu, 0) = \frac{n}{m}.
\end{equation}

This expression for the Drude weight as a second derivative of the pressure (Eq.~\eqref{eq:Dnnp}) is equivalent to the integral form derived in rapidity space (Eq.~\eqref{eq:Dnn}). Starting from the definition of the pressure (Eq.~\eqref{eq:p}), its second derivative with respect to $\lambda$ involves both the first and second derivatives of the pseudoenergy $\epsilon_1$:
\begin{equation}
\partial_{\lambda}^2 p = \int \frac{\mathrm{d}k}{2\pi} \left[ \frac{\tilde{\theta}_1}{T} (\partial_{\lambda} \epsilon_1)^2 - \theta_1 \partial_{\lambda}^2 \epsilon_1 \right],
\label{eq:palpha2}
\end{equation}
where $\tilde{\theta}_i = \theta_i (1 - \theta_i)$.

The second derivative of the pseudoenergies vector $\vec{\epsilon}$ with respect to $\lambda$ satisfies the integral equation:
\begin{equation}
\partial_{\lambda}^2 \vec{\epsilon} = \left\{ -\sum_{l=1}^{\infty} \phi_l * \left[ \frac{\tilde{\theta}_{2,l}}{T} (\partial_{\lambda} \epsilon_{2,l})^2 \right],\; -\phi_l * \left[ \frac{\tilde{\theta}_1}{T} (\partial_{\lambda} \epsilon_1)^2 \right] + \sum_{m=1}^{\infty} \Phi_{lm} * \left[ \frac{\tilde{\theta}_{2,m}}{T} (\partial_{\lambda} \epsilon_{2,m})^2 \right] \right\}^{\mathrm{dr}}.
\end{equation}
Using the identity (Eq.~\eqref{eq:identity}), the dressing operation on $\partial_{\lambda}^2 \vec{\epsilon}$ can be transferred to the vector $\vec{q}_n$. Applying the dressed equation for $\vec{q}_n$, the term containing the second derivative of $\vec{\epsilon}$ can be expressed in terms of its first derivative:
\begin{equation}
\int \mathrm{d}k\, \theta_1 \partial_{\lambda}^2 \epsilon_1 = -\int \mathrm{d}k\, \frac{\tilde{\theta}_1}{T} (\partial_{\lambda} \epsilon_1)^2 \left[ (\vec{q}_n^{\mathrm{dr}})_1 - 1 \right] - \sum_{l=1}^{\infty} \int \mathrm{d}k\, \frac{\tilde{\theta}_{2,l}}{T} (\partial_{\lambda} \epsilon_{2,l})^2 (\vec{q}_n^{\mathrm{dr}})_{2,l}.
\end{equation}
Substituting this result into Eq.~\eqref{eq:palpha2} yields:
\begin{equation}
\partial_{\lambda}^2 p = \int \frac{\mathrm{d}k}{2\pi} \left[ \frac{\tilde{\theta}_1}{T} (\partial_{\lambda} \epsilon_1)^2 (\vec{q}_n^{\mathrm{dr}})_1 + \sum_{l=1}^{\infty} \frac{\tilde{\theta}_{2,l}}{T} (\partial_{\lambda} \epsilon_{2,l})^2 (\vec{q}_n^{\mathrm{dr}})_{2,l} \right].
\end{equation}

Finally, using the relations $(\partial_{\lambda} \vec{\epsilon})_i = - (\vec{q}_k^{\mathrm{dr}})_i / 2m = - v_i (\vec{q}_n^{\mathrm{dr}})_i / 2$ and $(\vec{q}_n^{\mathrm{dr}})_i \theta_i / 2\pi = \rho_i^{\mathrm{tot}} \theta_i = \rho_i$, we recover the integral form of $D_{nn}$ given in Eq.~\eqref{eq:Dnn}.
This establishes the equivalence between the thermodynamic definition (second derivative of pressure) and the microscopic definition (integral over rapidity space) of the Drude weight.

\section{Compressibility and Drude Weight at Zero Temperature}
The compressibility matrix $\boldsymbol{\kappa}$ and Drude weight matrix $\mathbf{D}$ are defined as
\begin{equation}
\boldsymbol{\kappa} = 
\begin{pmatrix}
\dfrac{\partial n}{\partial \mu} & \dfrac{\partial n}{\partial H} \\[8pt]
\dfrac{\partial m^z}{\partial \mu} & \dfrac{\partial m^z}{\partial H}
\end{pmatrix}, \qquad
\mathbf{D} = 
\begin{pmatrix}
D_{nn} & D_{nm} \\
D_{mn} & D_{mm}
\end{pmatrix}.
\end{equation}
At zero temperature, their matrix elements are given by the dressed charge formalism
\begin{equation}
\kappa_{ij} = \frac{1}{\pi} \left( \frac{Z_{i1}Z_{j1}}{v_c} + \frac{Z_{i2}Z_{j2}}{v_s} \right), \qquad
D_{ij} = \frac{1}{\pi} \left( v_c Z_{i1}Z_{j1} + v_s Z_{i2}Z_{j2} \right), \quad i,j = n,m .
\label{eq:kappa}
\end{equation}
We introduce the dressed charge matrix $\mathbf{Z}$ and the diagonal velocity matrix $\mathbf{v}$:
\begin{equation}
\mathbf{Z} = 
\begin{pmatrix}
Z_{n1} & Z_{n2} \\
Z_{m1} & Z_{m2}
\end{pmatrix}, \qquad 
\mathbf{v} = 
\begin{pmatrix}
v_c & 0 \\
0 & v_s
\end{pmatrix}.
\end{equation}

To derive the matrix product forms, we first compute the product $\mathbf{Z} \mathbf{v}^{-1} \mathbf{Z}^\mathsf{T}$, where $\mathbf{v}^{-1} = \operatorname{diag}(1/v_c, 1/v_s)$.  The $(i,j)$-element is
\begin{align*}
\left( \mathbf{Z} \mathbf{v}^{-1} \mathbf{Z}^\mathsf{T} \right)_{ij} 
&= \sum_{\alpha=1,2} Z_{i\alpha} (\mathbf{v}^{-1})_{\alpha\alpha} Z_{j\alpha} \\
&= \frac{Z_{i1} Z_{j1}}{v_c} + \frac{Z_{i2} Z_{j2}}{v_s}.
\end{align*}
Multiplying by the factor $1/\pi$ reproduces exactly $\kappa_{ij}$ in (\ref{eq:kappa}), giving
\begin{equation}
\boldsymbol{\kappa} = \frac{1}{\pi} \, \mathbf{Z} \, \mathbf{v}^{-1} \, \mathbf{Z}^\mathsf{T}.
\end{equation}

Similarly, for the Drude weight we compute $\mathbf{Z} \mathbf{v} \mathbf{Z}^\mathsf{T}$:
\begin{align*}
\left( \mathbf{Z} \mathbf{v} \mathbf{Z}^\mathsf{T} \right)_{ij} 
&= \sum_{\alpha=1,2} Z_{i\alpha} v_{\alpha\alpha} Z_{j\alpha} \\
&= v_c Z_{i1} Z_{j1} + v_s Z_{i2} Z_{j2}.
\end{align*}
Again, multiplying by $1/\pi$ yields $D_{ij}$, so that
\begin{equation}
\mathbf{D} = \frac{1}{\pi} \, \mathbf{Z} \, \mathbf{v} \, \mathbf{Z}^\mathsf{T}.
\end{equation}

A useful relation between $\mathbf{D}$ and $\boldsymbol{\kappa}$ can be obtained by defining the matrix $\mathbf{V} \equiv \mathbf{Z} \mathbf{v} \mathbf{Z}^{-1}$ (assuming $\mathbf{Z}$ is invertible).  Then $\mathbf{V}^2 = (\mathbf{Z} \mathbf{v} \mathbf{Z}^{-1})^2 = \mathbf{Z} \mathbf{v}^2 \mathbf{Z}^{-1}$.  Using the expressions above,
\begin{align}
\mathbf{V}^2 \boldsymbol{\kappa} 
&= \left( \mathbf{Z} \mathbf{v}^2 \mathbf{Z}^{-1} \right) \left( \frac{1}{\pi} \mathbf{Z} \mathbf{v}^{-1} \mathbf{Z}^\mathsf{T} \right) \\
&= \frac{1}{\pi} \mathbf{Z} \mathbf{v}^2 \mathbf{v}^{-1} \mathbf{Z}^\mathsf{T} \\
&= \frac{1}{\pi} \mathbf{Z} \mathbf{v} \mathbf{Z}^\mathsf{T} = \mathbf{D}.
\end{align}
Thus we obtain the compact relation
\begin{equation}
\mathbf{D} = \mathbf{V}^2 \, \boldsymbol{\kappa}, \qquad 
\text{with } \mathbf{V}^2 = \mathbf{Z} \mathbf{v}^2 \mathbf{Z}^{-1}.
\end{equation}
Multiplying on the right by $\bm{\kappa}^{-1}$, we obtain the concise form:
\begin{equation}
\mathbf{D} \bm{\kappa}^{-1} = \mathbf{V}^2 = \mathbf{Z} \mathbf{v}^2 \mathbf{Z}^{-1}.
\label{eq:concise_form} 
\end{equation}

We first compute the inverse of the compressibility matrix:
\begin{equation}
\boldsymbol{\kappa}^{-1} = \left( \frac{1}{\pi} \, \mathbf{Z} \, \mathbf{v}^{-1} \, \mathbf{Z}^\mathsf{T} \right)^{-1} 
= \pi \, (\mathbf{Z}^\mathsf{T})^{-1} \, \mathbf{v} \, \mathbf{Z}^{-1}.
\end{equation}
Substituting this expression into the following formula yields
\begin{align}
\frac{1}{\pi^2} (\mathbf{Z}\mathbf{Z}^\mathsf{T}) \, \boldsymbol{\kappa}^{-1} \, (\mathbf{Z}\mathbf{Z}^\mathsf{T})
&= \frac{1}{\pi^2} (\mathbf{Z}\mathbf{Z}^\mathsf{T}) 
   \Bigl[ \pi \, (\mathbf{Z}^\mathsf{T})^{-1} \, \mathbf{v} \, \mathbf{Z}^{-1} \Bigr] 
   (\mathbf{Z}\mathbf{Z}^\mathsf{T}) \\
&= \frac{1}{\pi} \, \mathbf{Z} \, 
   \Bigl( \mathbf{Z}^\mathsf{T} (\mathbf{Z}^\mathsf{T})^{-1} \Bigr) \, 
   \mathbf{v} \, 
   \Bigl( \mathbf{Z}^{-1} \mathbf{Z} \Bigr) \, 
   \mathbf{Z}^\mathsf{T}.
\end{align}
Using the identities \(\mathbf{Z}^\mathsf{T} (\mathbf{Z}^\mathsf{T})^{-1} = \mathbf{I}\) and \(\mathbf{Z}^{-1} \mathbf{Z} = \mathbf{I}\), we obtain
\begin{align}
\frac{1}{\pi} \, \mathbf{Z} \, \mathbf{I} \, \mathbf{v} \, \mathbf{I} \, \mathbf{Z}^\mathsf{T}
= \frac{1}{\pi} \, \mathbf{Z} \, \mathbf{v} \, \mathbf{Z}^\mathsf{T}
= \mathbf{D}.
\end{align}
Thus, the following identity is established:
\begin{equation}
\mathbf{D} = \frac{1}{\pi^2} (\mathbf{Z}\mathbf{Z}^\mathsf{T}) \, \boldsymbol{\kappa}^{-1} \, (\mathbf{Z}\mathbf{Z}^\mathsf{T}).
\end{equation}

These matrix product forms provide a concise connection between thermodynamic response functions and the dressed charge parameters in integrable systems.

\section{Spin Drude Weight at Zero Temperature and Phase Boundaries}

The spin Drude weight can be expressed in integral form in rapidity space as:
\begin{equation}
D_{mi} = \frac{1}{T} \int \mathrm{d}k\, \rho_1 (1 - \theta_1) v_1^2 (\vec{q}_m^{\mathrm{dr}})_1 (\vec{q}_i^{\mathrm{dr}})_1 
+ \frac{1}{T} \sum_l \int \mathrm{d}k\, \rho_{2,l} (1 - \theta_{2,l}) v_{2,l}^2 (\vec{q}_m^{\mathrm{dr}})_{2,l} (\vec{q}_i^{\mathrm{dr}})_{2,l}.
\label{eq:Dmi}
\end{equation}
Using the derivative of the occupation number (Eq.~\eqref{eq:dtheta_k}), this expression simplifies to:
\begin{equation}
D_{mi} = -\frac{1}{2\pi} \int \mathrm{d}k\, \frac{\partial\theta_1}{\partial k} v_1 (\vec{q}_m^{\mathrm{dr}})_1 (\vec{q}_i^{\mathrm{dr}})_1 
- \frac{1}{2\pi} \sum_l \int \mathrm{d}k\, \frac{\partial\theta_{2,l}}{\partial k} v_{2,l} (\vec{q}_m^{\mathrm{dr}})_{2,l} (\vec{q}_i^{\mathrm{dr}})_{2,l}.
\label{eq:Dmi2}
\end{equation}

At zero temperature, only the $l=1$ string contribution remains significant. The occupation number equals unity within the Fermi sea and zero outside, making its derivative with respect to $k$ a positive (negative) delta function at the negative (positive) Fermi point. Consequently, the Drude weight receives contributions only from these Fermi points:
\begin{equation}
D_{mi} = \frac{1}{\pi} \left[ v_1 (\vec{q}_m^{\mathrm{dr}})_1 (\vec{q}_i^{\mathrm{dr}})_1 \right]_{k=k_0} + \frac{1}{\pi} \left[ v_{2,1} (\vec{q}_m^{\mathrm{dr}})_{2,1} (\vec{q}_i^{\mathrm{dr}})_{2,1} \right]_{k=\lambda_0},
\label{eq:DmiT0}
\end{equation}
where $k_0$ and $\lambda_0$ represent the positive boundaries of the rapidity distribution functions $\rho_1(k)$ and $\rho_{2,1}(k)$ at zero temperature, determined by $\epsilon_1(k_0) = \epsilon_{2,1}(\lambda_0) = 0$.

At zero temperature, the dressed equations (Eqs.~\eqref{eq:dressed_equ} simplify to:
\begin{align}
f_1^{\mathrm{dr}}(k) &= f_1(k) + \int_{-\lambda_0}^{\lambda_0} \phi_1(k-\lambda) f_{2,1}^{\mathrm{dr}}(\lambda) \mathrm{d}\lambda, \label{eq:dressed1_T0} \\
f_{2,1}^{\mathrm{dr}}(\lambda) &= f_{2,1}(\lambda) + \int_{-k_0}^{k_0} \phi_1(\lambda-k) f_1^{\mathrm{dr}}(k) \mathrm{d}k - \int_{-\lambda_0}^{\lambda_0} \phi_2(\lambda-\lambda') f_{2,1}^{\mathrm{dr}}(\lambda') d\lambda'. \label{eq:dressed2_T0}
\end{align}
Furthermore, the Yang-Yang integral equations imply $\vec{\epsilon} = \vec{q}_e^{\mathrm{dr}}$, which leads to $D_{me} = 0$ at zero temperature.

For the fully polarized Fermi gas ($P=1$), the Drude weight contains only the first term:
\begin{equation}
D_{mi} = \frac{1}{\pi} \left[ v_1 (\vec{q}_m^{\mathrm{dr}})_1 (\vec{q}_i^{\mathrm{dr}})_1 \right]_{k=k_0}, \label{eq:DmiP1}
\end{equation}
where $(\vec{q}_m^{\mathrm{dr}})_1 = 1/2$, $k_0 = \pi n = k_F$ is the Fermi momentum, the charge sound velocity $v_1(k_0) = k_F/m = v_F$ is the Fermi velocity, and $(\vec{q}_{e_0}^{\mathrm{dr}})_1(k_0) = k_F^2/2m = E_F$ is the Fermi energy. This yields:
\[
D_{mm} = \frac{v_F}{4\pi} = \frac{n}{4m}, \quad D_{me_0} = \frac{v_F E_F}{2\pi} = \frac{n E_F}{2m},
\]
which represent the upper limits at zero temperature.

For strong interactions, the spin sound velocity $v_2(\lambda_0)$ becomes negligible for any polarization, and the Drude weight is again dominated by the first term in Eq.~\eqref{eq:DmiP1}. A Taylor expansion of the integral kernel $\phi_n(k)$ for small $1/c$ gives:
\begin{equation}
D_{mm} = \left(\frac{P}{2}\right)^2 \frac{v_F}{\pi}, \quad D_{me_0} = \frac{v_F E_F}{\pi} \frac{P}{2} \left[ 1 - \frac{16}{3\gamma}(1-P) + \frac{20}{\gamma^2}(1-P)^2 \right],
\end{equation}
both increasing with polarization. Here dimensionless interaction strength $\gamma$ is defined as $\gamma=c/n$. For the non-interacting Fermi gas, the integral kernel $\phi_n(k)$ in dressed Eq.~\eqref {eq:dressed1_T0} and  ~\eqref {eq:dressed2_T0}  becomes a delta function, yielding $D_{mm} = v_F/4\pi$ (independent of polarization) and $D_{me_0} = \frac{v_F E_F}{\pi} \frac{(P^2+3)P}{8}$ (polarization-dependent).

For zero temperature, there are three phases, including the vacuum  state, fully polarized phase, and the mixed phase. The boundary between vacuum state and fully polarized phase is determined by 
\begin{equation}
\mu_\uparrow=\mu+H/2=0.
\label{eq:bound1}
\end{equation}
 If we define the energy unit $\epsilon_b=\hbar^2 c^2/2m$, the boundary between fully polarized phase and the mixed phase is determined by 
\begin{align}
H&=\frac{\epsilon_{b}}{\pi}\left\{ \left[2\tilde{\mu}_{\uparrow}+\frac{1}{2}\right]\arctan\left(2\sqrt{\tilde{\mu}_{\uparrow}}\right)-\sqrt{\tilde{\mu}_{\uparrow}}\right\} \\
\mu&=\tilde{\mu}_{\uparrow}\epsilon_{b}-H/2.
\label{eq:bound2}
\end{align}
with arbitrary non-negative $\tilde{\mu}_{\uparrow}$.

\section{Drude Weight at the Criticality}
We consider criticality of Drude weight near phase transition between the fully polarized state and mixed state.
In the low-temperature limit, only real-$k$ quasiparticles and length-1 $\lambda$-strings contribute to the Drude weights. For convenience, in the following discussion, we replace the subscript 2,1 with 2 for all quantities, e.g., $\epsilon_{2,1}(\lambda)=\epsilon_{2}(\lambda)$, $\theta_{2,1}=\theta_{2}$ and $f_{2,1}=f_{2}$. Accordingly, Yang-Yang integral equations \eqref{eq:YangYang} reduce to

\begin{equation}
\epsilon_{1}(k)	=\frac{k^{2}}{2m}-\mu-\frac{H}{2}-T\int\mathrm{d}\lambda\phi_{1}(k-\lambda)\ln\left(1+e^{-\epsilon_{2}(\lambda)/T}\right),
\label{eq.dressed-energy1}
\end{equation}
\begin{equation}
\epsilon_{2}(\lambda)	=H-T\int\mathrm{d}k\phi_{1}(\lambda-k)\ln\left(1+e^{-\epsilon_{1}(k)/T}\right)+T\int\mathrm{d}\lambda'\phi_{2}(\lambda-\lambda')\ln\left(1+e^{-\epsilon_{2}(\lambda')/T}\right).
\label{eq.dressed-energy2}
\end{equation}
 In the vicinity of the phase transition between the fully polarized and mixed phase, the critical behavior is dominated by the spin degrees of freedom. Consequently, $\lambda$ can be regarded as a small parameter, and the kernel functions appearing in Yang-Yang integral equations are expanded with respect to $\lambda$\ucite{6} up to $\mathcal{O}(\lambda^{2})$.
\begin{equation}
\phi_{1}(k-\lambda)\simeq\frac{A_{1}(k)}{2}+\frac{A_{2}(k)}{2}\lambda^{2}+C_{\phi_1},
\end{equation}
\begin{equation}
\phi_{2}(\lambda-\lambda')\simeq\left(\frac{1}{\pi c}-\frac{\lambda^{2}}{\pi c^{3}}\right)+\lambda'{}^{2}\left(-\frac{1}{\pi c^{3}}+\frac{6\lambda^{2}}{\pi c^{5}}\right)+C_{\phi_2}.
\end{equation}
Here $A_{1}=2\phi_{1}(k)$ and $A_{2}=\partial^2\phi_{1}(k)/\partial k^2$. $C_{\phi_1}$ and $C_{\phi_2}$ are linear functions of $\lambda$ or $\lambda'$ and thus odd functions, leading to a vanishing integral over the entire space.  Substituting above expansions into Eqs.\eqref{eq.dressed-energy1} and \eqref{eq.dressed-energy2}, we obtain the approximate pseudo-energies in the critical regime
\begin{equation}
\epsilon_{1}(k)	=\frac{k^{2}}{2m}-\mu-\frac{H}{2}-A_{1}(k)F_{1}-A_{2}(k)F_{2},
\end{equation}
\begin{equation}
\epsilon_{2}(\lambda)	=D_{1}+D_{2}\lambda^{2},
\end{equation}
with
\[
D_{1}=H-C_{1}+\frac{F_{1}}{\pi c}-\frac{F_{2}}{\pi c^{3}}, D_{2}=-C_{2}-\frac{F_{1}}{\pi c^{3}}+\frac{6F_{2}}{\pi c^{5}},
\]
in which
\[
\begin{aligned}
C_{1} &= T\int_{0}^{\infty}\mathrm{d}k\,A_{1}(k)
\ln\!\left(1+e^{-\epsilon_{1}(k)/T}\right), 
&\qquad
C_{2} &= T\int_{0}^{\infty}\mathrm{d}k\,A_{2}(k)
\ln\!\left(1+e^{-\epsilon_{1}(k)/T}\right),\\
F_{1} &= T\int_{0}^{\infty}\mathrm{d}\lambda\,
\ln\!\left(1+e^{-\epsilon_{2}(\lambda)/T}\right), 
&\qquad
F_{2} &= T\int_{0}^{\infty}\mathrm{d}\lambda\,\lambda^{2}
\ln\!\left(1+e^{-\epsilon_{2}(\lambda)/T}\right).
\end{aligned}
\]

Likewise, using Eqs.~\eqref{eq:TBA} and \eqref{eq:dressed_equ}, one obtains the low-temperature expressions for the total densities of states in the vicinity of the phase transition, together with the corresponding dressed functions, which take the form
\begin{align}
\rho_{1}^{\mathrm{tot}}(k) &= \frac{1}{2\pi}+A_{1}(k)R_{1}+A_{2}(k)R_{2}, \\
\rho_{2}^{\mathrm{tot}}(\lambda) &= \left(S_{2}+\frac{R_{1}}{\pi c^{3}}-\frac{6R_{2}}{\pi c^{5}}\right)\lambda^{2}
+\left(S_{1}-\frac{R_{1}}{\pi c}+\frac{R_{2}}{\pi c^{3}}\right), \label{eq.rho_2}\\
f_{1}^{\mathrm{dr}}(k) &= f_{1}(k)+A_{1}(k)V_{1}+A_{2}(k)V_{2}, \\
f_{2}^{\mathrm{dr}}(\lambda) &= \left(W_{2}+\frac{V_{1}}{\pi c^{3}}-\frac{6V_{2}}{\pi c^{5}}\right)\lambda^{2}
+\left(f_{2}(\lambda)+W_{1}-\frac{V_{1}}{\pi c}+\frac{V_{2}}{\pi c^{3}}\right).
\end{align}
Here $R_{1(2)},S_{1(2)},V_{1(2)}$ and $W_{1(2)}$ are defined as:
\[
\begin{aligned}
R_{1} &= \int_{0}^{\infty}\mathrm{d}\lambda\,
\rho_{2}^{\mathrm{tot}}(\lambda)\,\theta_{2}(\lambda),
&\qquad
R_{2} &= \int_{0}^{\infty}\mathrm{d}\lambda\,\lambda^{2}
\rho_{2}^{\mathrm{tot}}(\lambda)\,\theta_{}(\lambda),\\
S_{1} &= \int_{0}^{\infty}\mathrm{d}k\,A_{1}(k)
\rho_{1}^{\mathrm{tot}}(k)\,\theta_{1}(k),
&\qquad
S_{2} &= \int_{0}^{\infty}\mathrm{d}k\,A_{2}(k)
\rho_{1}^{\mathrm{tot}}(k)\,\theta_{1}(k),\\
V_{1} &= \int_{0}^{\infty}\mathrm{d}\lambda\,
f_{2}^{\mathrm{dr}}(\lambda)\,\theta_{2}(\lambda),
&\qquad
V_{2} &= \int_{0}^{\infty}\mathrm{d}\lambda\,\lambda^{2}
f_{2}^{\mathrm{dr}}(\lambda)\,\theta_{2}(\lambda),\\
W_{1} &= \int_{0}^{\infty}\mathrm{d}k\,A_{1}(k)
f_{1}^{\mathrm{dr}}(k)\,\theta_{1}(k),
&\qquad
W_{2} &= \int_{0}^{\infty}\mathrm{d}k\,A_{2}(k)
f_{1}^{\mathrm{dr}}(k)\,\theta_{1}(k).
\end{aligned}
\]

At the boundary of the phase transition for the zero-temperature limit, the Fermi boundary of the spin strings collapses to $\lambda_{0}=0$ and the corresponding spin pseudo-energy satisfies $\epsilon_{2}>0$. Under these conditions, the coefficients reduce to $R_{1}=R_{2}=V_{1}=V_{2}=0$. As a result, the total densities of states and the dressed functions simplify to
\begin{align}
\rho_{1}^{0} &= \frac{1}{2\pi},
&\qquad
\rho_{2}^{\mathrm{tot}}(0) &= \frac{1}{2\pi}\int_{0}^{k_{0}}\mathrm{d}k\,A_{1}(k),
\label{eq.rho01} \\
f_{1}^{0\mathrm{dr}}(k) &= f_{1}(k),
&\qquad
f_{2}^{0\mathrm{dr}}(0) &= f_{2}(0)+\int_{0}^{k_{0}}\mathrm{d}k\,A_{1}(k)f_{1}(k).
\label{eq.rho02}
\end{align}
Here $k_{0}$ denotes the charge Fermi momentum, which is determined by the condition $\epsilon_{1}(k_{0})=0$. At finite temperature, it is necessary to incorporate corrections arising from the spin sector into both the total density of states and the dressed equations. To this end, retaining the second-order contribution in a small-$\lambda$ expansion yields
\begin{align}
\rho_{2}^{\mathrm{tot}}(\lambda) & =\rho_{2}^{\mathrm{tot}}(0)+S_{2}\lambda^{2}=\rho_{2}^{\mathrm{tot}}(0)+\frac{\lambda^{2}}{2\pi}\int_{0}^{k_{0}}\mathrm{d}kA_{2}(k)\\
f_{2}^{\mathrm{dr}}(\lambda) & =f_{2}^{\mathrm{dr}}(0)+W_{2}\lambda^{2}=f_{2}^{0\mathrm{dr}}(0)+\lambda^{2}\int_{0}^{k_{0}}\mathrm{d}kA_{2}(k)f_{1}(k)
\end{align}
At the phase transition, the charge excitations remain in the low-temperature regime. Consequently, even at finite temperature, the dressed charge retains the same structure and can be written as
\begin{align}
f_{1}^{\mathrm{dr}}(k) & =f_{1}(k)+A_{1}(k)V_{1}\nonumber \\
 & =f_{1}(k)+A_{1}(k)\int_{0}^{\infty}\mathrm{d}\lambda f_{2}^{\mathrm{dr}}(\lambda)\theta_{2}(\lambda)\nonumber \\
 & \simeq f_{1}(k)+A_{1}(k)f_{2}^{0\mathrm{dr}}(0)\int_{0}^{\infty}\mathrm{d}\lambda\theta_{2}(\lambda)\nonumber \\
 & =f_{1}(k)+A_{1}(k)f_{2}^{0\mathrm{dr}}(0)I_{1/2}
  \label{eq.f1}
\end{align}
where 
\[
I_{\nu}\equiv\int_{0}^{\infty}d\lambda\,\lambda^{2\nu-1}\,\theta_{2}(\lambda)
=-\frac{\Gamma(\nu)}{2}\left(\frac{T}{D_{2}}\right)^{\nu}
\mathrm{Li}_{\nu}\!\left(-e^{-D_{1}/T}\right)
\]
Here, $\Gamma(x)$ denotes the Gamma function, which satisfies $\Gamma\!\left(\tfrac{1}{2}\right)=\sqrt{\pi}$ and $\Gamma\!\left(\tfrac{3}{2}\right)=\tfrac{\sqrt{\pi}}{2}$ and $\mathrm{Li}_{\nu}(x)$ denotes the Polylogarithm functions.
Similarly, the total density of states of the charge branch at finite temperature can be expressed as
\begin{align}
\rho_{1}^{\mathrm{tot}}(k) & =\rho_{1}^{0}+A_{1}(k)R_{1}\nonumber \\
 & \simeq \frac{1}{2\pi}+A_{1}(k)\rho_{2}^{\mathrm{tot}}(0)I
 \label{eq.rho1}
\end{align}

From Eq.~\eqref{eq:Dij}, the Drude weights can be decomposed into the sum of the charge contribution and the spin-string contribution,
\[
D_{ij}  =\sum_{\alpha=1,2}{D_{ij,\alpha}},
\]
with
\begin{align}
D_{ij,\alpha} & =\frac{1}{T}\int\mathrm{d}k\rho_{\alpha}(k)\bigl(1-\theta_{\alpha}(k)\bigr)v_{\alpha}^{2}(q_{i}^{\mathrm{dr}})_{\alpha}(q_{j}^{\mathrm{dr}})_{\alpha}.
\end{align}
In the following calculations, $(q_{\beta})_{1}$ and $(q_{\beta})_{2}$ ($\beta=i,j$) are bare charges independent of the $k$ and $\lambda$. Their values are fixed by the transport channel under consideration and correspond to the charge branch and the spin-string branch, respectively. For different conserved quantities, one take
\[
\{(q_{n})_{1},(q_{n})_{2}\}=\{1,0\},\qquad
\{(q_{m})_{1},(q_{m})_{2}\}=\left\{\tfrac{1}{2},-1\right\}.
\]

In the low-temperature limit, the charge contribution $D_{ij,1}$ admits a Sommerfeld expansion, yielding
\begin{align}
D_{ij,1} & =\frac{1}{\pi}\left.(q_{i}^{\mathrm{dr}})_{1}(q_{j}^{\mathrm{dr}})_{1}v_{1}(k)\right|_{k=k_{0}}+\frac{\pi T^{2}}{6}\left.\frac{\mathrm{d}^{2}}{\mathrm{d}\epsilon_{1}^{2}}\left[(q_{i}^{\mathrm{dr}})_{1}(q_{j}^{\mathrm{dr}})_{1}v_{1}(k)\right]\right|_{\epsilon_{1}=0}.
\label{eq.D1}
\end{align}
Using Eqs.~\eqref{eq.dressed-energy2} and \eqref{eq.rho_2}, the effective velocity of the quasi-particles in the charge sector can be written as 
\[
v_{1}(k_{0})=\frac{\epsilon_{1}'(k_{0})}{2\pi\rho_{1}^{\mathrm{tot}}(k_{0})}\simeq\epsilon_{1}^{'}(k_{0})\left[1-2\pi A_{1}(k_{0})\rho_{2}^{\mathrm{tot}}(0)I\right].
\]
Substituting this expression into the zero-temperature term of Eqs.~\eqref{eq.D1} and combined with Eq. ~\eqref{eq.f1}, we obtain the following approximation
\begin{align*}
	&\frac{1}{\pi}\left.(q_{i}^{\mathrm{dr}})_{1}(q_{j}^{\mathrm{dr}})_{1}v_{1}(k)\right|_{k=k_{0}}\\
	&\simeq\frac{1}{\pi}\epsilon_{1}'(k_{0})
	\left\{(q_{i})_{1}(q_{j})_{1}+A_{1}(k_{0})\left[(q_{j})_{1}(q_{i}^{\mathrm{dr}})_{2}^{0}(0)+(q_{i})_{1}(q_{j}^{\mathrm{dr}})_{2}^{0}(0)\right]I
	   -(q_{i})_{1}(q_{j})_{1}2\pi A_{1}(k_{0})\rho_{2}^{\mathrm{tot}}(0) I\right\}\\
	&=D(k_{0})+D(k_{0})\left\{-2\pi A_{1}(k_{0})\rho_{2}^{\mathrm{tot}}(0)+\frac{A_{1}(k_{0})}{(q_{i})_{1}(q_{j})_{1}}\left[(q_{j})_{1}(q_{i}^{\mathrm{dr}})_{2}^{0}(0)+(q_{i})_{1}(q_{j}^{\mathrm{dr}})_{2}^{0}(0)\right]\right\} I
\end{align*}
where only the leading contribution and its first correction from the spin sector are retained.
Here we have defined the zero-temperature Drude weight as $D(k_{0})=\frac{1}{\pi}\epsilon_{1}^{'}(k_{0})(q_{i})_{1}(q_{j})_{1}$. Noting that $\epsilon_{1}(k_{0})=0$ at the Fermi boundary, only the leading term contributes in the zero-temperature limit, and one has
 \[
(q_{i}^{\mathrm{dr}})_{1}(q_{j}^{\mathrm{dr}})_{1}v_{1}(k)\simeq\pi D(k).
\]
Accordingly, the second derivative appearing in Eq.~\eqref{eq.D1} should be understood as being taken with respect to the energy variable $\epsilon_{1}$ while keeping the full functional dependence on $k$ and only afterward evaluated at the Fermi point. In this sense, the second derivative can be equivalently written as
\begin{align}
\frac{\pi T^{2}}{6}\left.\frac{d^{2}}{d\epsilon_{1}^{2}}\left((q_{i}^{\mathrm{dr}})_{1}(q_{j}^{\mathrm{dr}})_{1}v_{1}(k)\right)\right|_{\epsilon_{1}=0} & \simeq\frac{\pi^{2}T^{2}}{6}\left.\frac{\mathrm{d}^{2}}{\mathrm{d}\epsilon_{1}^{2}}D(k)\right|_{k=k_{0}}.
\end{align}
As can be seen from Eqs.~\eqref{eq.rho01} and \eqref{eq.rho02}, in the zero-temperature limit the kernel $A_{1}(k_{0})$ can be replaced by the corresponding derivatives with respect to the Fermi momentum $k_0$, evaluated at the Fermi point,
\[
\left(f_{2}^{0\mathrm{dr}}(0)\right)'=A_{1}(k_{0})f_{1}(k_{0}),\qquad
\left(\rho_{2}^{\mathrm{tot}}(0)\right)'=\frac{A_{1}(k_{0})}{2\pi},\qquad
\rho_{1}^{0}=\frac{1}{2\pi}.
\]
Based on this substitution and upon collecting all contributions, the charge-branch contribution to the Drude weight can finally be written as
\begin{align}
D_{ij,1} & =D(k_{0})\left[1-I_{\frac{1}{2}}\sum_{\beta}\frac{\frac{1}{2}\rho_{2}^{\mathrm{tot}}(0)\left((q_{\beta}^{\mathrm{dr}})_{2}^{0}(0)\right)'-\left(\rho_{2}^{\mathrm{tot}}(0)\right)'(q_{\beta}^{\mathrm{dr}})_{2}^{0}(0)}{\rho_{1}^{\mathrm{0}}(q_{\beta}^{\mathrm{dr}})^{0}_{1}(0)}\right]+\frac{\pi^{2}T^{2}}{6}\left.\frac{\mathrm{d}^{2}}{\mathrm{d}\epsilon_{1}^{2}}D(k)\right|_{k=k_{0}}.
\end{align}

On the other hand, the spin-string contribution is given by
\begin{align}
D_{ij,2}=\frac{1}{T}\int_{-\infty}^{\infty}\mathrm{d}\lambda\theta_{2}(\lambda)\left(1-\theta_{2}(\lambda)\right)\frac{(q_{i}^{\mathrm{dr}})_{2}^{0}(q_{j}^{\mathrm{dr}})_{2}}{4\pi^{2}\rho_{2}^{\mathrm{tot}}(\lambda)}\left(\epsilon'_{2}(\lambda)\right)^{2}.
\end{align}
Near $\lambda=0$, the ratio $\frac{(q_{i}^{\mathrm{dr}})_{2}(\lambda)\,(q_{j}^{\mathrm{dr}})_{2}(\lambda)}{\rho_{2}^{\mathrm{tot}}(\lambda)}$ can be evaluated at its zeroth-order value, while the derivative of the pseudo energy behaves as
$\bigl(\epsilon'_{2}(\lambda)\bigr)^2 = 4D_2^{\,2}\lambda^2$.
Therefore, the spin-string contribution to the Drude weight is finally given by
\begin{align}
D_{ij,2} & =\frac{2}{T}\int_{0}^{\infty}\mathrm{d}\lambda\theta_{2}(\lambda)\left(1-\theta_{2}(\lambda)\right)\frac{(q_{i}^{\mathrm{dr}})_{2}^{0}(0)(q_{j}^{\mathrm{dr}})_{2}^{0}(0)}{4\pi^2\rho_{2}^{\mathrm{tot}}(0)}4D_{2}^{2}\lambda^{2}\nonumber \\
 & =8D_{2}^{2}\frac{(q_{i}^{\mathrm{dr}})_{2}^{0}(0)(q_{j}^{\mathrm{dr}})_{2}^{0}(0)}{4\pi^2\rho_{2}^{\mathrm{tot}}(0)}\frac{1}{T}\int_{0}^{\infty}\mathrm{d}\lambda\theta_{2}(\lambda)\left(1-\theta_{2}(\lambda)\right)\lambda^{2}\nonumber \\
 & =-2\frac{(q_{i}^{\mathrm{dr}})_{2}^{0}(0)(q_{j}^{\mathrm{dr}})_{2}^{0}(0)}{4\pi^2\rho_{2}^{\mathrm{tot}}(0)}4D_{2}^{2}\frac{\left(\pi T\right)^{1/2}}{4D_{2}^{3/2}}\mathrm{Li}_{1/2}\left(-e^{-D_{1}/T}\right)\nonumber \\
 & =-\frac{2}{\rho_{2}^{\mathrm{tot}}(0)}\left(\frac{(q_{i}^{\mathrm{dr}})_{2}^{0}}{2\pi}\right)\left(\frac{(q_{j}^{\mathrm{dr}})_{2}^{0}}{2\pi}\right)D_{2}^{1/2}\left(\pi T\right)^{1/2}\mathrm{Li}_{1/2}\left(-e^{-D_{1}/T}\right)\nonumber \\
 & =-\frac{2}{\rho_{2}^{\mathrm{tot}}(0)}\left(\frac{(q_{i}^{\mathrm{dr}})_{2}^{0}}{2\pi}\right)\left(\frac{(q_{j}^{\mathrm{dr}})_{2}^{0}}{2\pi}\right)\left(\frac{\varepsilon''_{2}(0)}{2}\right)^{1/2}\left(\pi T\right)^{1/2}\mathrm{Li}_{1/2}\left(-e^{-D_{1}/T}\right).
\end{align}

Collecting the above results, we obtain an analytic expression for the Drude weights in the vicinity of the phase transition,
\begin{align}
D_{ij}&=D_{\mathrm{bg}}+\left(\mathcal{A}_{1}+\mathcal{A}_{2}\right)T^{1/2}\mathrm{Li}_{\frac{1}{2}}\left(-e^{-\varepsilon_{2}(0)/T}\right),\: i,j\in\{n,m\},
\end{align}
where the background term is given by 
\[
D_{\mathrm{bg}}=D(k_0)+\frac{\pi^{2}T^{2}}{6}\left.\frac{\mathrm{d}^{2}}{\mathrm{d}\epsilon_{1}^{2}}D(k)\right|_{k=k_{0}}.
\]
The coefficients associated with the charge and spin contributions read
\begin{align}
\mathcal{A}_{1}&=\frac{\sqrt{\pi}}{2}D(k_{0})\left(\frac{\varepsilon_{2}''(0)}{2}\right)^{-1/2}\sum_{\beta=i,j}\frac{\frac{1}{2}\rho_{2}^{\mathrm{tot}}(0)\bigl((q_{\beta}^{\mathrm{dr}})_{2}^{0}(0)\bigr)'-\bigl(\rho_{2}^{\mathrm{tot}}(0)\bigr)'(q_{\beta}^{\mathrm{dr}})_{2}^{0}(0)}{\rho_{1}^{0}\,(q_{\beta}^{\mathrm{dr}})_{1}^{0}(0)}, \\
\mathcal{A}_{2}&=-\frac{2\sqrt{\pi}}{\rho_{2}^{\mathrm{tot}}(0)}\left(\frac{(q_{i}^{\mathrm{dr}})_{2}^{0}}{2\pi}\right)\left(\frac{(q_{j}^{\mathrm{dr}})_{2}^{0}}{2\pi}\right)\left(\frac{\varepsilon_{2}''(0)}{2}\right)^{1/2}.
\end{align}

However, when considering energy transport, namely \(\beta=e\), the dressed equation takes the form
\begin{equation}
\begin{aligned}
(q_{e}^{\mathrm{dr}})_{1}(k)
&=(q_{e})_{1}(k)+\int_{-\infty}^{\infty}\phi_{1}(k-\lambda)\theta_{2}(\lambda)(q_{e}^{\mathrm{dr}})_{2}(\lambda)\mathrm{d}\lambda,\\
(q_{e}^{\mathrm{dr}})_{2}(\lambda)
&=(q_{e})_{2}(\lambda)+\int_{-\infty}^{\infty}\phi_{1}(\lambda-k)\theta_{1}(k)(q_{e}^{\mathrm{dr}})_{1}(k)\mathrm{d}k
-\int_{-\infty}^{\infty}\phi_{2}(\lambda-\lambda')\theta_{2}(\lambda)(q_{e}^{\mathrm{dr}})_{2}(\lambda')\mathrm{d}\lambda'.
\end{aligned}
\end{equation}
Since the bare term $(q_{e})_{1}(k_{0})=\frac{k_{0}^{2}}{2m}-\mu-\frac{H}{2}$ is nearly zero in the vicinity of the phase transition. The lowest-order correction is no longer sufficient to capture the leading behavior, and higher-order terms in $\lambda$ must therefore be retained. Accordingly, one obtains
\begin{equation}
  (q_{e}^{\mathrm{dr}})_{1}(k)=(q_{e})_{1}(k)+A_{1}(k)(q_{e}^{\mathrm{dr}})_{2}^{0}(0) I_{1/2}+A_{1}(k)\left((q_{e}^{\mathrm{dr}})_{2}(0)\right)'' I_{3/2},
\end{equation}
where
\[
\left((q_{e}^{\mathrm{dr}})_{2}(0)\right)''=\int_{0}^{k_{0}}\mathrm{d}k\,A_{2}(k)(q_{e}^{\mathrm{dr}})_{1}(k).
\]

\begin{figure}
\centering
\includegraphics[scale=0.5]{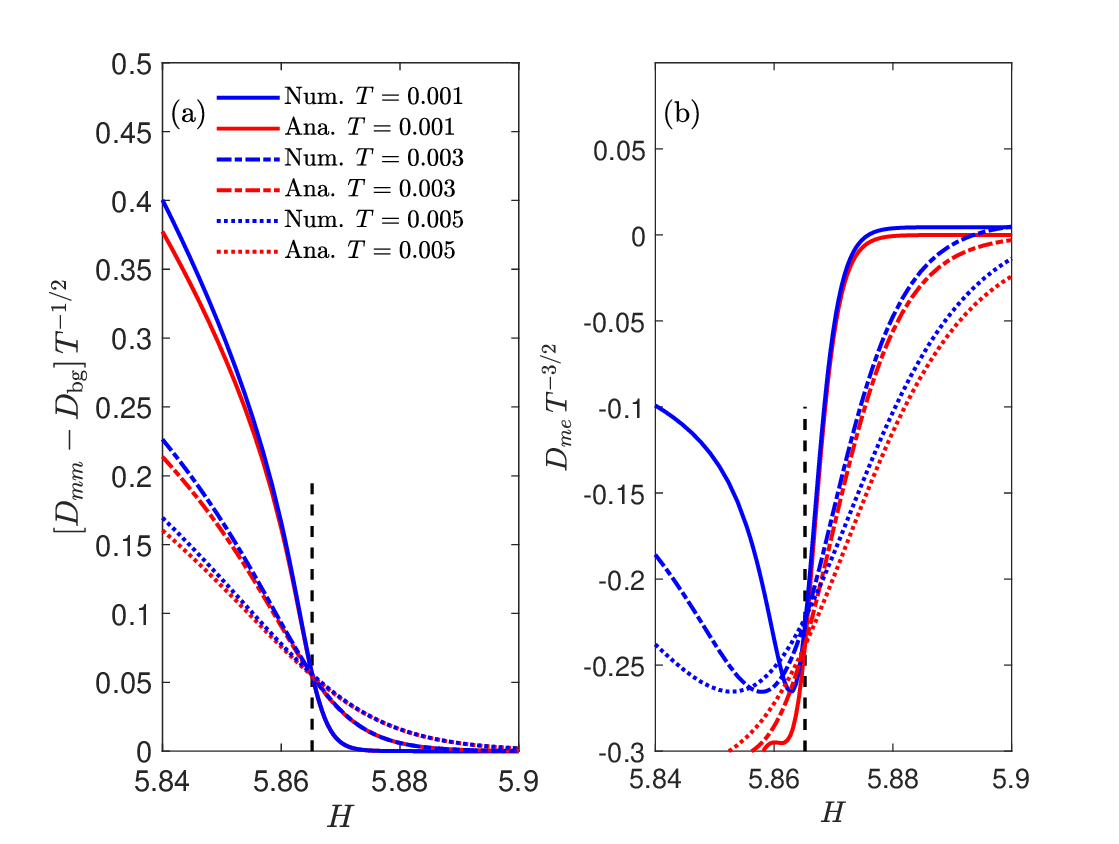}
\caption{\label{fig:Dmm} Scaling behavior of the rescaled spin DWs $[D_{mm}-D(k_0)]\,T^{-1/2}$ (a) and $D_{me}\,T^{-3/2}$  (b) as a function of the magnetic field $H$ for different temperatures $T$. The solid lines represent numerical results obtained from the TBA equations, while the symbols denote the corresponding analytical predictions. The vertical dashed line indicates the critical magnetic field $H_c$. Other parameters are fixed at $c=1$, $\mu=4$ and $m=1$.}
\end{figure}

Accordingly, higher-order corrections must also be retained when evaluating the DW. The contribution from the charge branch can be expressed as
\begin{equation}
\begin{aligned}
D_{ie,1} & =D_{ie,1}^{\mathrm{bare}}-D(k_{0})\frac{\left((q_{e}^{\mathrm{dr}})_{2}(0)\right)''}{(q_{e})_{1}(k_{0})}\frac{\sqrt{\pi}T^{3/2}}{4}
\left(\frac{\varepsilon_{2}''(0)}{2}\right)^{-3/2}
\operatorname{Li}_{3/2}\!\left(-e^{-D_1/T}\right),
\end{aligned}
\end{equation}
where $D_{ie,1}^{\mathrm{bare}}$ denotes the part already obtained above
\[
D_{ie,1}^{\mathrm{bare}}=D(k_{0})\left[1-I_{\frac{1}{2}}\sum_{\beta=i,e}\frac{\frac{1}{2}\rho_{2}^{\mathrm{tot}}(0)\left((q_{\beta}^{\mathrm{dr}})_{2}^{0}(0)\right)'-\left(\rho_{2}^{\mathrm{tot}}(0)\right)'(q_{\beta}^{\mathrm{dr}})_{2}^{0}(0)}{\rho_{1}^{\mathrm{0}}(q_{\beta}^{\mathrm{dr}})^{0}_{1}(0)}\right]+\frac{\pi^{2}T^{2}}{6}\left.\frac{\mathrm{d}^{2}}{\mathrm{d}\epsilon_{1}^{2}}D(k)\right|_{k=k_{0}}.
\]
By contrast, the contribution originating from the spin branch reads
\begin{equation}
D_{ie,2}
=D_{ie,2}^{\mathrm{bare}}-
\frac{3\sqrt{\pi}}{\rho_{2}^{\mathrm{tot}}(0)}
\left(\frac{(q_{i}^{\mathrm{dr}})_{2}^{0}}{2\pi}\right)
\left[\frac{\left((q_{e}^{\mathrm{dr}})_{2}(0)\right)''}{2\pi}\right]
\left(\frac{\varepsilon_{2}''(0)}{2}\right)^{-1/2}
T^{3/2}\mathrm{Li}_{3/2}\!\left(-e^{-D_{1}/T}\right),
\end{equation}
where $D_{ie,2}^{\mathrm{bare}}$ takes the form
\[
D_{ie,2}^{\mathrm{bare}}=-\frac{2}{\rho_{2}^{\mathrm{tot}}(0)}\left(\frac{(q_{i}^{\mathrm{dr}})_{2}^{0}}{2\pi}\right)\left(\frac{(q_{e}^{\mathrm{dr}})_{2}^{0}}{2\pi}\right)\left(\frac{\varepsilon''_{2}(0)}{2}\right)^{1/2}\left(\pi T\right)^{1/2}\mathrm{Li}_{1/2}\left(-e^{-D_{1}/T}\right).
\]

In Fig.~\ref{fig:Dmm}, we show the scaling function for $D_{mm}$ and $D_{me}$ as a function of the magnetic field near the phase transition. The analytical results are in excellent agreement with numerical solutions of the TBA equations.

\section{Drude Weight under Strong Interactions}

The kernel function in integral equations including interaction strength $c$ takes the form
\begin{equation}
\phi_{n}\left(k\right)=\frac{1}{\pi}\frac{2nc}{\left(nc\right)^{2}+4k^{2}}.
\end{equation}
In the regime of strong interactions,   $k/c$ and $q/c$ are more less than one, and then we expand the kernel function with respect to $1/c$ up to order $O(1/c^{3})$:
\begin{align}
\phi_{n}(k-q)
&=\frac{1}{\pi}\frac{2nc}{(nc)^{2}+4(k-q)^{2}} \nonumber\\
&\approx \frac{1}{\pi}\left(\frac{2}{nc}-\frac{q^{2}-2kq+k^{2}}{(nc/2)^{3}}\right).
\end{align}

At low temperatures with only one string $l=1$, substituting the above
approximated kernel function into the Yang-Yang Thermodynamic equations in equilibrium
yields
\begin{align}
\epsilon_{1}\left(k\right)
&=\frac{k^{2}}{2m}-\mu-\frac{H}{2}
 -T\int_{-\infty}^{\infty}\mathrm{d}q\,\phi_{1}\left(k-q\right)
 \ln\!\left[1+\exp\!\left(-\frac{\epsilon_{2}\left(q\right)}{T}\right)\right]
\nonumber\\
&\approx \beta k^{2}-A_{\epsilon}, \label{eq:epsilon1_def} \\
\epsilon_{2}\left(k\right)
&=H
 -T\int_{-\infty}^{\infty}\mathrm{d}q\,\phi_{1}\left(k-q\right)
 \ln\!\left[1+\exp\!\left(-\frac{\epsilon_{1}\left(q\right)}{T}\right)\right]
\nonumber\\
&\quad + T\int_{-\infty}^{\infty}\mathrm{d}q\,\phi_{2}\left(k-q\right)
 \ln\!\left[1+\exp\!\left(-\frac{\epsilon_{2}\left(q\right)}{T}\right)\right]
\nonumber\\
&\approx \alpha k^{2}-A_{\phi}. \label{eq:epsilon2_def}
\end{align}
 where we define the coefficients 
\begin{align}
\alpha &= \frac{16P_{1}-2P_{2}}{c^{3}}, \label{eq:alpha_def}\\
\beta &= \frac{1}{2m}+\frac{16P_{2}}{c^{3}}, \label{eq:beta_def}\\
A_{\epsilon} &= \mu+\frac{H}{2}+\frac{4}{c}P_{2}-\frac{16}{c^{3}}R_{2}, \label{eq:A_eps_def}\\
A_{\phi} &= -H-\frac{2P_{2}-4P_{1}}{c}-\frac{16R_{1}-2R_{2}}{c^{3}}, \label{eq:A_phi_def}\\
P_{i} &= \frac{T}{2\pi}\int_{-\infty}^{\infty}
\ln\!\left[1+\exp\!\left(-\frac{\epsilon_{i}\left(k\right)}{T}\right)\right]\mathrm{d}k, \label{eq:P_i_def}\\
R_{i} &= \frac{T}{2\pi}\int_{-\infty}^{\infty}q^{2}
\ln\!\left[1+\exp\!\left(-\frac{\epsilon_{i}\left(q\right)}{T}\right)\right]\mathrm{d}q, \label{eq:R_i_def}
\end{align}
with $i=1,2.$ To further simplify the expressions for the coefficients, we introduce the following general thermal integrals: 
\begin{align}
B_{\epsilon_{n}} & =\int_{-\infty}^{\infty}k^{2n-1}\theta_{1}(k)\mathrm{d}k=-\Gamma(n)\left(\frac{T}{\beta}\right)^{n}\mathrm{Li}_{n}(-e^{A_{\epsilon}/T}), \label{B_epsilon}\\
B_{\phi_{n}} & =\int_{-\infty}^{\infty}k^{2n-1}\theta_{2}(k)\mathrm{d}k=-\Gamma(n)\left(\frac{T}{\alpha}\right)^{n}\mathrm{Li}_{n}(-e^{A_{\phi}/T}),\label{B_phi}
\end{align}
where $\theta_i(k) = 1/(1+e^{\epsilon_i(k)/T})$, and $n\in\{\frac{1}{2},\frac{3}{2},\frac{5}{2}\}$. Here $\Gamma(n)$ denotes the Gamma function, and $\mathrm{Li}_{s}(z)$ is the polylogarithm function, which is directly related to the Fermi-Dirac integral $F_{s-1}(\eta)$ via
\[
-\mathrm{Li}_{s}(-e^{\eta})=F_{s-1}(\eta)=\frac{1}{\Gamma(s)}\int_{0}^{\infty}\frac{x^{s-1}}{e^{x-\eta}+1}\mathrm{d}x.
\]
Using the quadratic forms $\epsilon_{1}(k)=\beta k^{2}-A_{\epsilon}$ and $\epsilon_{2}(k)=\alpha k^{2}-A_{\phi}$, the integrals $P_{i}$ and $R_{i}$ can be expressed in terms of the thermal integrals $B_{\epsilon_{n}}$ and $B_{\phi_{n}}$:
\begin{align}
P_{1} &= \frac{\beta}{\pi}B_{\epsilon_{3/2}}, \quad
P_{2} = \frac{\alpha}{\pi}B_{\phi_{3/2}}, \\
R_{1} &= \frac{\beta}{3\pi}B_{\epsilon_{5/2}}, \quad
R_{2} = \frac{\alpha}{3\pi}B_{\phi_{5/2}}.
\end{align}
Substituting these back into Eqs.~(\ref{eq:alpha_def})--(\ref{eq:A_phi_def}) gives a closed set of self-consistent equations:
\begin{align}
\alpha &= \frac{16\beta B_{\epsilon_{3/2}}-2\alpha B_{\phi_{3/2}}}{\pi c^{3}}, \\
\beta &= \frac{1}{2m}+\frac{16\alpha B_{\phi_{3/2}}}{\pi c^{3}}, \\
A_{\epsilon} &= \mu+\frac{H}{2}
+\frac{4\alpha}{\pi c}B_{\phi_{3/2}}
-\frac{16\alpha}{3\pi c^{3}}B_{\phi_{5/2}}, \\
A_{\phi} &= -H
-\frac{2\alpha B_{\phi_{3/2}}-4\beta B_{\epsilon_{3/2}}}{\pi c}
-\frac{16\beta B_{\epsilon_{5/2}}-2\alpha B_{\phi_{5/2}}}{3\pi c^{3}}.
\end{align}

To compute the Drude weight, we derive the dressed charges within the $1/c^{3}$ approximation.
The fundamental dressed charge $\vec{q}_{n}^{\mathrm{dr}}=\{1,0\}^{\mathrm{dr}}$ satisfies
\begin{align}
(\vec{q}_{n}^{\mathrm{dr}})_{1}(k)
&=1+\int_{-\infty}^{\infty}\mathrm{d}q\,\phi_{1}\left(k-q\right)
(\vec{q}_{n}^{\mathrm{dr}})_{2}(q)\theta_{2}
\approx \beta_{\rho_{0}}k^{2}+A_{\rho_{0}}, \label{qn1_c} \\ 
(\vec{q}_{n}^{\mathrm{dr}})_{2}(q)
&=\int_{-\infty}^{\infty}\mathrm{d}k\,\phi_{1}\left(k-q\right)
(\vec{q}_{n}^{\mathrm{dr}})_{1}(k)\theta_{1}
\nonumber\\
&\quad -\int_{-\infty}^{\infty}\mathrm{d}k\,\phi_{2}\left(k-q\right)
(\vec{q}_{n}^{\mathrm{dr}})_{2}(k)\theta_{2}
\approx \beta_{\sigma_{0}}q^{2}+A_{\sigma_{0}}, \label{qn2_c}
\end{align}
where the coefficients are defined by
\begin{align}
\beta_{\rho_{0}} &= -\frac{8F_{\sigma_{0}}}{\pi c^{3}}, \quad
A_{\rho_{0}} = 1+\frac{2}{\pi c}F_{\sigma_{0}}, \\
\beta_{\sigma_{0}} &= \frac{F_{\sigma_{0}}-8F_{\rho_{0}}}{\pi c^{3}}, \\
A_{\sigma_{0}} &= \frac{2F_{\rho_{0}}-F_{\sigma_{0}}}{\pi c}
-\frac{8}{\pi c^{3}}B_{\epsilon_{3/2}},
\end{align}
and
\begin{equation}
F_{\rho_{0}}=\int_{-\infty}^{\infty}(\vec{q}_{n}^{\mathrm{dr}})_{1}(k)\theta_{1}\mathrm{d}k, \quad
F_{\sigma_{0}}=\int_{-\infty}^{\infty}(\vec{q}_{n}^{\mathrm{dr}})_{2}(k)\theta_{2}\mathrm{d}k.  \label{FF}
\end{equation}
This coupled integral system is solved iteratively up to order $o(1/c^{3})$.
Substituting Eqs.~(\ref{qn1_c})--(\ref{qn2_c}) into Eqs.~(\ref{FF}) yields
\begin{align}
F_{\rho_{0}}  & =B_{\epsilon_{\frac{1}{2}}}+\frac{2}{\pi c}F_{\sigma_{0}}B_{\epsilon_{\frac{1}{2}}}-\frac{8F_{\sigma_{0}}}{\pi c^{3}}B_{\epsilon_{\frac{3}{2}}} \\
F_{\sigma_{0}} & =\frac{2F_{\rho_{0}}}{\pi c}B_{\phi_{\frac{1}{2}}}-\frac{F_{\sigma_{0}}}{\pi c}B_{\phi_{\frac{1}{2}}}-\frac{8}{\pi c^{3}}B_{\epsilon_{\frac{3}{2}}}B_{\phi_{\frac{1}{2}}}+\frac{F_{\sigma_{0}}}{\pi c^{3}}B_{\phi_{\frac{3}{2}}}-\frac{8F_{\rho_{0}}}{\pi c^{3}}B_{\phi_{\frac{3}{2}}}
\end{align}
Up to the second-order approximation, $F_{\rho_{0}}=B_{\epsilon_{\frac{1}{2}}}+\frac{4}{(\pi c)^{2}}B_{\epsilon_{\frac{1}{2}}}^{2}B_{\phi_{\frac{1}{2}}}\text{and }F_{\sigma_{0}}=\frac{2}{\pi c}B_{\epsilon_{\frac{1}{2}}}B_{\phi_{\frac{1}{2}}}-\frac{2}{(\pi c)^{2}}B_{\epsilon_{\frac{1}{2}}}B_{\phi_{\frac{1}{2}}}^{2}.$
From these, the coefficients can be approximated as 
\begin{align}
\beta_{\rho_{0}} & =-\frac{8F_{\sigma_{0}}}{\pi c^{3}}\approx0,\\
\beta_{\sigma_{0}} & =\left(\frac{F_{\sigma_{0}}}{\pi c^{3}}-\frac{8F_{\rho_{0}}}{\pi c^{3}}\right)\approx-\frac{8B_{\epsilon_{\frac{1}{2}}}}{\pi c^{3}},\\
A_{\rho_{0}} & =1+\frac{2}{\pi c}F_{\sigma_{0}}\approx1+\frac{4}{(\pi c)^{2}}B_{\epsilon_{\frac{1}{2}}}B_{\phi_{\frac{1}{2}}}-\frac{4}{(\pi c)^{3}}B_{\epsilon_{\frac{1}{2}}}B_{\phi_{\frac{1}{2}}}^{2},\\
A_{\sigma_{0}} & =\frac{2F_{\rho_{0}}-F_{\sigma_{0}}}{\pi c}-\frac{8}{\pi c^{3}}B_{\epsilon_{\frac{3}{2}}}\approx\frac{2B_{\epsilon_{\frac{1}{2}}}}{\pi c}-\frac{2B_{\epsilon_{\frac{1}{2}}}B_{\phi_{\frac{1}{2}}}}{(\pi c)^{2}}+\frac{8B_{\epsilon_{\frac{1}{2}}}^{2}B_{\phi_{\frac{1}{2}}}+2B_{\epsilon_{\frac{1}{2}}}B_{\phi_{\frac{1}{2}}}^{2}}{(\pi c)^{3}}-\frac{8}{\pi c^{3}}B_{\epsilon_{\frac{3}{2}}}.
\end{align}
Substituting these into the dressed charge definitions yields the explicit forms
\begin{align}
(\vec{q}_{n}^{\mathrm{dr}})_{1}(k) & =1+\frac{4}{(\pi c)^{2}}B_{\epsilon_{\frac{1}{2}}}B_{\phi_{\frac{1}{2}}}-\frac{4}{(\pi c)^{3}}B_{\epsilon_{\frac{1}{2}}}B_{\phi_{\frac{1}{2}}}^{2},\\
(\vec{q}_{n}^{\mathrm{dr}})_{2}(q) & =-\frac{8B_{\epsilon_{\frac{1}{2}}}}{\pi c^{3}}q^{2}+\frac{2B_{\epsilon_{\frac{1}{2}}}}{\pi c}-\frac{2B_{\epsilon_{\frac{1}{2}}}B_{\phi_{\frac{1}{2}}}}{(\pi c)^{2}}+\frac{8B_{\epsilon_{\frac{1}{2}}}^{2}B_{\phi_{\frac{1}{2}}}+2B_{\epsilon_{\frac{1}{2}}}B_{\phi_{\frac{1}{2}}}^{2}-8\pi^{2}B_{\epsilon_{\frac{3}{2}}}}{(\pi c)^{3}}.
\end{align}
 Using the same method, the other dressed charges ($\vec{q}_{k}^{\mathrm{dr}}=\{k,0\}^{\mathrm{dr}},\vec{q}_{m}^{\mathrm{dr}}=\{1/2,-1\}^{\mathrm{dr}},\vec{q}_{e}^{\mathrm{dr}}=\{\frac{k^{2}}{2m}-\mu-\frac{H}{2},H\}^{\mathrm{dr}}$)
are calculated. The results are listed here
\begin{align}
(\vec{q}_{k}^{\mathrm{dr}})_{1}(k) & =k,\\
(\vec{q}_{k}^{\mathrm{dr}})_{2}(q) & =\frac{16q}{\pi c^{3}}B_{\epsilon_{\frac{3}{2}}},\\
(\vec{q}_{m}^{\mathrm{dr}})_{1}(k) & =\frac{8B_{\phi_{\frac{1}{2}}}}{\pi c^{3}}k^{2}+\frac{1}{2}+\frac{T_{m_{C1}}}{\pi c}+\frac{T_{m_{C2}}}{(\pi c)^{2}}+\frac{T_{m_{C3}}}{(\pi c)^{3}},\\
(\vec{q}_{m}^{\mathrm{dr}})_{2}(q) & =-\left(\frac{B_{\phi_{\frac{1}{2}}}+4B_{\epsilon_{\frac{1}{2}}}}{\pi c^{3}}\right)q^{2}-1+\frac{T_{m_{S1}}}{\pi c}+\frac{T_{m_{S2}}}{(\pi c)^{2}}+\frac{T_{m_{S3}}}{(\pi c)^{3}},\\
(\vec{q}_{e}^{\mathrm{dr}})_{1}(k) & =\left(\frac{1}{2m}-\frac{8HB_{\phi_{\frac{1}{2}}}}{\pi c^{3}}\right)k^{2}-\mu-\frac{H}{2}+\frac{T_{e_{C1}}}{\pi c}+\frac{T_{e_{C2}}}{(\pi c)^{2}}+\frac{T_{e_{C3}}}{(\pi c)^{3}},\\
(\vec{q}_{e}^{\mathrm{dr}})_{2}(q) & =\left(\frac{HB_{\phi_{\frac{1}{2}}}-\frac{4}{m}B_{\epsilon_{\frac{3}{2}}}+(8\mu+4H)B_{\epsilon_{\frac{1}{2}}}}{\pi c^{3}}\right)q^{2}+H+\frac{T_{e_{S1}}}{\pi c}-\frac{T_{e_{S2}}}{(\pi c)^{2}}+\frac{T_{e_{S3}}}{(\pi c)^{3}},
\end{align}
in which, 
\begin{align}
T_{m_{C1}}= & -2B_{\phi_{\frac{1}{2}}},\\
T_{m_{C2}}= & 2B_{\epsilon_{\frac{1}{2}}}B_{\phi_{\frac{1}{2}}}+2B_{\phi_{\frac{1}{2}}}^{2},\\
T_{m_{C3}}= & 8\pi^{2}B_{\phi_{\frac{3}{2}}}-10B_{\epsilon_{\frac{1}{2}}}B_{\phi_{\frac{1}{2}}}^{2}-2B_{\phi_{\frac{1}{2}}}^{3},\\
T_{m_{S1}}= & B_{\epsilon_{\frac{1}{2}}}+B_{\phi_{\frac{1}{2}}},\\
T_{m_{S2}}= & -B_{\phi_{\frac{1}{2}}}^{2}-5B_{\epsilon_{\frac{1}{2}}}B_{\phi_{\frac{1}{2}}},\\
T_{m_{S3}}= & 4B_{\epsilon_{\frac{1}{2}}}^{2}B_{\phi_{\frac{1}{2}}}+9B_{\phi_{\frac{1}{2}}}^{2}B_{\epsilon_{\frac{1}{2}}}+B_{\phi_{\frac{1}{2}}}^{3}-\pi^{2}(4B_{\epsilon_{\frac{3}{2}}}+B_{\phi_{\frac{3}{2}}}),\\
T_{e_{C1}}= & 2HB_{\phi_{\frac{1}{2}}},\\
T_{e_{C2}}= & \frac{2}{m}B_{\epsilon_{\frac{3}{2}}}B_{\phi_{\frac{1}{2}}}-(4\mu+2H)B_{\epsilon_{\frac{1}{2}}}B_{\phi_{\frac{1}{2}}}-2HB_{\phi_{\frac{1}{2}}}^{2},\\
T_{e_{C3}}= & 2HB_{\phi_{\frac{1}{2}}}^{3}+(4\mu+10H)B_{\epsilon_{\frac{1}{2}}}B_{\phi_{\frac{1}{2}}}^{2}-\frac{2}{m}B_{\epsilon_{\frac{3}{2}}}B_{\phi_{\frac{1}{2}}}^{2}-8\pi^{2}HB_{\phi_{\frac{3}{2}}},\\
T_{e_{S1}}= & \frac{1}{m}B_{\epsilon_{\frac{3}{2}}}-(2\mu+H)B_{\epsilon_{\frac{1}{2}}}-HB_{\phi_{\frac{1}{2}}},\\
T_{e_{S2}}= & HB_{\phi_{\frac{1}{2}}}^{2}+(2\mu+5H)B_{\epsilon_{\frac{1}{2}}}B_{\phi_{\frac{1}{2}}}-\frac{1}{m}B_{\epsilon_{\frac{3}{2}}}B_{\phi_{\frac{1}{2}}},\\
T_{e_{S3}}= & \frac{4}{m}B_{\epsilon_{\frac{3}{2}}}B_{\epsilon_{\frac{1}{2}}}B_{\phi_{\frac{1}{2}}}-(10\mu+9H)B_{\epsilon_{\frac{1}{2}}}^{2}B_{\phi_{\frac{1}{2}}}-4HB_{\phi_{\frac{1}{2}}}^{2}B_{\epsilon_{\frac{1}{2}}},\\
 & -HB_{\phi_{\frac{1}{2}}}^{3}+\frac{1}{m}B_{\epsilon_{\frac{3}{2}}}B_{\phi_{\frac{1}{2}}}^{2}-\frac{4\pi^{2}}{m}B_{\epsilon_{\frac{5}{2}}}+(8\pi^{2}\mu+4\pi^{2}H)B_{\epsilon_{\frac{3}{2}}}+\pi^{2}HB_{\phi_{\frac{3}{2}}}.
\end{align}

At low temperature, the Drude weight $D_{nn}$ from formula (S48)  simplifies via integration by parts to
\begin{align}
D_{nn}= & \frac{1}{2\pi m}\int\mathrm{d}k\theta_{1}(k)\frac{\mathrm{d}}{\mathrm{d}k}\left[(\vec{q}_{k}^{\mathrm{dr}})_{1}(k)\cdot(\vec{q}_{n}^{\mathrm{dr}})_{1}(k)\right]+\frac{1}{2\pi m}\int\mathrm{d}k\theta_{2}(k)\frac{\mathrm{d}}{\mathrm{d}k}\left[(\vec{q}_{k}^{\mathrm{dr}})_{2}(k)\cdot(\vec{q}_{n}^{\mathrm{dr}})_{2}(k)\right].
\end{align}
Finally, substituting above dressed charge into the Drude weight formula, we obtain
the explicit expressions
\begin{align}
D_{nn}= & \frac{1}{2\pi m}\int\mathrm{d}k\theta_{1}(k)\frac{\mathrm{d}}{\mathrm{d}k}\left[k\cdot\left(1+\frac{4}{(\pi c)^{2}}B_{\epsilon_{\frac{1}{2}}}B_{\phi_{\frac{1}{2}}}-\frac{4}{(\pi c)^{3}}B_{\epsilon_{\frac{1}{2}}}B_{\phi_{\frac{1}{2}}}^{2}\right)\right]+\frac{1}{2\pi m}\int\mathrm{d}k\theta_{2}(k)\frac{\mathrm{d}}{\mathrm{d}k}\left[\frac{16k}{\pi c^{3}}B_{\epsilon_{\frac{3}{2}}}\cdot\frac{2}{\pi c}B_{\epsilon_{\frac{1}{2}}}\right]\\
= & \frac{1}{2\pi m}\int\mathrm{d}k\theta_{1}(k)\left(1+\frac{4}{(\pi c)^{2}}B_{\epsilon_{\frac{1}{2}}}B_{\phi_{\frac{1}{2}}}-\frac{4}{(\pi c)^{3}}B_{\epsilon_{\frac{1}{2}}}B_{\phi_{\frac{1}{2}}}^{2}\right)+o(1/c^{4})\\
= & \frac{1}{2\pi m}\left(B_{\epsilon_{\frac{1}{2}}}+\frac{4}{(\pi c)^{2}}B_{\epsilon_{\frac{1}{2}}}^{2}B_{\phi_{\frac{1}{2}}}-\frac{4}{(\pi c)^{3}}B_{\epsilon_{\frac{1}{2}}}^{2}B_{\phi_{\frac{1}{2}}}^{2}\right).
\end{align}
Following the same procedure, we obtain the expressions for the remaining Drude weights.

To $D_{nm}:$ 
\begin{align}
D_{nm}= & \frac{1}{2\pi m}\int\mathrm{d}k\theta_{1}(k)\frac{\mathrm{d}}{\mathrm{d}k}\left[(\vec{q}_{m}^{\mathrm{dr}})_{1}(k)(\vec{q}_{k}^{\mathrm{dr}})_{1}(k)\right]+\frac{1}{2\pi m}\int\mathrm{d}k\theta_{2}(k)\frac{\mathrm{d}}{\mathrm{d}k}\left[(\vec{q}_{m}^{\mathrm{dr}})_{2}(k)(\vec{q}_{k}^{\mathrm{dr}})_{2}(k)\right],\\
= & \frac{1}{2\pi m}\left[\frac{1}{2}B_{\epsilon_{\frac{1}{2}}}+\frac{T_{nm_{1}}}{\pi c}+\frac{T_{nm_{2}}}{(\pi c)^{2}}+\frac{T_{nm_{3}}}{(\pi c)^{3}}\right].
\end{align}
in which,
\begin{align*}
T_{nm_{1}}= & -2B_{\phi_{\frac{1}{2}}}B_{\epsilon_{\frac{1}{2}}},\\
T_{nm_{2}}= & 2B_{\epsilon_{\frac{1}{2}}}^{2}B_{\phi_{\frac{1}{2}}}+2B_{\phi_{\frac{1}{2}}}^{2}B_{\epsilon_{\frac{1}{2}}},\\
T_{nm_{3}}= & 8\pi^{2}\left(B_{\phi_{\frac{3}{2}}}B_{\epsilon_{\frac{1}{2}}}+B_{\phi_{\frac{1}{2}}}B_{\epsilon_{\frac{3}{2}}}\right)-10B_{\epsilon_{\frac{1}{2}}}^{2}B_{\phi_{\frac{1}{2}}}^{2}-2B_{\phi_{\frac{1}{2}}}^{3}B_{\epsilon_{\frac{1}{2}}}.
\end{align*}
To $D_{mm}:$ 
\begin{align}
D_{mm}= & \frac{1}{2\pi m}\int\mathrm{d}k\theta_{1}(k)\frac{\mathrm{d}}{\mathrm{d}k}\left[\frac{(\vec{q}_{m}^{\mathrm{dr}})_{1}(k)(\vec{q}_{m}^{\mathrm{dr}})_{1}(k)}{(\vec{q}_{n}^{\mathrm{dr}})_{1}(k)}(\vec{q}_{k}^{\mathrm{dr}})_{1}(k)\right]+\frac{1}{2\pi m}\int\mathrm{d}k\theta_{2}(k)\frac{\mathrm{d}}{\mathrm{d}k}\left[\frac{(\vec{q}_{m}^{\mathrm{dr}})_{2}(k)(\vec{q}_{m}^{\mathrm{dr}})_{2}(k)}{(\vec{q}_{n}^{\mathrm{dr}})_{2}(k)}(\vec{q}_{k}^{\mathrm{dr}})_{2}(k)\right],\\
= & \frac{1}{2\pi m}\left[\frac{1}{4}B_{\epsilon_{\frac{1}{2}}}+\frac{T_{mm_{1}}}{\pi c}+\frac{T_{mm_{2}}}{(\pi c)^{2}}+\frac{T_{mm_{3}}}{(\pi c)^{3}}\right],
\end{align}
in which,
\begin{align*}
T_{mm_{1}}= & -2B_{\phi_{\frac{1}{2}}}B_{\epsilon_{\frac{1}{2}}},\\
T_{mm_{2}}= & B_{\epsilon_{\frac{1}{2}}}^{2}B_{\phi_{\frac{1}{2}}}+6B_{\epsilon_{\frac{1}{2}}}B_{\phi_{\frac{1}{2}}}^{2}+8\pi^{2}B_{\epsilon_{\frac{3}{2}}}B_{\phi_{\frac{1}{2}}}/B_{\epsilon_{\frac{1}{2}}},\\
T_{mm_{3}}= & -9B_{\epsilon_{\frac{1}{2}}}^{2}B_{\phi_{\frac{1}{2}}}^{2}-10B_{\epsilon_{\frac{1}{2}}}B_{\phi_{\frac{1}{2}}}^{3}-8\pi^{2}(B_{\phi_{\frac{1}{2}}}B_{\epsilon_{\frac{3}{2}}}+B_{\phi_{\frac{3}{2}}}B_{\epsilon_{\frac{1}{2}}}-B_{\epsilon_{\frac{3}{2}}}B_{\phi_{\frac{1}{2}}}^{2}/B_{\epsilon_{\frac{1}{2}}}).
\end{align*}
To $D_{ne}:$ 
\begin{align}
D_{ne}= & \frac{1}{2\pi m}\int\mathrm{d}k\theta_{1}(k)\frac{\mathrm{d}}{\mathrm{d}k}\left[(\vec{q}_{e}^{\mathrm{dr}})_{1}(k)(\vec{q}_{k}^{\mathrm{dr}})_{1}(k)\right]+\frac{1}{2\pi m}\int\mathrm{d}k\theta_{2}(k)\frac{\mathrm{d}}{\mathrm{d}k}\left[(\vec{q}_{e}^{\mathrm{dr}})_{2}(k)(\vec{q}_{k}^{\mathrm{dr}})_{2}(k)\right],\\
= & \frac{1}{2\pi m}\left[T_{ne_{0}}+\frac{T_{ne_{1}}}{\pi c}+\frac{T_{ne_{2}}}{(\pi c)^{2}}+\frac{T_{ne_{3}}}{(\pi c)^{3}}\right],
\end{align}
in which,
\begin{align*}
T_{ne_{0}}= & \left(\frac{3}{2m}-\frac{24HB_{\phi_{\frac{1}{2}}}}{\pi c^{3}}\right)B_{\epsilon_{\frac{3}{2}}}-(\mu+\frac{H}{2})B_{\epsilon_{\frac{1}{2}}},\\
T_{ne_{1}}= & 2HB_{\epsilon_{\frac{1}{2}}}B_{\phi_{\frac{1}{2}}},\\
T_{ne_{2}}= & \frac{2}{m}B_{\epsilon_{\frac{3}{2}}}B_{\epsilon_{\frac{1}{2}}}B_{\phi_{\frac{1}{2}}}-(4\mu+2H)B_{\epsilon_{\frac{1}{2}}}^{2}B_{\phi_{\frac{1}{2}}}-2HB_{\phi_{\frac{1}{2}}}^{2}B_{\epsilon_{\frac{1}{2}}},\\
T_{ne_{3}}= & 2\left(HB_{\phi_{\frac{1}{2}}}^{3}B_{\epsilon_{\frac{1}{2}}}+(2\mu+5H)B_{\epsilon_{\frac{1}{2}}}^{2}B_{\phi_{\frac{1}{2}}}^{2}-\frac{1}{m}B_{\epsilon_{\frac{3}{2}}}B_{\phi_{\frac{1}{2}}}^{2}B_{\epsilon_{\frac{1}{2}}}+4\pi^{2}H(2B_{\epsilon_{\frac{3}{2}}}B_{\phi_{\frac{1}{2}}}-B_{\phi_{\frac{3}{2}}}B_{\epsilon_{\frac{1}{2}}})\right).
\end{align*}
To $D_{me}:$
\begin{align}
D_{me}= & \frac{1}{2\pi m}\int\mathrm{d}k\theta_{1}(k)\frac{\mathrm{d}}{\mathrm{d}k}\left[\frac{(\vec{q}_{m}^{\mathrm{dr}})_{1}(k)(\vec{q}_{e}^{\mathrm{dr}})_{1}(k)}{(\vec{q}_{n}^{\mathrm{dr}})_{1}(k)}(\vec{q}_{k}^{\mathrm{dr}})_{1}(k)\right]+\frac{1}{2\pi m}\int\mathrm{d}k\theta_{2}(k)\frac{\mathrm{d}}{\mathrm{d}k}\left[\frac{(\vec{q}_{m}^{\mathrm{dr}})_{2}(k)(\vec{q}_{e}^{\mathrm{dr}})_{2}(k)}{(\vec{q}_{n}^{\mathrm{dr}})_{2}(k)}(\vec{q}_{k}^{\mathrm{dr}})_{2}(k)\right],\\
= & \frac{1}{2\pi m}\left[T_{me_{0}}+\frac{T_{me_{1}}}{\pi c}+\frac{T_{me_{2}}}{(\pi c)^{2}}+\frac{T_{me_{3}}}{(\pi c)^{3}}\right],
\end{align}
in which, 
\begin{align*}
T_{me_{0}}= & \frac{3}{4m}B_{\epsilon_{\frac{3}{2}}}-(\frac{\mu}{2}+\frac{H}{4})B_{\epsilon_{\frac{1}{2}}},\\
T_{me_{1}}= & (2\mu+2H)B_{\epsilon_{\frac{1}{2}}}B_{\phi_{\frac{1}{2}}}-\frac{3}{m}B_{\epsilon_{\frac{3}{2}}}B_{\phi_{\frac{1}{2}}},\\
T_{me_{2}}= & \frac{1}{m}B_{\epsilon_{\frac{3}{2}}}B_{\phi_{\frac{1}{2}}}B_{\epsilon_{\frac{1}{2}}}-(2\mu+H)B_{\epsilon_{\frac{1}{2}}}^{2}B_{\phi_{\frac{1}{2}}}-(6H+2\mu)B_{\phi_{\frac{1}{2}}}^{2}B_{\epsilon_{\frac{1}{2}}}+\frac{3}{m}B_{\epsilon_{\frac{3}{2}}}B_{\phi_{\frac{1}{2}}}^{2}-8\pi^{2}HB_{\epsilon_{\frac{3}{2}}}B_{\phi_{\frac{1}{2}}}/B_{\epsilon_{\frac{1}{2}}},\\
T_{me_{3}}= & \frac{20\pi^{2}}{m}B_{\epsilon_{\frac{5}{2}}}B_{\phi_{\frac{1}{2}}}+(10H+2\mu)B_{\phi_{\frac{1}{2}}}^{3}B_{\epsilon_{\frac{1}{2}}}-8\pi^{2}\left(\mu+H\right)\left(B_{\epsilon_{\frac{3}{2}}}B_{\phi_{\frac{1}{2}}}+B_{\phi_{\frac{3}{2}}}B_{\epsilon_{\frac{1}{2}}}\right)-\frac{5}{m}B_{\epsilon_{\frac{3}{2}}}B_{\epsilon_{\frac{1}{2}}}B_{\phi_{\frac{1}{2}}}^{2}\\
 & +(10\mu+9H)B_{\epsilon_{\frac{1}{2}}}^{2}B_{\phi_{\frac{1}{2}}}^{2}+\frac{3}{m}\left(4\pi^{2}B_{\epsilon_{\frac{3}{2}}}B_{\phi_{\frac{3}{2}}}-B_{\epsilon_{\frac{3}{2}}}B_{\phi_{\frac{1}{2}}}^{3}\right)-8\pi^{2}\left(\frac{1}{m}B_{\epsilon_{\frac{3}{2}}}^{2}B_{\phi_{\frac{1}{2}}}-HB_{\epsilon_{\frac{3}{2}}}B_{\phi_{\frac{1}{2}}}^{2}\right)/B_{\epsilon_{\frac{1}{2}}}.
\end{align*}

\begin{figure}
\centering
\includegraphics[scale=0.5]{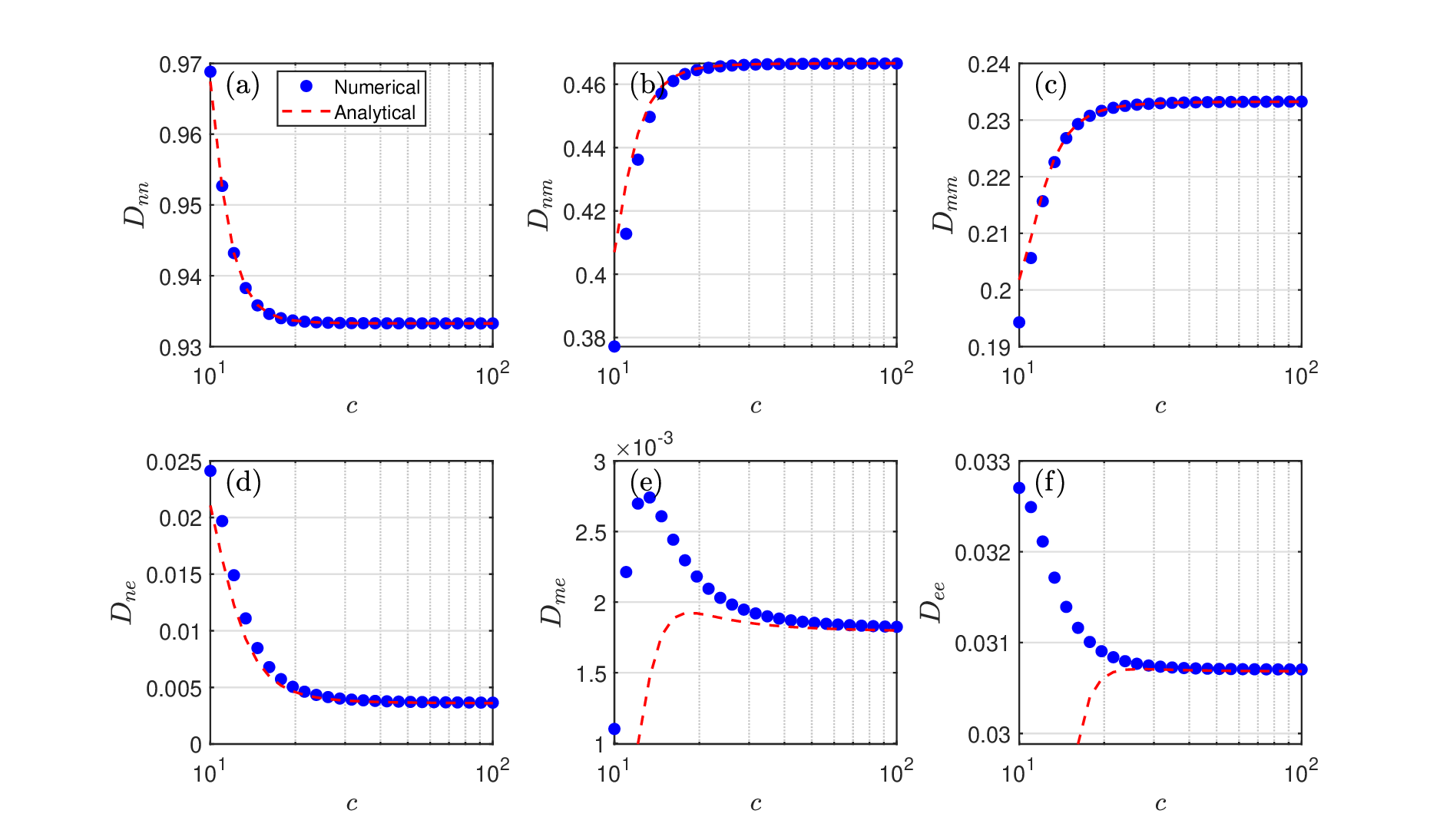}
\caption{\label{fig:strong}Comparison between the exact numerical TBA results (blue solid dots)
and the analytical strong-coupling asymptotic expressions (red dashed line)
for the Drude weights.
Parameters: $\mu=3.8$, $H=1.0$, $T=0.1$.}
\end{figure}

To $D_{ee}:$ 
\begin{align}
D_{ee}= & \frac{1}{2\pi m}\int\mathrm{d}k\theta_{1}(k)\frac{\mathrm{d}}{\mathrm{d}k}\left[\frac{(\vec{q}_{e}^{\mathrm{dr}})_{1}(k)(\vec{q}_{e}^{\mathrm{dr}})_{1}(k)}{(\vec{q}_{n}^{\mathrm{dr}})_{1}(k)}(\vec{q}_{k}^{\mathrm{dr}})_{1}(k)\right]+\frac{1}{2\pi m}\int\mathrm{d}k\theta_{2}(k)\frac{\mathrm{d}}{\mathrm{d}k}\left[\frac{(\vec{q}_{e}^{\mathrm{dr}})_{2}(k)(\vec{q}_{e}^{\mathrm{dr}})_{2}(k)}{(\vec{q}_{n}^{\mathrm{dr}})_{2}(k)}(\vec{q}_{k}^{\mathrm{dr}})_{2}(k)\right],\\
= & \frac{1}{2\pi m}\left[T_{ee_{0}}+\frac{T_{ee_{1}}}{\pi c}+\frac{T_{ee_{2}}}{(\pi c)^{2}}+\frac{T_{ee_{3}}}{(\pi c)^{3}}\right],
\end{align}
in which, 
\begin{align*}
T_{ee_{0}}= & \frac{5}{4m^{2}}B_{\epsilon_{\frac{5}{2}}}-\frac{3}{m}(\mu+\frac{H}{2})B_{\epsilon_{\frac{3}{2}}}+(\mu+\frac{H}{2})^{2}B_{\epsilon_{\frac{1}{2}}},\\
T_{ee_{1}}= & \frac{6}{m}HB_{\epsilon_{\frac{3}{2}}}B_{\phi_{\frac{1}{2}}}-(4\mu+2H)HB_{\epsilon_{\frac{1}{2}}}B_{\phi_{\frac{1}{2}}},\\
T_{ee_{2}}= & \frac{1}{m^{2}}\left(6B_{\epsilon_{\frac{3}{2}}}^{2}B_{\phi_{\frac{1}{2}}}-5B_{\epsilon_{\frac{1}{2}}}B_{\epsilon_{\frac{5}{2}}}B_{\phi_{\frac{1}{2}}}\right)-\frac{1}{m}(4\mu+2H)B_{\epsilon_{\frac{3}{2}}}B_{\epsilon_{\frac{1}{2}}}B_{\phi_{\frac{1}{2}}}+(2\mu+H)^{2}B_{\epsilon_{\frac{1}{2}}}^{2}B_{\phi_{\frac{1}{2}}}\\
 & -\frac{6}{m}HB_{\epsilon_{\frac{3}{2}}}B_{\phi_{\frac{1}{2}}}^{2}+(4\mu+6H)HB_{\epsilon_{\frac{1}{2}}}B_{\phi_{\frac{1}{2}}}^{2}+8\pi^{2}H^{2}B_{\epsilon_{\frac{3}{2}}}B_{\phi_{\frac{1}{2}}}/B_{\epsilon_{\frac{1}{2}}},\\
T_{ee_{3}}= & \frac{6}{m}HB_{\epsilon_{\frac{3}{2}}}B_{\phi_{\frac{1}{2}}}^{3}-\frac{6}{m^{2}}B_{\epsilon_{\frac{3}{2}}}^{2}B_{\phi_{\frac{1}{2}}}^{2}-\frac{48\pi^{2}}{m}HB_{\epsilon_{\frac{3}{2}}}B_{\phi_{\frac{3}{2}}}-\frac{80\pi^{2}}{m}HB_{\epsilon_{\frac{5}{2}}}B_{\phi_{\frac{1}{2}}}+\frac{1}{m}(4\mu+10H)B_{\epsilon_{\frac{3}{2}}}B_{\epsilon_{\frac{1}{2}}}B_{\phi_{\frac{1}{2}}}^{2}\\
 & +16\pi^{2}(2\mu+H)H\left(B_{\epsilon_{\frac{1}{2}}}B_{\phi_{\frac{3}{2}}}+2B_{\epsilon_{\frac{3}{2}}}B_{\phi_{\frac{1}{2}}}\right)-(4\mu+10H)HB_{\epsilon_{\frac{1}{2}}}B_{\phi_{\frac{1}{2}}}^{3}-(2\mu+H)(2\mu+9H)B_{\epsilon_{\frac{1}{2}}}^{2}B_{\phi_{\frac{1}{2}}}^{2}\\
 & +\frac{5}{m^{2}}B_{\epsilon_{\frac{5}{2}}}B_{\epsilon_{\frac{1}{2}}}B_{\phi_{\frac{1}{2}}}^{2}+\frac{16\pi^{2}H}{m}B_{\epsilon_{\frac{3}{2}}}^{2}B_{\phi_{\frac{1}{2}}}/B_{\epsilon_{\frac{1}{2}}}-8\pi^{2}H^{2}B_{\epsilon_{\frac{3}{2}}}B_{\phi_{\frac{1}{2}}}^{2}/B_{\epsilon_{\frac{1}{2}}}.
\end{align*}

We  compare the analytical expressions derived above with exact numerical solutions of the TBA equations.
As seen in  Fig.~\ref{fig:strong}, the analytical and numerical results agree remarkably well in the strong-interaction regime, validating the approximation scheme used in this work.

It can be seen from Eq. ~\eqref{B_epsilon} and \eqref{B_phi}  that $B_{\epsilon_n}$($B_{\phi_n}$ ) are associated with the charge excitation(spin excitation). In the limit of infinite interaction strength, the Drude weight under consideration solely originates from charge excitations, while the dominant contribution to spin excitations arises from the $1/c$ correction term.

\section{Drude Weight from Bipartition Setup}

We now consider the bipartition protocol with a small chemical potential bias. Following the derivation logic of the Lieb-Liniger model\ucite{4}, we present the derivation process for the two-component Fermi gas. In this scenario, the system is prepared in an inhomogeneous initial state composed of two homogeneous, equilibrated macroscopic regions brought into contact at time $t=0$. Specifically, we impose a small chemical potential drop at the origin: $\mu^{L} = \mu + \delta\mu/2$ for the left reservoir and $\mu^{R} = \mu - \delta\mu/2$ for the right reservoir, while keeping other potentials fixed. The subsequent time evolution follows GHD equations (Eqs.~\eqref{eq:GHD}.

Within the bipartition framework, the GHD equations yield solutions for the occupation numbers along each characteristic ray $\zeta = x/t$:
\begin{equation}
\theta_i(k) = \theta_i^L(k) \Theta(v_i(k) - \zeta) + \theta_i^R(k) \Theta(\zeta - v_i(k)), \quad i=1 \text{ and } 2,l,
\label{eq:thetai}
\end{equation}
where $\theta_i^L(k)$ and $\theta_i^R(k)$ represent the occupation numbers in the left and right equilibrium reservoirs, respectively, and $\Theta(z)$ is the Heaviside step function.

The Drude weight can be defined as the asymptotic current growth rate in the limit of vanishing bias $\delta\mu$:
\begin{equation}\label{Dnj_bias_mu}
D_{nj} = \lim_{\delta\mu \rightarrow 0} \frac{\partial}{\partial \delta\mu} \lim_{t \rightarrow \infty} \frac{J_j(t)}{t} = \lim_{\delta\mu \rightarrow 0} \frac{\partial}{\partial \delta\mu} \int \mathrm{d}\zeta\, j_j(\zeta) = \lim_{\delta\mu \rightarrow 0} \frac{\int \mathrm{d}\zeta\, j_j(\zeta)}{\delta\mu},
\end{equation}
where the local current $j_j(\zeta)$ is defined by Eqs.~\eqref{eq:jn} and \eqref{eq:je0}.

Following established methods\ucite{4}, we demonstrate that this definition of the Drude weight is consistent with the integral form derived previously. The local current can also be expressed as:
\begin{equation}
j_j(\zeta) = \frac{1}{2\pi} \int \mathrm{d}k\, \frac{k}{m} \theta_1 (\vec{q}_j^{\mathrm{dr}})_1.
\label{eq:jj}
\end{equation}
Using the identity
\begin{equation}
\partial_a \sum_i \int \mathrm{d}k\, f_i \theta_i (\vec{q}_j^{\mathrm{dr}})_i = \sum_i \int \mathrm{d}k\, f_i^{\mathrm{dr}} \partial_a \theta_i (\vec{q}_j^{\mathrm{dr}})_i,
\end{equation}
which holds for arbitrary functions $f_i$ and conjugate variable $a$, the derivative of the local current with respect to the chemical potential bias becomes:
\begin{equation}
\frac{\partial}{\partial \delta\mu} j_j(\zeta) = \frac{1}{2\pi m} \sum_i \int \mathrm{d}k\, (\vec{q}_k^{\mathrm{dr}})_i \partial_{\delta\mu} \theta_i(k) (\vec{q}_j^{\mathrm{dr}})_i.
\label{eq:jmu}
\end{equation}

In the limit $\delta\mu \rightarrow 0$, the derivative of the occupation number for each ray $\zeta$ can be expressed in terms of equilibrium quantities:
\begin{equation}
\partial_{\delta\mu} \theta_i = \partial_{\delta\mu} \theta_i^L \Theta(v_i - \zeta) + \partial_{\delta\mu} \theta_i^R \Theta(\zeta - v_i).
\end{equation}
For an equilibrium state, we have $\partial_{\delta\mu} \theta_i^{L,R} = -\theta_i^{L,R}(1 - \theta_i^{L,R}) \partial_{\delta\mu} \epsilon_i^{L,R}/T$, with $\partial_{\delta\mu} \epsilon_i^L/T = \frac{1}{2T} \partial_{\mu^L} \epsilon_i^L$ and $\partial_{\delta\mu} \epsilon_i^R/T = -\frac{1}{2T} \partial_{\mu^R} \epsilon_i^R$. As $\delta\mu \rightarrow 0$, $\theta_i^L(k) = \theta_i^R(k) = \theta_i(k)$, leading to:
\begin{equation}
\partial_{\delta\mu} \theta_i = \frac{1}{T} \theta_i (1 - \theta_i) \frac{1}{2} \partial_\mu \epsilon_i \, \mathrm{sgn}(\zeta - v_i),
\end{equation}
where $\mathrm{sgn}(z)$ is the sign function. Substituting this result into Eq.~\eqref{eq:jmu} yields:
\begin{equation}
\frac{\partial}{\partial \delta\mu} j_j(\zeta) = \frac{1}{2T} \sum_i \int \mathrm{d}k\, g_i \, \mathrm{sgn}(\zeta - v_i),
\label{eq.j_deltamu}
\end{equation}
with
\[
g_i = \frac{1}{2\pi m} (\vec{q}_k^{\mathrm{dr}})_i \theta_i (1 - \theta_i) \partial_\mu \epsilon_i (\vec{q}_j^{\mathrm{dr}})_i = v_i \rho_i (1 - \theta_i) \partial_\mu \epsilon_i (\vec{q}_j^{\mathrm{dr}})_i.
\]
Using the integral identity $\int_{-\infty}^{\infty} d\zeta\, \mathrm{sgn}(\zeta - v_i) = -2v_i$, we obtain the Drude weight:
\begin{equation}
D_{nj} = -\frac{1}{T} \sum_i \int \mathrm{d}k\, g_i v_i,
\end{equation}
which is consistent with Eqs.~\eqref{eq:Dnn}, \eqref{eq:Dnm}, and \eqref{eq:Dne0} upon recognizing that $-\partial_\mu \epsilon_i = (\vec{q}_n^{\mathrm{dr}})_i$.

\begin{figure}
\centering
\includegraphics[scale=0.5]{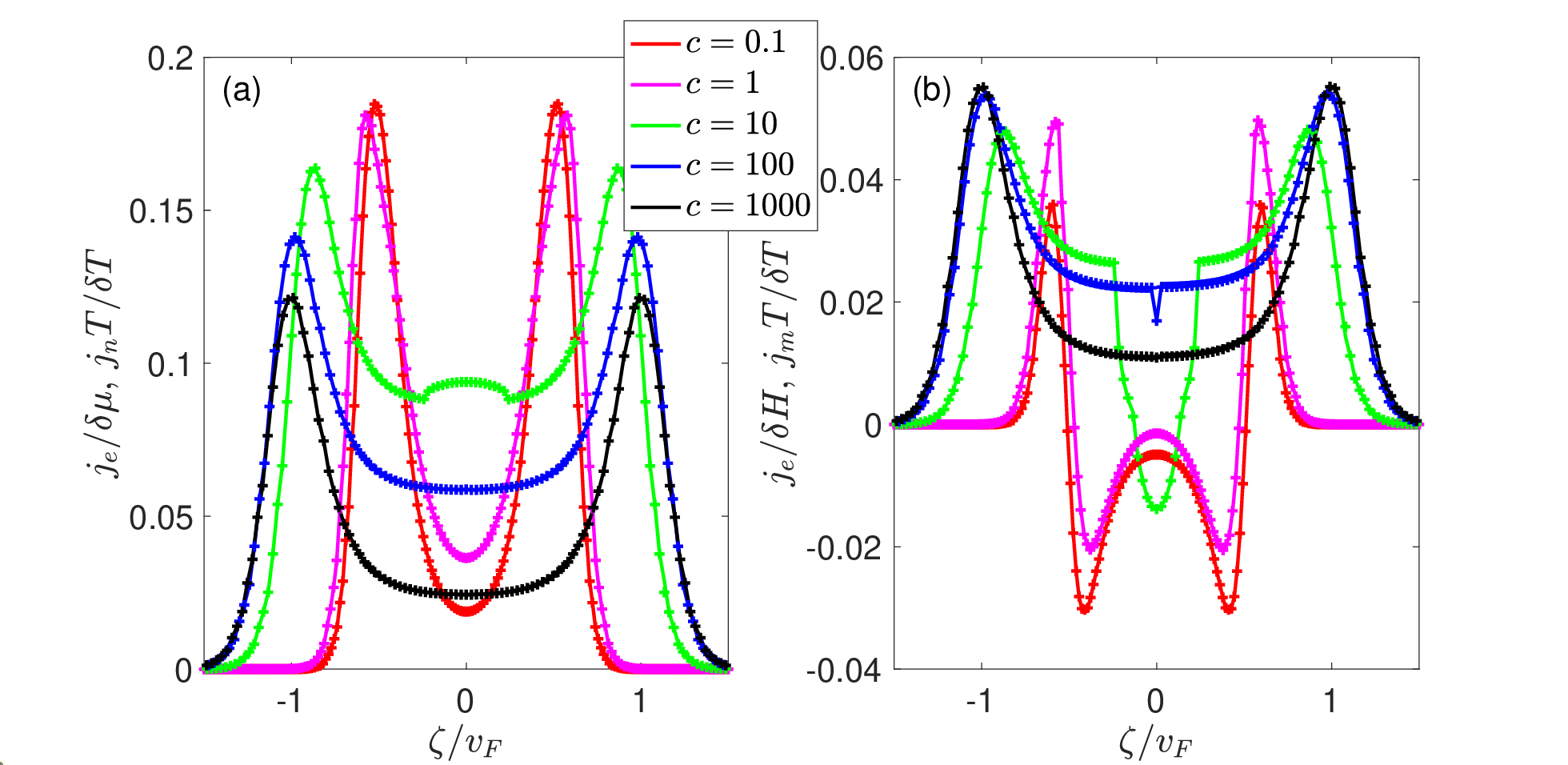}
\caption{\label{fig:current}Currents of energy, particle density, and magnetization density as functions of ray $\zeta$ for different interaction strengths $c$.
Panel (a) shows the energy current (solid lines) under a bias in chemical potential with $\delta\mu=0.05$,
while panel (b) shows the energy current (solid lines) under a bias in magnetic field with $\delta H=0.05$.
The particle density current $j_{n}$ and magnetization density current $j_{m}$ (plus markers) are obtained from a temperature bias setup with $\delta T=0.01$.
Other parameters are fixed as $T=1$, $\mu=5$, $H=2$, and $m=1$.}
\end{figure}

This approach can be extended to other biases. For a magnetic field bias with $H^L = H + \delta H/2$ and $H^R = H - \delta H/2$ (while keeping other potentials fixed), the spin-related Drude weight is defined as:
\begin{equation}
D_{mj} = \lim_{\delta H \rightarrow 0} \frac{\partial}{\partial \delta H} \lim_{t \rightarrow \infty} \frac{J_j(t)}{t} = \lim_{\delta H \rightarrow 0} \frac{\partial}{\partial \delta H} \int \mathrm{d}\zeta\, j_j(\zeta) = \lim_{\delta H \rightarrow 0} \frac{\int \mathrm{d}\zeta\, j_j(\zeta)}{\delta H},
\label{eq:Dmij}
\end{equation}
which is consistent with the integral form in Eq.~\eqref{eq:Dmi}.

For a temperature bias with $T^L = T + \delta T/2$ and $T^R = T - \delta T/2$, the key difference lies in the derivative of $\epsilon_i/T$:
\begin{equation}
\partial_{\delta T} \frac{\epsilon_i^L(k)}{T^L} = -\frac{1}{2T^2} \partial_{1/T^L} \frac{\epsilon_i^L(k)}{T^L} = -\frac{1}{2T^2} (\vec{q}_e^{\mathrm{dr}})_i,
\end{equation}
\begin{equation}
\partial_{\delta T} \frac{\epsilon_i^R(k)}{T^R} = \frac{1}{2T^2} \partial_{1/T^R} \frac{\epsilon_i^R(k)}{T^R} = \frac{1}{2T^2} (\vec{q}_e^{\mathrm{dr}})_i,
\end{equation}
introducing an additional factor of $1/T$ compared to the chemical potential case. The energy-related Drude weight is then defined as:
\begin{equation}
D_{ej} = T \lim_{\delta T \rightarrow 0} \frac{\partial}{\partial \delta T} \lim_{t \rightarrow \infty} \frac{J_j(t)}{t} = T \lim_{\delta T \rightarrow 0} \frac{\int \mathrm{d}\zeta\, j_j(\zeta)}{\delta T},
\end{equation}
which agrees with the integral form in Eq.~\eqref{eq:Dne}.

As shown in the numerical results of Fig.~\ref{fig:current}, the normalized energy current $j_{e}/\delta\mu$ under a chemical potential bias agrees well with the normalized particle current $j_{n}T/\delta T$ under a temperature bias in the small-bias limit, namely, 
\begin{equation}
\lim_{\delta \mu \rightarrow 0}\frac{\partial}{\partial \delta\mu} j_e(\zeta) = T\lim_{\delta T \rightarrow 0}\frac{\partial}{\partial \delta T} j_n(\zeta).
\label{eq.j_relation1}
\end{equation}
This relationship can also be derived by comparing the derivative of the current in Eq.~\eqref{eq.j_deltamu} for the biased chemical potential configuration with that for the corresponding biased temperature configuration. Similarly, the normalized energy current $j_{e}/\delta H$ under a magnetic field bias matches the normalized magnetization current $j_{m}T/\delta T$ induced by a temperature bias when the bias is sufficiently small, namely, 
\begin{equation}
\lim_{\delta H \rightarrow 0}\frac{\partial}{\partial \delta H} j_e(\zeta) = T\lim_{\delta T \rightarrow 0}\frac{\partial}{\partial \delta T} j_m(\zeta).
\label{eq.j_relation2}
\end{equation}
Given that these current integrals are related to the Drude weight, their agreement implies the equivalence of the corresponding Drude weights.  

\begin{figure}
\centering
\includegraphics[scale=0.5]{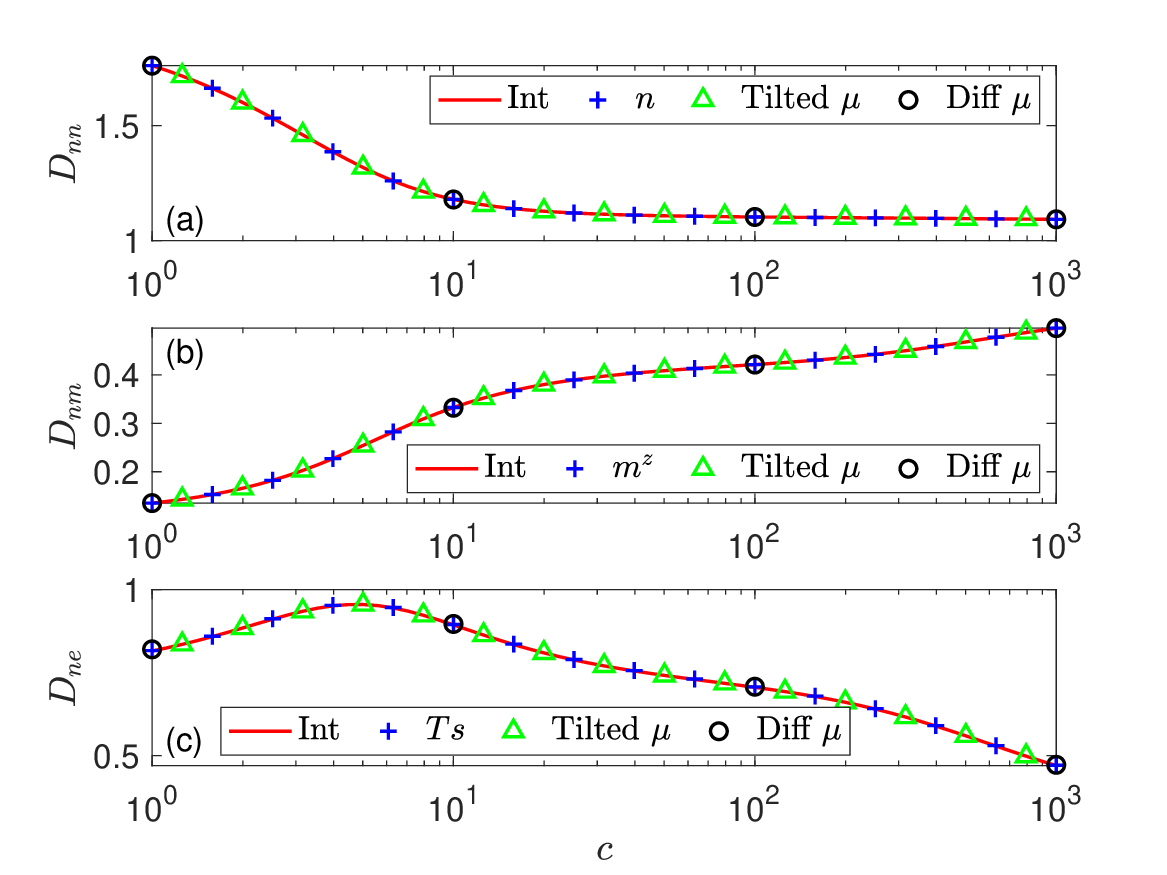}
\caption{\label{fig:Dn} Comparison of charge-related Drude weights among different methods. Here $c$ denotes the interaction strength. The red solid lines correspond to results from the integral form \eqref{eq:Dij} in rapidity space, the blue triangles to the tilted chemical potential protocol with \eqref{eq:Dni_tilted_mu}, and the black circles to the partitioning protocol with two different chemical potentials \eqref{Dnj_bias_mu}. The green pluses represent particle density (a), magnetization density (b), and the product of temperature and entropy density (c), respectively. Other parameters are fixed at $T=1$, $\mu=5$, $H=2$, and $m=1$.}
\end{figure}

\begin{figure}
\centering
\includegraphics[scale=0.5]{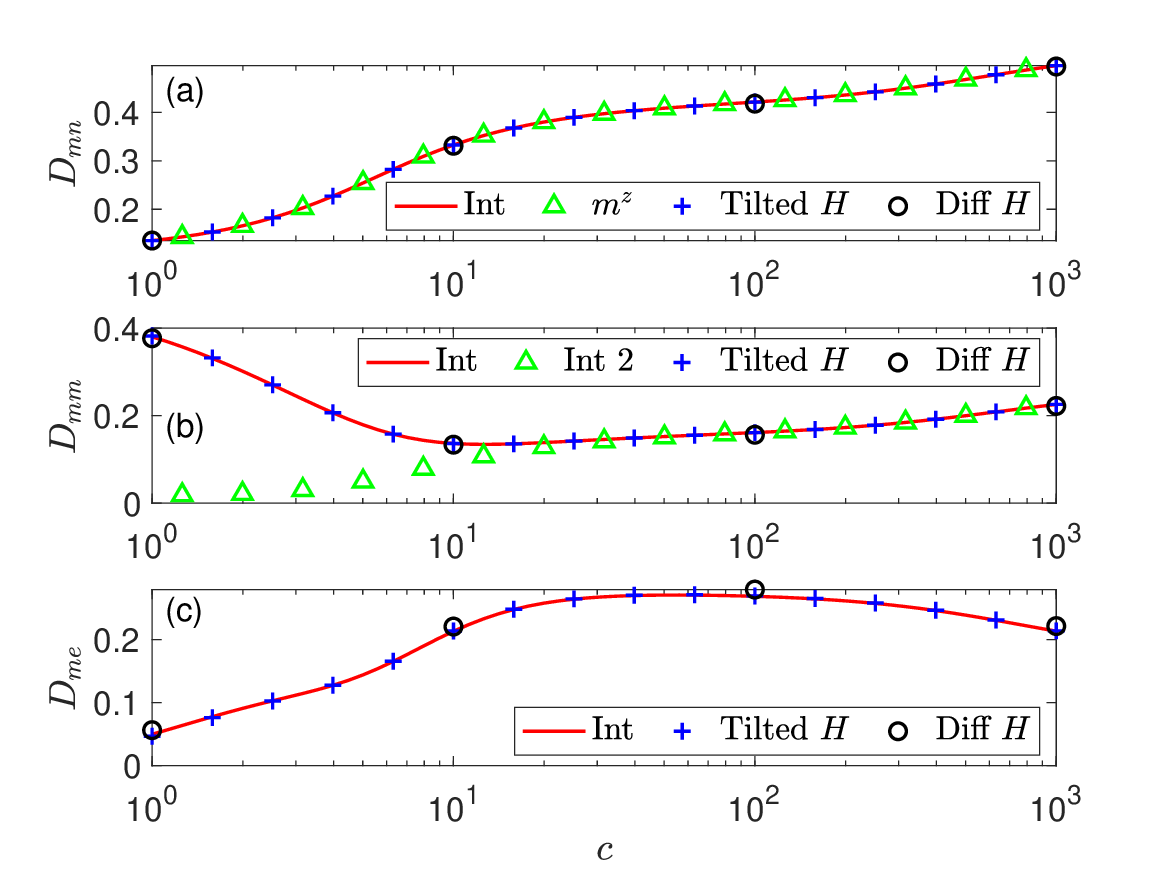}
\caption{\label{fig:Dm} Comparison of spin-related Drude weights among different methods. Here $c$ denotes the interaction strength. The red solid lines represent results from the integral form \eqref{eq:Dij} in rapidity space, the blue pluses correspond to the tilted magnetic field protocol, and the black circles to the partitioning protocol with two different magnetic fields \eqref{eq:Dmij}. The green triangles indicate the magnetization density (a) and the integral expression $ \int \mathrm{d}k\, \rho_1 (\vec{q}_m^{\mathrm{dr}})_1^2 /m (\vec{q}_n^{\mathrm{dr}})_1^2$ (b), respectively. Other parameters are fixed at $T=1$, $\mu=5$, $H=2$, and $m=1$.}
\end{figure}

\section{Numerical Check}

Through numerical calculations, Figure 4 demonstrates that the charge-related Drude weights from the integral form in rapidity space are consistent with  the thermodynamic expressions: \( D_{nn} = n/m \), \( D_{nm} = m^z/m \), and \( D_{ne} = Ts/m \). Additionally, the results obtained from the tilted chemical potential protocol and the partitioning protocol with two different chemical potentials also agree with the integral form.

 Figure 5 demonstrates that the spin-related Drude weights obtained from the tilted magnetic field protocol and the partitioning protocol with two different magnetic fields also agree with the integral form. The integral expression $D_{mm}=\int \mathrm{d}k\, \rho_1 (\vec{q}_m^{\mathrm{dr}})_1^2 /m (\vec{q}_n^{\mathrm{dr}})_1^2$  provides an accurate description for strong interaction strengths.

\end{document}